\documentclass[reprint,amsmath,amssymb,prd,nofootinbib]{revtex4-2}

\usepackage{xcolor}
\usepackage{graphicx}

\usepackage{url}
\usepackage{xurl}

\usepackage{color,soul}

\usepackage[colorlinks = true,
            linkcolor = blue,
            urlcolor  = blue,
            citecolor = blue,
            anchorcolor = blue]{hyperref}

\usepackage{bm}
\usepackage{braket}
\usepackage{amsmath}
\usepackage{amsfonts}
\DeclareMathOperator{\Tr}{Tr}

\usepackage{tabularx}
\usepackage{multirow}
\newcolumntype{Y}{>{\centering\arraybackslash}X}

\def\xbf{\mathbf{x}}
\def\thetabf{\boldsymbol{\theta}}
\def\UENC{U_{\text{ENC}}}
\def\UPARAM{U_{\text{PARAM}}}

\def\ZNR{Z^\text{NR}}
\def\IIR{I^\text{IR}}
\def\XIR{X^\text{IR}}

\def\I22{\left[\begin{matrix}1 & 1\\ 1 & 1\end{matrix}\right]}

\usepackage{xspace}
\newcommand{\PennyLane}{\textit{PennyLane}\xspace}
\newcommand{\Top}{\textsc{Top}\xspace}
\newcommand{\JetNet}{\textsc{JetNet}\xspace}

\urldef{\urlgithub}\url{https://github.com/NTUHEP-QML/QCGNN}
\urldef{\urlamplitude}\url{https://pennylane.ai/qml/glossary/quantum_embedding/#amplitude-embedding}
\urldef{\urlmpgnn}\url{https://pytorch-geometric.readthedocs.io/en/latest/tutorial/create_gnn.html}
\urldef{\urlibmqcal}\url{https://drive.google.com/drive/folders/17FiHB1bkYkCIC4Nvx3nyT3HkWwpi7WKE?usp=share_link}
\urldef{\urllightning}\url{https://pennylane.ai/blog/2022/07/lightning-fast-simulations-with-pennylane-and-the-nvidia-cuquantum-sdk/}

\begin{document}

\date{\today}

\title{Jet Discrimination with Quantum Complete Graph Neural Network}

\author{Yi-An Chen}
\email{maplexworkx0302@gmail.com}
\author{Kai-Feng Chen}
\affiliation{Department of Physics, National Taiwan University, Taipei, Taiwan}

\begin{abstract}
    Machine learning, particularly deep neural networks, has been widely used in high-energy physics, demonstrating remarkable results in various applications. Furthermore, the extension of machine learning to quantum computers has given rise to the emerging field of quantum machine learning. In this paper, we propose the Quantum Complete Graph Neural Network (QCGNN), which is a variational quantum algorithm based model designed for learning on complete graphs. QCGNN with deep parametrized operators offers a polynomial speedup over its classical and quantum counterparts, leveraging the property of quantum parallelism. We investigate the application of QCGNN with the challenging task of jet discrimination, where the jets are represented as complete graphs. Additionally, we conduct a comparative analysis with classical models to establish a performance benchmark. The code is available at \urlgithub
\end{abstract}

\maketitle

\section{Introduction}

The proton-proton collisions at the Large Hadron Collider (LHC) produce jets from hard scattering events. Jets are collimated sprays of particles formed through the hadronization of elementary particles. Jet discrimination, i.e., identifying the type of elementary particle that initiates the jet, is one of the most challenging tasks in particle physics.

Deep neural networks (DNNs), celebrated for their architectural flexibility and expressive power, have been widely adopted in high-energy physics (HEP) \cite{hepmllivingreview,hepml1,hepml2}. Designing a DNN model tailored for jet discrimination poses a significant challenge due to the variable number of constituent particles within jets. Various data representations and DNN models have been proposed for jet discrimination, including images \cite{2pcnn,jet_image1,jet_image2,jet_image3,jet_image4,jet_image5,jet_image6,jet_image7}, sequences \cite{jet_seq1,jet_seq2,jet_seq3,jet_seq4,jet_seq5,jet_seq6,jet_seq7}, trees \cite{jet_tree1,jet_tree2}, graphs \cite{jet_graph1,jet_graph2,jet_graph3,jet_graph4,jet_graph5,jet_graph6,jet_graph7,jet_graph8,jet_graph9,jet_graph10,jet_graph11,jet_graph12,jet_graph13}, and sets \cite{pfn,ptcnet,part,jet_set1,jet_set2,jet_set3,jet_set4,jet_set5,jet_set6,jet_set7,jet_set8}. Jet images are typically two-dimensional representations (e.g., pseudorapidity $\eta$ versus azimuthal angle $\phi$) where particle information is encoded in a discretized two-dimensional grid. Sequences or trees order particles according to specific criteria (e.g., by transverse momentum or distance parameter). Despite the simplicity of these representations, they often lose information about individual particles, lack translational or rotational invariance, or disregard permutation invariance. To preserve the information of each constituent particle and the relevant symmetries, graphs or sets are widely used for jet representation, with each particle represented as a node in a graph or an element of a set.

In the upcoming high-luminosity LHC (HL-LHC), the data volume is expected to increase by several orders of magnitude compared to the LHC. The increased luminosity and event complexity due to pile-up will make data analysis even more challenging. Consequently, efficient methodologies and novel technologies in data analysis, such as parallel computing and machine learning, are in high demand. Furthermore, quantum computing has made significant strides in recent decades, leading to the development of quantum machine learning (QML) \cite{qml1,qml2,qml3,qml4,qml5}. QML leverages the unique properties of quantum systems, such as superposition and entanglement, to potentially achieve learning capabilities unattainable with classical computers. QML has been explored in several HEP analyses \cite{qml_hep1}, including reconstruction \cite{qml_reco1,qml_reco2,qml_reco3}, classification \cite{qml_cf1,qml_cf2,qml_cf3,qml_cf4,qml_cf5,qml_cf6,qml_cf7,qml_cf8,qml_cf9,qml_cf10,qml_cf11,qml_cf12,qml_cf13,qml_cf14}, anomaly detection \cite{qml_ad1,qml_ad2,qml_ad3,qml_ad4,qml_ad5}, and data generation \cite{qml_gen1,qml_gen2,qml_gen3,qml_gen4,qml_gen5}.

In this paper, we introduce the Quantum Complete Graph Neural Network (QCGNN), a variational quantum algorithm \cite{vqc1,vqc2,vqc3} based model specifically designed for learning complete graphs \cite{completegraph}. For QCGNN with deep parametrized operators, it has a polynomial speedup over its classical counterparts by utilizing the property of quantum parallelism. The application of QCGNN is studied through jet discrimination using two public datasets: the \Top dataset \cite{zenodo_top,dataset_top} for binary classification and the \JetNet dataset \cite{zenodo_jetnet,dataset_jetnet} for multi-class classification.

The structure of this paper is as follows: In Sec.~\ref{sec_method}, we describe the architectures of QCGNN and the classical graph neural networks used for benchmarking, as well as discuss the computational complexity of learning with classical and quantum models. Sec.~\ref{sec_setup} details the experimental setup for jet discrimination, with the results presented in Sec.~\ref{sec_result}. Finally, we summarize our findings in Sec.~\ref{sec_summary}.

\section{Methodology} \label{sec_method}

\subsection{Graph Neural Network} \label{sec_gnn}
Graphs are ubiquitous data structures that represent relationships and connections between entities. Analyzing and extracting valuable information from graph-structured data is a fundamental challenge in modern data science and machine learning. To address this challenge, Graph Neural Networks (GNNs) have emerged as a powerful and versatile framework for learning from graph-structured data \cite{gnnex1,gnnex2,gnnex3}.

A graph $G$ is described by its set of nodes and edges. Let $N$ denote the number of nodes, and $E_{ij}$ the edge from the $i$-th node to the $j$-th node. Throughout this paper, the graphs are assumed to be \textit{undirected} ($E_{ij}=E_{ji}$) and \textit{unweighted} (all edges have equal weights). Furthermore, a graph is considered \textit{complete} \cite{completegraph} if all pairs of nodes are connected. A complete graph that is undirected and unweighted can also be viewed as a set. Each node is associated with a feature vector $\xbf\in \mathbb{R}^{d}$, where $d$ is the dimension of the features.

Graphs are permutation invariant, meaning that reordering the indices of the nodes does not alter the information contained in the graph. The general structure of permutation-invariant models has been studied in \cite{deepset,pfn}. In this work, we focus on a specific type of GNN, the Message Passing Graph Neural Network (MPGNN) \cite{mpn1}, which provides a simple and intuitive approach to designing GNNs. The MPGNN is constructed by iterating the following formula\footnote{The formula is adapted from the PyTorch Geometric documentation \urlmpgnn. The $e_{j,i}$ term is omitted here as we only consider undirected and unweighted graphs.}
\begin{equation} \label{eqmp}
    \xbf^{(k)}_i=\gamma^{(k)}\left(\xbf^{(k-1)}_i,\bigoplus_{\substack{j\in \mathcal{N}(i)}}\Phi^{(k)}(\xbf^{(k-1)}_i,\xbf^{(k-1)}_j)\right),
\end{equation}
where $\Phi^{(k)}$ extracts information from the neighboring nodes $\mathcal{N}(i)$ of the $i$-th node, and $\gamma^{(k)}$ updates the node features in the $k$-th iteration. As pointed out in \cite{deepset}, summation over a set is the sufficient and necessary condition for satisfying permutation invariance. This concept can be generalized by the aggregation function $\bigoplus$, which is typically chosen as \textbf{SUM}, \textbf{MEAN}, \textbf{MAX}, or \textbf{MIN}\footnote{Note that \textbf{MAX} and \textbf{MIN} can be expressed as summation over the $p$-norm, which can be absorbed into $\Phi$.}, among others. A brief discussion on the relation between state-of-the-art models and the MPGNN is provided in Appendix \ref{app_mpgnn}.

\subsection{Quantum Complete Graph Neural Network} \label{secqcgnn}

\begin{figure*}[htbp]
    \centering
    \includegraphics[width=\textwidth]{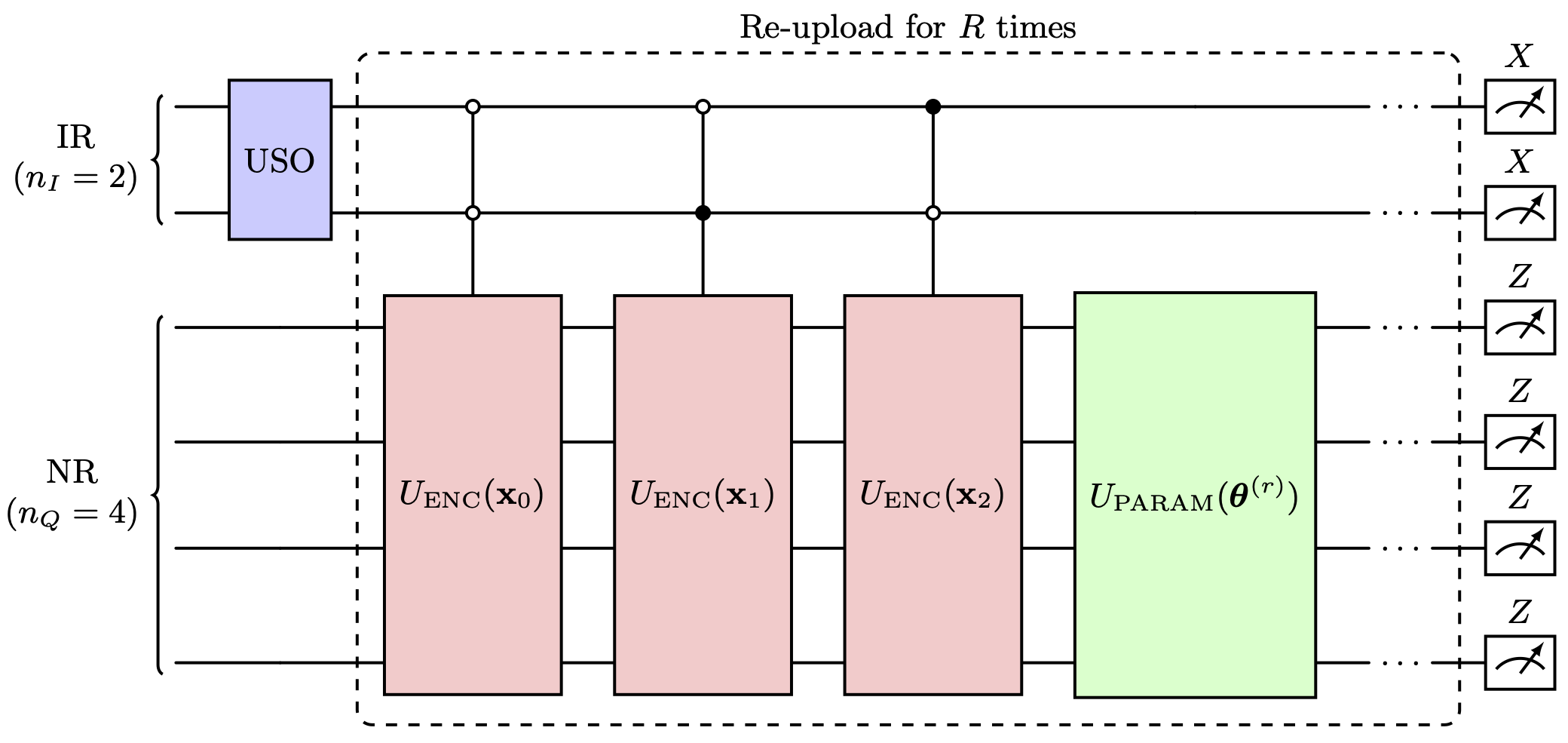}
    \caption{Example of a QCGNN ansatz for learning a 3-node complete graph, with $n_I=2$ and $n_Q=4$. The quantum state is initially prepared in a uniform quantum state as described in Eq.~\ref{psi_0} via a uniform state oracle (blue block). The dashed box contains the encoding (red blocks) and parametrized operators (green blocks), which may be re-uploaded $R$ times \cite{reupload}. The encoding operators consist of multi-controlled operators that correspond to Eq.~\ref{mc_gates}, transforming the quantum state to Eq.~\ref{psi_1}, or, in this example, Eq.~\ref{psi_1_ex3}. The empty ($\circ$) and filled ($\bullet$) circles in the IR represent the controlled conditions for the controlled qubits: the encoding operators are applied if the corresponding qubit in IR is in state $\ket{0}$ and $\ket{1}$, respectively. Since the encoding  operators are operated with different controlled conditions, the red controlled encoding blocks are mutually commutable, i.e., the quantum state stays the same if one first encode $\xbf_1$ then $\xbf_0$ with their corresponding controlled condition. The quantum state then evolves to Eq.~\ref{psi_2} through the parametrized operators. If the dashed box is re-uploaded $R$ times, the quantum state evolves to Eq.~\ref{psi_final}. Finally, the qubits in the IR are measured in the $X$ basis as described in Eq.~\ref{Jdecompose}, while the NR qubits are measured using Pauli-$Z$ observables to calculate the result in Eq.~\ref{eq_JP}. The details of the encoding and parametrized operators used are provided in Sec.~\ref{sec_cq_model_setup}.}
    \label{fig:exqcgnn}
\end{figure*}

In the noisy intermediate-scale quantum (NISQ) era \cite{nisq}, variational quantum algorithms \cite{vqc1,vqc2,vqc3} present an intuitive approach to implementing quantum neural networks. These networks utilize a variational quantum circuit (VQC) with tunable parameters updated via classical optimization routines. Typically, an $n$-qubit VQC can be expressed as:
\begin{equation*}
    f(\xbf,\thetabf) = \bra{0}^{\otimes n}U^\dagger(\xbf,\thetabf)PU(\xbf,\thetabf)\ket{0}^{\otimes n}
\end{equation*}
for some Pauli string observable $P$ and unitary operator $U$. The unitary operator encodes the data $\xbf$ into the quantum state and includes several tunable parameters $\thetabf$. These parameters are optimized via gradient descent \cite{psr1,psr2,psr3,psr4} using an appropriate loss function. Typically, $U$ can be divided into encoding and parametrized operators, denoted as $\UENC$ and $\UPARAM$, respectively.

The QCGNN architecture comprises two qubit registers: an index register (IR) and a neural network register (NR), with $n_I$ and $n_Q$ qubits, respectively. Fig.~\ref{fig:exqcgnn} illustrates an example of a QCGNN circuit with $n_I=2$ and $n_Q=4$. To encode the information of an undirected and unweighted complete graph with $N$ nodes, we set $n_I=\lceil \log_2(N)\rceil$. For graphs with different number of nodes, $n_I$ could be different, e.g., a 4-particle jet needs $n_I=2$ qubits, while a 5-particle jet requires $n_I=3$. We will show that the output of QCGNN justifies that QCGNN can handle variable size of graphs. The quantum state in the IR is initialized to uniform basis states through a uniform state oracle (USO) and evolves as
\begin{equation} \label{psi_0}
    \ket{\psi_0}=\frac{1}{\sqrt{N}}\sum^{N-1}_{i=0}\ket{i}\ket{0}^{\otimes n_Q}, 
\end{equation}
where $\ket{0}^{\otimes n_Q}$ is the initial quantum state in the NR with all qubits in $\ket{0}$ state. The decimal representation, e.g., $\ket{0}$, $\ket{1}$, $\ket{2}$, $\ket{3}$, is used here, equivalent to the binary representation $\ket{00}$, $\ket{01}$, $\ket{10}$, $\ket{11}$. The context should clarify whether the decimal or binary representation is being used. If $N$ is a power of 2, i.e., $n_I=\log_2 N$, the USO can be constructed using Hadamard gates. Otherwise, other methods described in \cite{uso1,uso2} can be employed.

The node features $\xbf_i$ of the $i$-th node are encoded through a series of unitary operators $\UENC(\xbf_i)$, controlled by the corresponding basis state in IR, where the control condition is the binary representation of $i$ with $n_I$ digits, denoted as $C^{n_I}_{\text{bin}(i)} \Bigl( \UENC(\xbf_i) \Bigr)$. This controlled operator acts on the quantum state as follows:
\begin{equation} \label{mc_gates}
    \begin{split}
        C^{n_I}_{\text{bin}(i)} & \Bigl( \UENC(\xbf_i) \Bigr) \Bigl[ \ket{j}\ket{0}^{\otimes n_Q} \Bigr] = \\
        &\delta_{ij}\ket{j}\UENC(\xbf_i)\ket{0}^{\otimes n_Q} + (1-\delta_{ij})\ket{j}\ket{0}^{\otimes n_Q},
    \end{split}
\end{equation}
where $\delta_{ij}$ is the Kronecker delta. The controlled operator $C^{n_I}_{\text{bin}(i)} \Bigl( \UENC(\xbf_i) \Bigr)$ on the left-hand side acts on both the IR and NR, while $\UENC(\xbf_i)$ on the right-hand side acts only on the NR. After encoding the information of all $N$ nodes, the quantum state evolves as
\begin{equation} \label{psi_1}
    \begin{split}
        \ket{\psi_0} &\rightarrow
        \ket{\psi_1} = \Biggl[\bigotimes^{N-1}_{i=0}C^{n_I}_{\text{bin}(i)} \Bigl( \UENC(\xbf_i) \Bigr)\Biggr]\ket{\psi_0} \\
        &= \frac{1}{\sqrt{N}}\sum^{N-1}_{j=0}\Biggl[\bigotimes^{N-1}_{i=0}C^{n_I}_{\text{bin}(i)} \Bigl( \UENC(\xbf_i) \Bigr)\Biggr] \Bigl( \ket{j}\ket{0}^{\otimes n_Q} \Bigr) \\
        &= \frac{1}{\sqrt{N}}\sum^{N-1}_{i=0}\ket{i}\UENC(\xbf_i)\ket{0}^{\otimes n_Q}, \\
    \end{split}
\end{equation}
where it is used that $\ket{j}$ must be acted upon by one of the $C^{n_I}_{\text{bin}(i)} \Bigl( \UENC(\xbf_i) \Bigr)$ as $i$ ranges from $0$ to $N-1$. For example, in Fig.~\ref{fig:exqcgnn}, where $N=3$, $n_Q=4$, and $n_I=\lceil \log_2 N \rceil=2$, the quantum state $\ket{\psi_1}$ can be expressed as
\begin{equation} \label{psi_1_ex3}
    \begin{split}
        \ket{\psi^{N=3}_1} &= \frac{1}{\sqrt{3}} \biggr[ \ket{0}\UENC(\xbf_0)\ket{0000} \\
        &+\ket{1}\UENC(\xbf_1)\ket{0000} + \ket{2}\UENC(\xbf_2)\ket{0000} \biggr] ,
    \end{split}
\end{equation}
where $\ket{0000}$ is equivalent to $\ket{0}^{\otimes 4}$. At first glance, Eq.~\ref{psi_1} and Eq.~\ref{psi_1_ex3} appear to depend on the node ordering. However, as we will demonstrate, by selecting the appropriate observables, the final output of QCGNN is permutation invariant.

Next, a series of parametrized unitary operators $\UPARAM(\thetabf)$ with tunable parameters $\thetabf$ are applied to the NR, evolving the quantum state to
\begin{equation} \label{psi_2}
    \begin{split}
        \ket{\psi_1} \rightarrow
        \ket{\psi_2} &= \UPARAM(\thetabf)\ket{\psi_1} \\
        &= \frac{1}{\sqrt{N}}\sum^{N-1}_{i=0}\ket{i}\UPARAM(\thetabf)\UENC(\xbf_i)\ket{0}^{\otimes n_Q}.
    \end{split}
\end{equation}

To increase the expressive power of VQCs, the \textit{data re-uploading technique} \cite{reupload} can be employed. The idea of data re-uploading is to encode the data multiple times, allowing the quantum state to interact with the data in a more complex manner. The data re-uploading technique can be implemented by alternately applying the encoding and parametrized operators several times, but with different parameters. As the encoding operators in $\ket{\psi_2}$ correspond to the node indices $\ket{i}$ in the IR, the data re-uploading technique can be implemented straightforwardly without making each particle information entangled. After re-uploading $R$ times, the final quantum state $\ket{\psi}$ evolves as
\begin{equation} \label{psi_final}
    \ket{\psi} = \frac{1}{\sqrt{N}}\sum^{N-1}_{i=0}\ket{i}\Biggl[\bigotimes^{R}_{r=1}\UPARAM(\thetabf^{(r)})\UENC(\xbf_i)\Biggr]\ket{0}^{\otimes n_Q}\\,
\end{equation}
where the parameters $\thetabf$ may differ across each $\UPARAM$ operator, denoted by the superscript $r$. We denote the quantum state of the NR as
\begin{equation*}
    \ket{\xbf_i,\thetabf}=\Biggl[\bigotimes^{R}_{r=1}\UPARAM(\thetabf^{(r)})\UENC(\xbf_i)\Biggr]\ket{0}^{\otimes n_Q}.
\end{equation*}

Given a Pauli string observable $P$ applied to the NR, the expectation value of the measurement is given by
\begin{equation} \label{eq_self}
    \begin{split}
        \braket{\psi|P|\psi}&=\frac{1}{N}\sum^{N-1}_{i=0}\sum^{N-1}_{j=0}\braket{i|j}\bra{\xbf_i, \thetabf}P\ket{\xbf_j, \thetabf}\\
        &=\frac{1}{N}\sum^{N-1}_{i=0}\sum^{N-1}_{j=0}\delta_{ij}\bra{\xbf_i, \thetabf}P\ket{\xbf_j, \thetabf}\\
        &=\frac{1}{N}\sum^{N-1}_{i=0}\bra{\xbf_i, \thetabf}P\ket{\xbf_i, \thetabf},
    \end{split}
\end{equation}
which results in a sum over each node. Now, consider an additional $2^{n_I}\times2^{n_I}$ Hermitian matrix $J$ filled with ones, used as an observable of the IR. Observing that
\begin{equation}\label{eq_JP}
    \begin{split}
        \braket{\psi|J\otimes P|\psi} &= \frac{1}{N}\sum^{N-1}_{i=0}\sum^{N-1}_{j=0}\bra{i}J\ket{j}\bra{\xbf_i, \thetabf}P\ket{\xbf_j, \thetabf} \\
        &= \frac{1}{N}\sum^{N-1}_{i=0}\sum^{N-1}_{j=0}\bra{\xbf_i, \thetabf}P\ket{\xbf_j, \thetabf} \\
        &= \frac{1}{N}\sum^{N-1}_{i=0}\sum^{N-1}_{j=0} h(\xbf_i, \xbf_j; P),
    \end{split}
\end{equation}
where $\bra{i}J\ket{j}=J_{ij}=1$, and we define
\begin{equation*}
    h(\xbf_i, \xbf_j; P) = \frac{1}{2} \Bigl[ \bra{\xbf_i, \thetabf}P\ket{\xbf_j, \thetabf} + \bra{\xbf_j, \thetabf}P\ket{\xbf_i, \thetabf} \Bigr],   
\end{equation*}
which is symmetric, i.e., $h(\xbf_i, \xbf_j; P)=h(\xbf_j, \xbf_i; P)$. Notice that Eq.~\ref{eq_JP} computes the average value over all possible pairs, leading to the \textbf{permutation invariance} of the final output. In practice, the observable $J$ can be decomposed as
\begin{equation}\label{Jdecompose}
    J=\bigotimes^{n_I}_{q=1}\I22=\bigotimes^{n_I}_{q=1}(\IIR_q+\XIR_q),
\end{equation}
where $\XIR_q$ refers to the Pauli-X observable of the $q$-th qubit in IR, and $\IIR_q$ is the $2 \times 2$ identity matrix. The expansion of Eq.~\ref{Jdecompose} on the right-hand side yields $2^{n_I}\approx N$ different combinations of Pauli string observables. The value of Eq.~\ref{eq_JP} can be obtained by summing the expectation values of all combinations of Pauli strings in Eq.~\ref{Jdecompose}, corresponding to the \textbf{SUM} operation in the classical MPGNN's aggregation function. In certain cases, one might wish to exclude contributions from nodes themselves, these values can be simply removed by considering
\begin{equation*}
    J\longrightarrow J-\bigotimes^{n_I}_{q=1}\IIR_q, 
\end{equation*} 
which is equivalent to subtracting the value of Eq.~\ref{eq_self} from Eq.~\ref{eq_JP}.


\subsection{Computational Complexity} \label{sec_complexity}

In this subsection, we examine the computational complexity of both MPGNN and QCGNN in the context of complete graphs. We focus on the simplest case where both models map a set of $d$-dimensional features to a scalar, i.e., $\mathbb{R}^d\rightarrow\mathbb{R}$. For instance, the MPGNN might employ a feed-forward neural network terminating in a single neuron, while the QCGNN could utilize one qubit in NR, measured in the $Z$ basis ($n_Q=1$ and $P=Z$). Additionally, we assume that the computational cost of obtaining a scalar output in classical models is roughly equivalent to the cost of measuring a Pauli string observable in quantum models.

To compute all pairwise information, MPGNN requires $O(N^2)$ computations, with each pair passing through the neural network once. In contrast, due to the quantum parallelism inherent in QCGNN, $\UPARAM$ can process all pairs of nodes simultaneously. To aggregate the final result, QCGNN requires only measuring on $O(2^{n_I}) \approx O(N)$ Pauli string observables, as indicated by Eq.~\ref{Jdecompose}. This suggests that QCGNN could offer a polynomial speedup over MPGNN. However, in the case of QCGNN, additional costs associated with multi-controlled operators and the USO should be taken into account. Although various methods exist for decomposing multi-controlled operators, such as those discussed in \cite{q_decomp1,q_decomp2,q_decomp3}, for simplicity, we adhere to a basic approach outlined in \cite{NC}, which requires an additional $O(n_I)\approx O(\log_2 N)$ ancilla qubits and Toffoli gates. Based on the results in \cite{uso1,uso2}, preparing a uniform quantum state necessitates $O(\log_2 N)$ gates. If the parametrized operators sufficiently deep, these additional costs may become negligible, enabling QCGNN to achieve an $O(N)$ speedup over MPGNN.

Certain traditional VQC ansatz, such as quantum kernel methods \cite{q_kernel1,q_kernel2}, also share similarities with QCGNN. Quantum kernel methods compute the kernel function $k(\xbf_i, \xbf_j)=|\braket{\xbf_i, \thetabf|\xbf_j, \thetabf}|^2$ and are typically constructed using $\UENC$ and $\UPARAM$. These methods have a computational complexity of $O(N^2)$, as they still require computing each pair individually. Again, if $\UPARAM$ is sufficiently deep, the additional costs from data encoding may be negligible, allowing QCGNN to achieve an $O(N)$ speedup over quantum kernel methods. This advantage arises because QCGNN computes $O(N^2)$ pairwise information simultaneously, with only $O(2^{n_I}) \approx O(N)$ additional measurement costs, as given by Eq.~\ref{Jdecompose}.

\subsection{Extending QCGNN to General Graphs}

QCGNN can also be extended to weighted graphs, but the additional cost might render it impractical. Consider a simple case, where an undirected, weighted graph has an adjacency matrix $A$ that can be expressed as the outer product of a vector $\ket{w}=\sum_i w_i\ket{i}$, with edge weight $A_{ij}=w_iw_j$. Instead of initializing the IR uniformly, we initialize the quantum state as
\begin{equation*}
    \sum^{N-1}_{i=0}\frac{1}{\sqrt{N}}\ket{i}\ket{0}^{\otimes n_Q}
    \longrightarrow
    \sum^{N-1}_{i=0}\frac{w_i}{\sqrt{\braket{w|w}}}\ket{i}\ket{0}^{\otimes n_Q},
\end{equation*}
so that the terms in Eq.~\ref{eq_JP} are modified as
\begin{equation*}
    \frac{1}{N} h(\xbf_i, \xbf_j; P)
    \rightarrow
    \frac{w_iw_j}{\braket{w|w}}h(\xbf_i, \xbf_j; P)
    =
    \frac{A_{ij}}{\Tr(A)}h(\xbf_i, \xbf_j; P).
\end{equation*}
Instead of initializing with a uniform state, the quantum state $\ket{w}$ can be initialized using \textbf{AMPLITUDE EMBEDDING}, where the information of the weights $w_i$ is embedded in the amplitude of $\ket{i}$.

To generalize to directed and weighted graphs, note that any matrix can be decomposed into symmetric and skew-symmetric matrices. Since both types are normal matrices, they are diagonalizable according to the spectral theorem. One can apply the method described above for each eigenbasis individually and multiply by a factor proportional to the corresponding eigenvalue. However, the additional computational cost associated with diagonalizing matrices and \textbf{AMPLITUDE EMBEDDING} might be substantial, potentially negating the advantages of QCGNN.

For practical applications, we primarily consider the use of QCGNN for undirected, unweighted, and complete graphs. The added complexity of handling weighted, directed, and incomplete graphs may diminish the computational benefits of QCGNN, making it less feasible for real-world applications without further optimizations.

\section{Experimental Setup} \label{sec_setup}

\subsection{Dataset for Jet Discrimination} \label{sec_data_setup}
We demonstrate the feasibility of QCGNN using two publicly available Monte Carlo simulated datasets\footnote{In the first published version, we used the dataset generated by ourselves using \cite{mg5,delphes1,delphes2,pythia1,hvt}. For the revised version, we switched to other existing public datasets since the number of data is much more sufficient.} for jet discrimination: the \Top dataset \cite{zenodo_top} and the \JetNet dataset \cite{zenodo_jetnet}. The jets in both datasets are clustered using the anti-$k_T$ algorithm \cite{antikt,fastjet} with a distance parameter $R=0.8$.

The \Top dataset \cite{zenodo_top} is used for binary classification, distinguishing signal jets from top quarks (Top) and background jets from mixed quark-gluon interactions (QCD). The transverse momentum of the jets is in the range $[550, 650]$ GeV. The dataset is divided into 1.2 million training samples, 400 thousand validation samples, and 400 thousand testing samples. Further details of the \Top dataset can be found in \cite{dataset_top}.

The \JetNet dataset \cite{zenodo_jetnet} is used for multi-class classification, with jets originating from gluons (g), top quarks (t), light quarks (q), $W$ bosons (w), and $Z$ bosons (z). Each class of jet has a transverse momentum of approximately 1 TeV, with around 170 thousand samples. For each jet event, only the top 30 particles with the highest transverse momentum are retained if the number of particles exceeds 30. Further details of the \JetNet dataset can be found in \cite{dataset_jetnet}

In our approach, each jet is represented as a complete graph. Each node corresponds to a particle in the jet, with node features related to particle flow information. For the $i$-th particle, the input features $\xbf^{(0)}_i$ include the transverse momentum fraction $z_i={p_T}_i/{p_T}_{jet}$, the relative pseudorapidity $\Delta\eta_i=\eta_i-\eta_{jet}$, and the relative azimuthal angle $\Delta\phi_i=\phi_i-\phi_{jet}$. For QCGNN, these input features are further preprocessed as follows:
\begin{equation} \label{QCGNN_feature}
    \begin{split}
        z_i &\longrightarrow \tan^{-1}(z_i), \\
        \Delta\eta_i &\longrightarrow \frac{\pi}{2}\frac{\Delta\eta_i}{R}, \\
        \Delta\phi_i &\longrightarrow \frac{\pi}{2}\frac{\Delta\phi_i}{R},
    \end{split}
\end{equation}
since rotation gates are used for data encoding (see Sec.~\ref{sec_cq_model_setup}). Note that the indices of the particles are arbitrary due to the use of permutation-invariant models for graphs.

\subsection{Classical and Quantum Models} \label{sec_cq_model_setup}

\begin{figure}[htbp]
    \centering
    \includegraphics[width=0.475\textwidth]{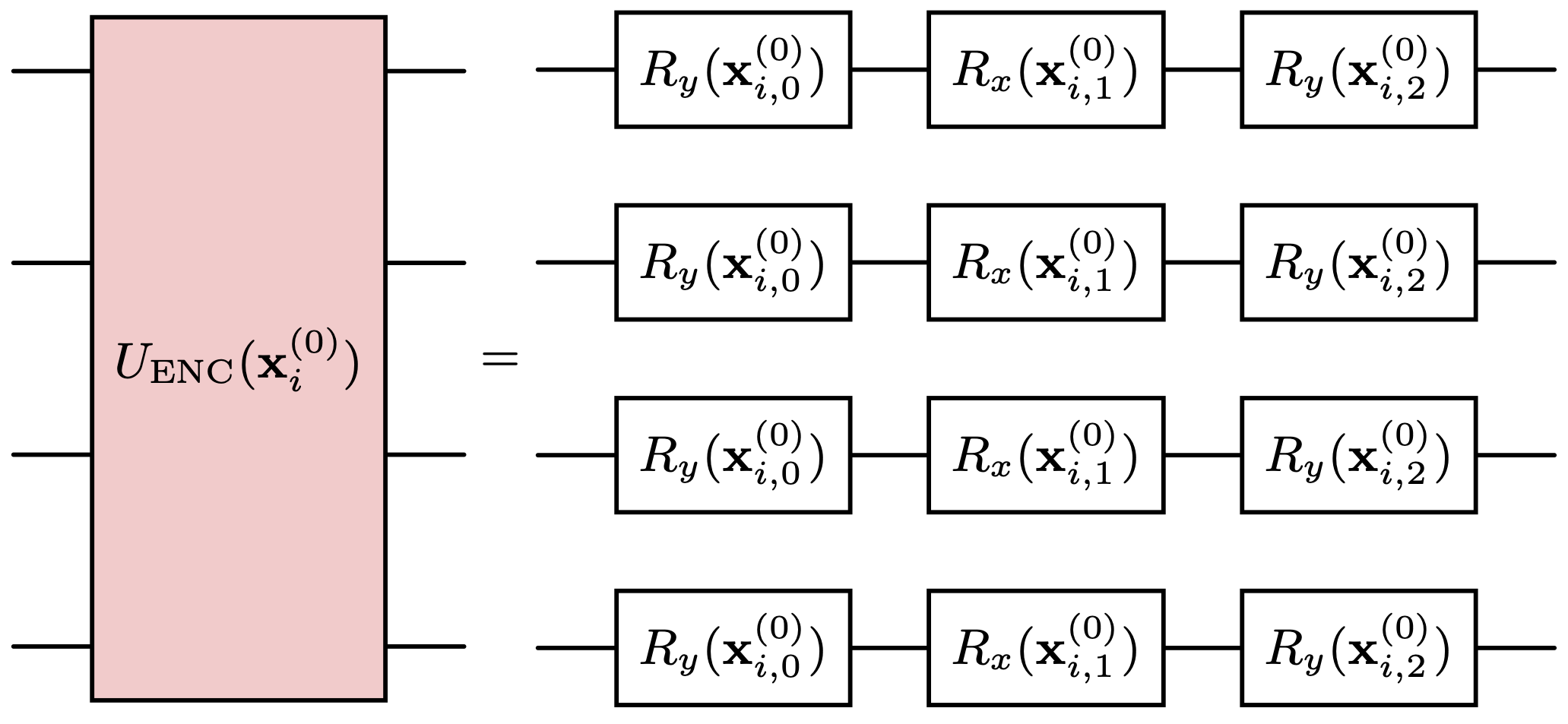}
    \caption{The ansatz of encoding operator used in QCGNN with $n_Q=4$. The rotation gates $R_x$ and $R_y$ are defined in Eq.~\ref{sin_rotgate}. The rotation angle $\xbf^{(0)}_{i,j}$ corresponds to the $j$-th feature of the $i$-th particle, with the three features being the transformed particle flow information described in Eq.~\ref{QCGNN_feature}. Since the data encoding method used in each NR qubit is identical, this ansatz can be generalized to any number of qubits.}
    \label{fig:exenc}
\end{figure}

\begin{figure}[htbp]
    \centering
    \includegraphics[width=0.475\textwidth]{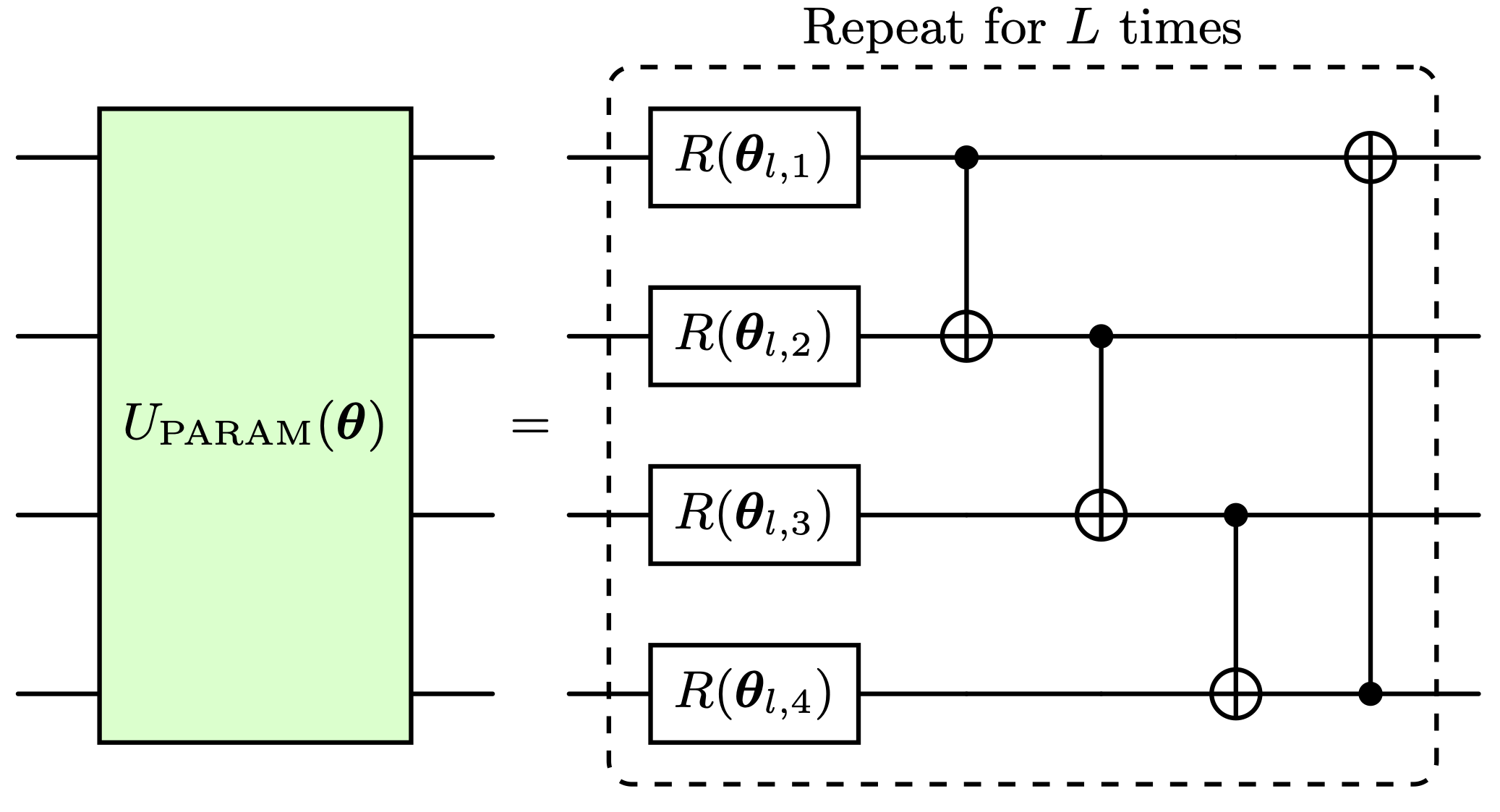}
    \caption{The ansatz of strongly entangling layers \cite{strong_ent} used in QCGNN is illustrated with $n_Q=4$ as an example. The three-angle rotation gate $R$ is defined in Eq.~\ref{tri_rotgate}. The parameters $\thetabf$ are tunable, with distinct parameters for each repetition $l$. Specifically, $\thetabf_{l,1}$, $\thetabf_{l,2}$, $\thetabf_{l,3}$, and $\thetabf_{l,4}$ are all three-dimensional parameters corresponding to the arguments of $R(\alpha, \beta, \gamma)$ in Eq.~\ref{tri_rotgate}. Note that this ansatz can be naturally generalized for $n_Q \ge 3$. For $n_Q < 3$, alternative ansatz should be considered.}
    \label{fig:exvqc}
\end{figure}

\begin{figure*}[htbp]
    \centering
    \includegraphics[width=\textwidth]{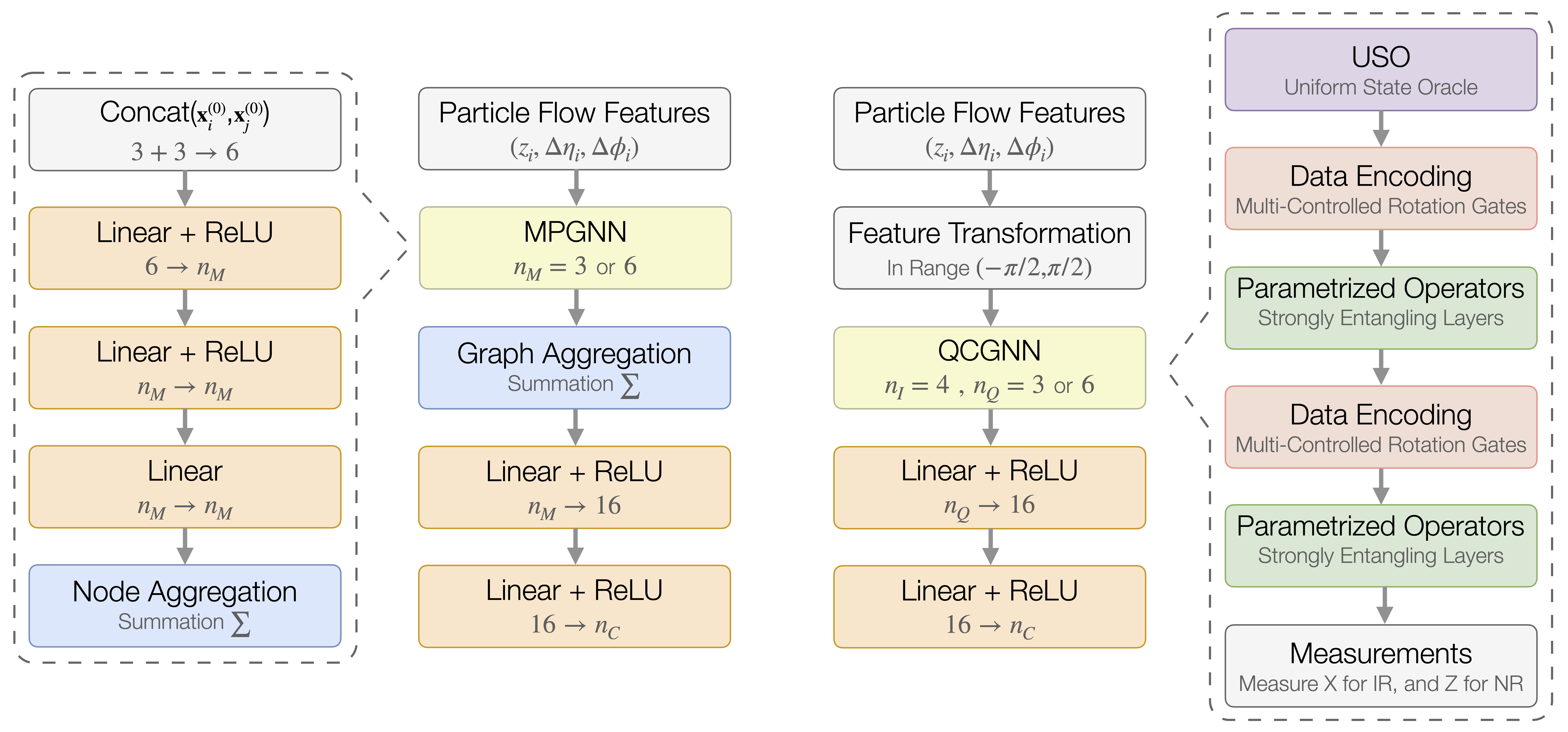}
    \caption{The detailed structure of the quantum model based on QCGNN and the classical model based on MPGNN used for benchmarking are described. The particle flow features are defined in Sec.~\ref{sec_data_setup}, and the hyperparameters of the models are discussed in Sec.~\ref{sec_train}. The classical model on the left is based on MPGNN, with the aggregation function chosen to be \textbf{SUM}. The number of hidden neurons in MPGNN is denoted as $n_M$, set equivalently to $n_Q$ (3 or 6) for comparison with QCGNN. The quantum model on the right is based on QCGNN, with feature preprocessed as described in Eq.~\ref{QCGNN_feature}. The ansatz for the parametrized operators follows the pattern depicted in Fig.~\ref{fig:exvqc}. Note that the data re-upload technique introduced in \cite{reupload} is used, which involves encoding the data twice and applying the parametrized operators with different parameters each time. The dimension of the final output is denoted as $n_C$, with $n_C=1$ for the \Top dataset and $n_C=5$ for the \JetNet dataset. The final output is passed through a Sigmoid function for binary classification ($n_C=1$) or a Softmax function for multi-class classification ($n_C=5$).}
    \label{fig:models}
\end{figure*}

The classical model for benchmarking is based on MPGNN from Eq.~\ref{eqmp}, with the aggregation function chosen to be \textbf{SUM}. The function $\Phi$ is implemented as a feed-forward neural network consisting of linear layers and the ReLU activation functions \cite{relu}, while $\gamma$ is simply the summation of $\Phi$ only, i.e., $\xbf^{(1)}_i=\sum_{j \neq i}\Phi(\xbf^{(0)}_i,\xbf^{(0)}_{j})$, where $j$ ranges from 0 to $N-1$ except $i$. The input to $\Phi$ is simply the concatenation of $\xbf^{(0)}_i$ and $\xbf^{(0)}_j$, requiring only 6 neurons in the input layer of $\Phi$. Consequently, the graph feature, denoted as $\xbf^C$, is computed through
\begin{equation} \label{XC}
    \begin{split}
        \xbf^C &= \sum^{N-1}_{i=0}\xbf^{(1)}_i \\
        &= \sum^{N-1}_{i=0}\sum_{j \neq i}\Phi(\xbf^{(0)}_i,\xbf^{(0)}_j) \\
        &= \sum^{N-1}_{i=0}\sum^{N-1}_{j=0}\Phi(\xbf^{(0)}_i,\xbf^{(0)}_j) - \sum^{N-1}_{i=0}\Phi(\xbf^{(0)}_i,\xbf^{(0)}_i)
    \end{split}
\end{equation}
The structure of MPGNN is similar to the Particle Flow Network (PFN) in \cite{pfn}, with the distinction that PFN calculates $\Phi(\xbf^{(0)}_i)$ for each particle, whereas MPGNN calculates the pairwise information $\Phi(\xbf^{(0)}_i, \xbf^{(0)}_j)$ between particles.

The quantum model is based on QCGNN, which consists of encoding operators and parametrized operators. The \textit{data re-uploading} technique \cite{reupload} is employed 2 times before the final measurements (indicated by the dashed box in Fig.~\ref{fig:exqcgnn} with $R=2$). For simplicity, we use single-angle rotation gates, defined as
\begin{equation} \label{sin_rotgate}
    \begin{split}
        R_x(\theta) &= 
        \begin{bmatrix}
            \cos(\theta/2) & -i\sin(\theta/2)\\
            -i\sin(\theta/2) & \cos(\theta/2)
        \end{bmatrix},\\
        R_y(\theta) &= 
        \begin{bmatrix}
            \cos(\theta/2) & -\sin(\theta/2)\\
            \sin(\theta/2) & \cos(\theta/2)
        \end{bmatrix},\\
        R_z(\theta) &= 
        \begin{bmatrix}
            e^{-i\theta/2} & 0\\
            0 & e^{i\theta/2}
        \end{bmatrix},\\
    \end{split}
\end{equation}
and triple-angle rotation gates, defined as
\begin{equation} \label{tri_rotgate}
    \begin{split}
        R(\alpha, \beta, \gamma) &= R_z(\gamma)R_y(\beta)R_z(\alpha) \\
        &=
        \begin{bmatrix}
            e^{-i\frac{\alpha+\gamma}{2}}\cos(\frac{\beta}{2}) & -e^{-i\frac{\alpha-\gamma}{2}}\sin(\frac{\beta}{2})\\
            e^{-i\frac{\alpha-\gamma}{2}}\sin(\frac{\beta}{2}) & e^{-i\frac{\alpha+\gamma}{2}}\cos(\frac{\beta}{2})
        \end{bmatrix}
    \end{split}
\end{equation}
to encode the particle flow information. The parametrized operators are constructed with strongly entangling layers \cite{strong_ent} using rotation gates and CNOT gates. The ansatz for encoding $i$-th particle features and the strongly entangling layers are shown in Fig.~\ref{fig:exenc} and Fig.~\ref{fig:exvqc} respectively. The $q$-th component of the graph feature $\xbf^Q\in\mathbb{R}^{n_Q}$ is computed by
\begin{equation} \label{XQ}
    \begin{split}
        \xbf^Q_q &= N \Bigl[ \braket{\psi|J\otimes \ZNR_q|\psi} - \braket{\psi|\ZNR_q|\psi} \Bigr] \\
        &= \sum^{N-1}_{i=0}\sum^{N-1}_{j=0} h(\xbf^{(0)}_i, \xbf^{(0)}_j; \ZNR_q) -  \sum^{N-1}_{i=0} h(\xbf^{(0)}_i, \xbf^{(0)}_i; \ZNR_q)\\
    \end{split}
\end{equation}
where the observable $\ZNR_q$ refers to the Pauli-Z measurement of the $q$-th qubit in NR only, and the summations are computed through Eq.~\ref{eq_self} and Eq.~\ref{eq_JP}. To be clarified, the subscript $i$ of $\xbf^{(0)}_i$ corresponds to the $i$-th node, while the subscript $q$ of $\xbf^Q_q$ refers to the $q$-th component of QCGNN output. Note that all qubits in NR can be measured simultaneously, but the measurement output from other qubits is ignored when calculating the expectation value over $\ZNR_q$. This setup can be thought of as a classical feed-forward neural network with $n_Q$ neurons in the output layer. Notice how Eq.~\ref{XQ} resembles Eq.~\ref{XC}, indicating the permutation invariance of the final output.

Eventually, both $\xbf^C$ and $\xbf^Q$ are followed by another feed-forward neural network consisting of linear layers and ReLU activation functions. The full setup of the classical and quantum models is depicted in Fig.~\ref{fig:models}. For binary classification using the \Top dataset, the output layer has a single neuron followed by a Sigmoid function and is trained with binary cross-entropy loss. For multi-class classification using the \JetNet dataset, the output layer has five neurons followed by a Softmax function and is trained with multi-class cross-entropy loss.

\subsection{Training Setup and Model Hyperparameters} \label{sec_train}

\begin{figure*}[htbp]
    \centering
    \includegraphics[width=\textwidth]{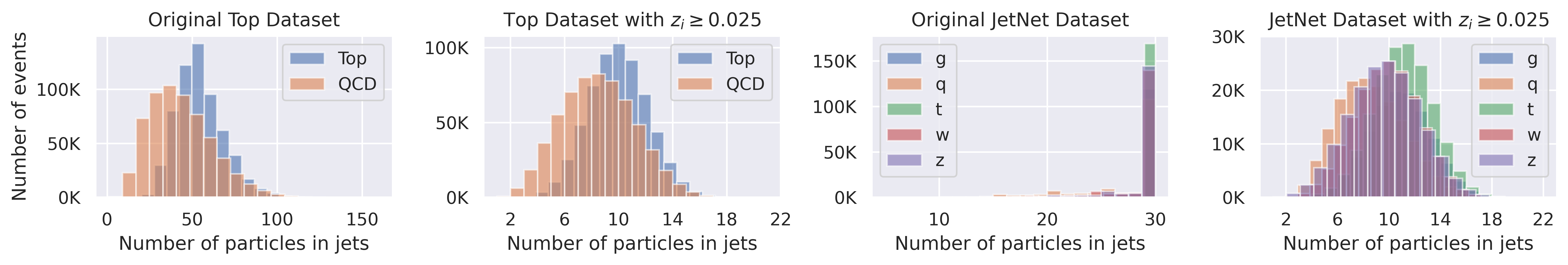}
    \caption{Histograms of the number of particles per jet for the \Top and the \JetNet datasets. A detailed description of both datasets is provided in Sec.~\ref{sec_data_setup}. The original \JetNet dataset exhibits a sharp distribution at 30 particles, as the original data retain only the first 30 particles with the highest transverse momentum. The histograms for particles with a relative transverse momentum $z_i \ge 0.025$ are also given. This threshold is selected to ensure that the majority of the distribution of the number of particles per jet falls between 4 and 16.}
    \label{fig:hist}
\end{figure*}

The complete training process was conducted with 5 different random seeds, each for 30 epochs. The \Top and the \JetNet datasets comprise 2 and 5 classes, respectively. For each class, we selected 25,000 training samples, 2,500 validation samples, and 2,500 testing samples. This limited data selection is due to the extensive training time required for the QCGNN, as discussed in Sec.~\ref{sec_train_QCGNN}. For each random seed, the data were randomly sampled from the original dataset. To balance between demonstrating the training performance and computational demands, particles with transverse momentum less than 2.5\% of ${p_T}_{jet}$, i.e., $z_i < 0.025$, were discarded, so that the majority of the distribution of the number of particles per jet is between 4 and 16. The histograms of the number of particles per jet for the original and preprocessed \Top and \JetNet datasets are shown in Fig.~\ref{fig:hist}. To mitigate the extensive training time associated with simulating QML, events with fewer than 4 or more than 16 particles were discarded. These choices strike a balance between performance and the amount of training data, as discussed in Appendix \ref{app_num_data}.

Due to limited computational resources, the number of qubits $n_Q$ in NR was tested with $n_Q=3$ and $n_Q=6$, with the number of strongly entangling layers in each parametrized operator ($\UPARAM$) set to $n_Q/3$. Given that the maximum number of particles in a jet is 16, we used $n_I=\lceil \log_2 16 \rceil=4$ qubits for IR. The number of hidden neurons $n_M$ in the MPGNN was set to 3 or 6 for comparison with the QCGNN, ensuring that both models have a comparable number of parameters, as discussed in Appendix \ref{app_num_param}.

We also evaluated the performance of classical state-of-the-art models, including the Particle Flow Network (PFN) \cite{pfn}, Particle Net (PNet) \cite{ptcnet}, and Particle Transformer (ParT) \cite{part}. The structure and hyperparameters of PFN, PNet, and ParT were configured according to their respective original publications. Notably, we excluded mass information from the interaction matrix of ParT, as only particle flow information $(p_T,\Delta\eta,\Delta\phi)$ was used. The $k$-nearest neighbor method used in the original PNet was configured with $k=3$, given that the minimum number of particles per jet was 4.

The classical models were implemented using \textit{PyTorch} \cite{pytorch} and \textit{PyTorch Geometric} \cite{pygeo}, while the quantum circuit of QCGNN was simulated using \PennyLane \cite{pennylane}. The cross-entropy loss was optimized using the \textsc{Adam} optimizer \cite{adam} with a learning rate of $10^{-3}$ for all models. The batch size was set to 64, the maximum allowable due to memory constraints, as simulating quantum circuits requires substantial memory resources.

\subsection{Implementing QCGNN with Simulators} \label{sec_train_QCGNN}

Unlike classical models, parameter gradients for real quantum computers cannot be computed using traditional methods such as finite difference methods. Instead, the parameter-shift rule (PSR) \cite{psr1,psr2,psr3,psr4} can be employed to calculate gradients. However, applying PSR on quantum computers necessitates extensive requests and long queue times with actual quantum devices. Furthermore, the noise in current quantum computers is insufficiently low to enable stable training of quantum neural networks, often resulting in training failures.

To circumvent these issues during the NISQ era \cite{nisq}, we trained the QCGNNs on classical computers using \PennyLane \cite{pennylane} quantum circuit simulators with zero noise. Nonetheless, simulating quantum circuits is highly time-consuming, even for a few qubits. Although \PennyLane supports QML on GPUs, speed improvements over CPUs are significant only with many qubits, typically more than 20 qubits\footnote{The benchmark of quantum simulation on GPU can be found in PennyLane's blog: "Lightning-fast simulations with PennyLane and the NVIDIA cuQuantum SDK"}. For this study, we used CPUs to train the QCGNNs. Training a 10-qubit ($n_I=4$ and $n_Q=6$) QCGNN with 10,000 samples and a batch size of 64 takes approximately 1,000 seconds per epoch. Consequently, the training over 5 random seeds for 30 epochs required nearly a month.

\section{Results} \label{sec_result}

\subsection{Performance of Classical and Quantum Models} \label{sec_performance}

\begin{table*}[htbp]
    \centering
    \begin{tabularx}{\textwidth}{ l Y Y Y Y Y Y }
      \hline
      \hline
      \multirow{2}{*}{Model} & \multicolumn{3}{c}{\centering \Top Dataset (2 classes)} & \multicolumn{3}{c}{\centering \JetNet Dataset (5 classes)} \\
      & \# params  & AUC & Accuracy & \# params & AUC & Accuracy \\
      \hline

      Particle Transformer & 2.2M & 0.946$\pm$0.005 & 0.868$\pm$0.009 & 2.2M & 0.889$\pm$0.002 & 0.656$\pm$0.006 \\
      Particle Net & 177K & 0.953$\pm$0.003 & 0.885$\pm$0.006 & 178K & 0.896$\pm$0.003 & 0.669$\pm$0.004 \\
      Particle Flow Network & 72.3K & 0.954$\pm$0.004 & 0.885$\pm$0.005 & 72.7K & 0.900$\pm$0.003 & 0.675$\pm$0.005 \\
      MPGNN - $n_M=64$ & 13K & 0.961$\pm$0.003 & 0.896$\pm$0.003 & 13.3K & 0.903$\pm$0.002 & 0.683$\pm$0.007 \\
      \hline
      MPGNN - $n_M=6$ & 255 & 0.924$\pm$0.006 & 0.866$\pm$0.006 & 323 & 0.865$\pm$0.004 & 0.615$\pm$0.010 \\
      MPGNN - $n_M=3$ & 126 & 0.922$\pm$0.005 & 0.864$\pm$0.006 & 194 & 0.757$\pm$0.110 & 0.475$\pm$0.141 \\
      QCGNN - $n_Q=6$ & 201 & 0.932$\pm$0.004 & 0.868$\pm$0.005 & 269 & 0.822$\pm$0.003 & 0.543$\pm$0.006 \\
      QCGNN - $n_Q=3$ & 99 & 0.919$\pm$0.006 & 0.864$\pm$0.005 & 167 & 0.796$\pm$0.009 & 0.505$\pm$0.014 \\
      
      \hline
      \hline
    
    \end{tabularx}
    \caption{The performance of different models on the \Top and the \JetNet datasets. As detailed in Sec.~\ref{sec_cq_model_setup}, $n_M$ denotes the number of hidden neurons in the MPGNN, and $n_Q$ represents the number of qubits in the NR of the QCGNN. The number of parameters for each model is provided, along with the area under the ROC curve (AUC) and accuracy. The AUC score is computed as the average of all possible pairwise combinations of classes, while the accuracy is calculated across all classes simultaneously. The results are averaged over 5 random seeds, with one standard deviation shown.}
    \label{table:result}
\end{table*}

 The performance of the classical and quantum models on the \Top and the \JetNet datasets is summarized in Table.~\ref{table:result}. The inference scores of the MPGNN and QCGNN are comparable when the number of parameters is in roughly the same order of amount. We anticipate that QCGNN has the potential to achieve performance on par with MPGNN as the number of qubits increases. When training with a smaller number of parameters on the multi-class classification \JetNet dataset, we observe that QCGNN is more stable than MPGNN, with the latter exhibiting a larger standard deviation.
 
 Among the state-of-the-art models, MPGNN with $n_M = 64$ has higher inference scores compared to others. This is likely due to the information of jets being lost during the data preprocessing, where only 4 to 16 particles per jet are utilized. When training with the full information from the original jet dataset, i.e., without discarding information from soft particles, other state-of-the-art models can compete with MPGNN, even better. Details of the state-of-the-art models trained on the original jet dataset are provided in Appendix \ref{app_num_data}.

\subsection{Executing Pre-trained QCGNN on IBMQ} \label{sec_ibmq}

\begin{figure*}[htbp]
    \centering
    \includegraphics[width=\textwidth]{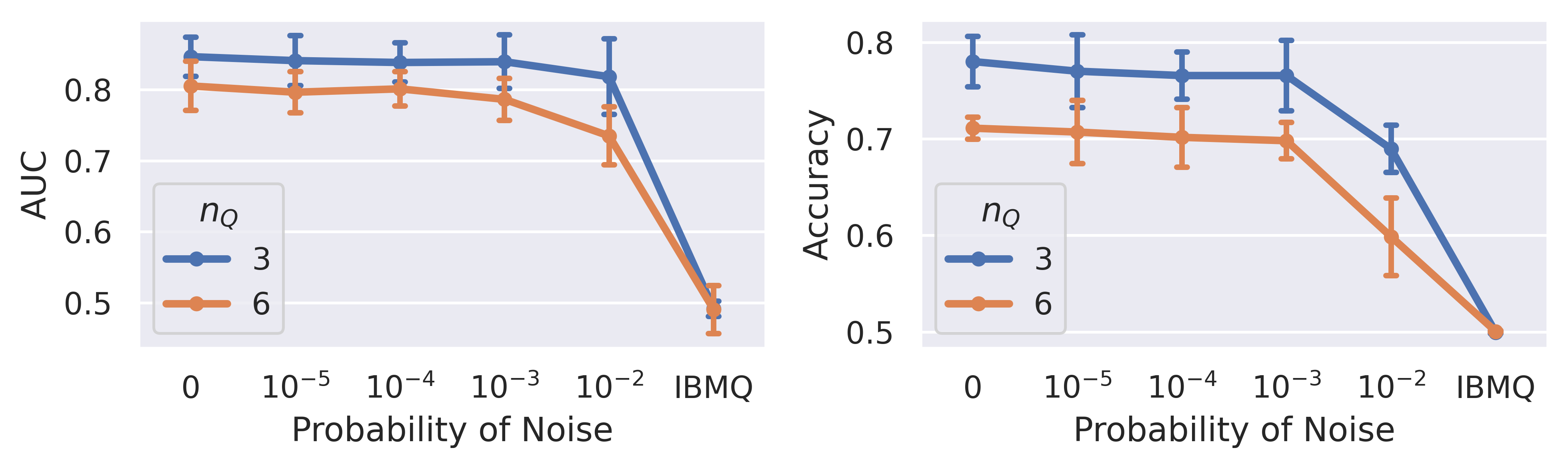}
    \caption{This figure illustrates the extrapolation of noise probabilities (depolarizing error and amplitude damping) using quantum circuit simulators with pre-trained QCGNNs. An ideal quantum computer corresponds to a noise probability of zero on the $x$-axis, while "IBMQ" refers to results obtained from \textit{ibm\_brussels}. The $y$-axis shows the area under the ROC curve (AUC) and accuracy for 400 jet data samples, each containing only 4 particles, which requires $n_I = 2$ qubits for IR. The error bars represent one standard deviation across 5 full executions with different random seeds.}
    \label{fig:noise_extrapolation}
\end{figure*}

Although implementing full training on quantum computers is impractical in the NISQ era, we can still evaluate the performance of the pre-trained QCGNN on IBMQ real devices \cite{ibmq}. To minimize the noise effects caused by real quantum gates, we select events with only four particles from the \Top dataset, i.e., using $n_I=2$ qubits in IR, thereby reducing the number of gates required for initial state preparation and data encoding. In this setup, the USO can be efficiently implemented using Hadamard gates for each qubit in IR. On IBMQ real devices, only 1-qubit and 2-qubit gates are available, and the multi-controlled gates used in data encoding are decomposed using methods described in \cite{NC}.

We selected \textit{ibm\_brussels} with 1024 shots to test the performance of QCGNN on an IBMQ real device. However, the quantum computers in the NISQ era are currently too noisy to yield usable results. For binary classification, the inference of QCGNN on \textit{ibm\_brussels} results in approximately $0.5$ AUC and $0.5$ accuracy, which equates to random guessing. To assess how noise affects the performance of QCGNN, we perform an extrapolation over noise using \PennyLane simulators, with the results shown in Fig.~\ref{fig:noise_extrapolation}. We simulate quantum noise, including \textit{depolarizing error} and \textit{amplitude damping}, occurring after each quantum operation with a certain probability. As indicated in Fig.~\ref{fig:noise_extrapolation}, the noise probability must be reduced to below $10^{-3}$ to achieve reliable results\footnote{The noise probability here corresponds to the simulated noise in the quantum circuit simulation.}.

\subsection{QCGNN Runtime on IBMQ} \label{sec_runtime}

\begin{table}[htbp]
    \centering
    \begin{tabularx}{0.475\textwidth}{ X >{\hsize=.5\hsize}Y >{\hsize=.75\hsize}Y >{\hsize=.75\hsize}Y}
      \hline
      \hline
      IBMQ Backend & N & $T_{\text{ENC}}$  & $T_{\text{PARAM}}$ \\
      \hline
      \multirow{3}{*}{ibm\_nazca} & 2 & 2.567 & 0.209 \\ 
      & 4 & 5.352 & 0.197 \\ 
      & 8 & 10.551 & 0.219 \\
      \hline
      \multirow{3}{*}{ibm\_strasbourg} & 2 & 2.595 & 0.217 \\ 
      & 4 & 5.416 & 0.197 \\ 
      & 8 & 11.085 & 0.211 \\
      \hline
      \hline
    \end{tabularx}
    \caption{The runtime of quantum circuit gate operations for encoding layers and parametrized layers on different IBMQ backends is analyzed. The number of particles per jet is denoted as $N$. The runtimes $T_{\text{ENC}}$ and $T_{\text{PARAM}}$ are defined in Sec.~\ref{sec_runtime} and are presented in seconds.}
    \label{table:runtime}
\end{table}

To validate the time complexity analysis discussed in Sec.~\ref{sec_complexity}, we initialized untrained QCGNNs and executed them on various IBMQ backends, including \textit{ibm\_nazca} and \textit{ibm\_strasbourg}, with $n_Q=100$ and 1024 shots. We set the number of nodes to 2, 4, and 8, such that only Hadamard gates are required for the initial state preparation. To determine the quantum gate runtime for encoding and parametrized operators, we first ran QCGNN without any operators to measure the runtime $T_{0}$ for quantum state initialization and measurement. We then applied encoding operators with 10 times of re-uploading to obtain the runtime $T_{1}$. Finally, we applied parametrized operators, constructed with 10 strongly entangling layers and 10 times of re-uploading (resulting in 100 strongly entangling layers in total), to measure the runtime $T_{2}$. Each runtime measurement was averaged over 10 executions. The runtime of encoding operators was computed as follows:
\begin{equation*}
    T_{\text{ENC}} = \frac{T_{1}-T_{0}}{10},
\end{equation*}
and the runtime of each strongly entangling layer in parametrized operators is calculated by
\begin{equation*}
    T_{\text{PARAM}} = \frac{T_{2}-T_{1}}{100}.
\end{equation*}
The results presented in Table.~\ref{table:runtime} indicate that the runtime of encoding operators scales approximately linearly with the number of particles per jet, while the runtime of parametrized operators remains approximately constant, as expected. As discussed in Sec.~\ref{sec_complexity}, when the parametrized operators are sufficiently deep, the runtime will be dominated by these operators, making the additional computational cost associated with data encoding negligible.

\section{Summary} \label{sec_summary}

The representation of jets as graphs, leveraging the property of permutation invariance, has been widely utilized in particle physics. However, constructing graphs from particle jets in a physically meaningful manner remains an unresolved challenge. In the absence of specific physical assumptions, we adopt a straightforward approach by representing jets as complete graphs with undirected, unweighted edges. Motivated by the structure of complete graphs, we propose the Quantum Complete Graph Neural Network (QCGNN) for learning through aggregation using \textbf{SUM} or \textbf{MEAN} operations. When training on $N$ particles, QCGNN exhibits $O(N)$ computational complexity if the parametrized operators are sufficiently deep, offering a polynomial speedup over classical models that require $O(N^2)$.

To demonstrate the practicality of QCGNN, we conduct experiments on jet discrimination. Sec.~\ref{sec_performance} shows that QCGNN performs comparably to classical models with a similar number of parameters. Moreover, QCGNN displays a more stable training process across different random seeds. Although the pre-trained QCGNN has been tested on IBMQ real devices, the noise in quantum circuits remains too significant to yield reliable results. To assess the impact of noise in the NISQ era, we perform noise extrapolation using simulators, as detailed in Sec.~\ref{sec_ibmq}. We also conducted a series of executions on IBMQ quantum devices to estimate the runtime of QCGNN, as discussed in Sec.~\ref{sec_runtime}. The time costs of encoding and parametrized operators are approximately linear and constant to the number of particles per jet, respectively.

In conclusion, QCGNN provides a more efficient method for learning unstructured jets using QML. The additional computational costs associated with quantum state initialization and data encoding are negligible when the parametrized operators are sufficiently deep, as discussed in Sec.~\ref{sec_complexity}. However, it remains an open question whether QML provides a definitive quantum advantage in HEP. Moreover, developing more expressive and suitable methods for HEP data encoding continues to be an intriguing and ongoing area of research.

\section*{Acknowledgement}

The authors thank Chiao-Hsuan Wang for helpful discussions and suggestions about quantum computation. The accessibility of IBMQ resources is supported by the IBM Quantum Hub at National Taiwan University.

\appendix

\section{Relation Between State-of-the-Art Models and MPGNN} \label{app_mpgnn}

In Sec.~\ref{sec_gnn}, we introduce the MPGNN. Here, we show that some state-of-the-art models can be considered as a special case of MPGNN, i.e., in the form of
\begin{equation}
    \xbf^{(k)}_i=\gamma^{(k)}\left(\xbf^{(k-1)}_i,\bigoplus_{j \in \mathcal{N}(i)}\Phi^{(k)}(\xbf^{(k-1)}_i,\xbf^{(k-1)}_j)\right).
    \tag{\ref{eqmp}}
\end{equation}

\subsection{Particle Flow Networks (PFN) as MPGNN}
The PFN introduced in \cite{pfn} first transforms the particle features into a latent space via a feed-forward neural network $\Phi$, followed by a summation. Then another feed-forward neural network $\gamma$ will be applied to get the final score $F$ of jet discrimination. In the form of MPGNN, the PFN can be written as
\begin{equation*}
    F=\gamma\left(\sum_{\xbf_i \in G}\Phi(\xbf_i)\right).
\end{equation*}

\subsection{Particle Net (PNet) as MPGNN}
The PNet introduced in \cite{ptcnet} turns jets into graphs by dynamically determining the edges through the distance in feature space or latent space. The EdgeConv of PNet can be written as
\begin{equation*}
    \xbf^{(k)}_i=\gamma^{(k)}\left(\bigoplus_{j \in \mathcal{N}(i)}\Phi^{(k)}(\xbf^{(k-1)}_i,\xbf^{(k-1)}_i-\xbf^{(k-1)}_j)\right),
\end{equation*}
where the neighbors are dynamically determined through the $k$-nearest neighbor method. The EdgeConv also calculates the difference between features, then passes through either a convolutional neural network or a feed-forward neural network, captured by $\gamma$ and $\Phi$.

\subsection{Particle Transformer (ParT) as MPGNN}
The ParT introduced in \cite{part} uses the transformer architecture to learn the jet features. The structure of the transformer is rather complicated, but each attention block can still be written in the form of MPGNN. As ParT considers all pairs of particle information without positional embedding, the ParT can be seen as dealing with complete graphs. The queries (Q) and the keys (K) in the attention mechanism are captured by $\Phi$, with the aggregation function chosen to be \textbf{SOFTMAX} (can be seen as a summation over a particular transformation that can be absorbed into $\Phi$), and the values (V) in the attention mechanism are captured by $\gamma$. Note the functions in the transformer such as GeLU or LayerNorm can also be absorbed into $\Phi$ and $\gamma$.

\section{Performance of Classical Models on Different Number of Training Samples} \label{app_num_data}

\begin{figure*}[htbp]
    \centering
    \includegraphics[width=\textwidth]{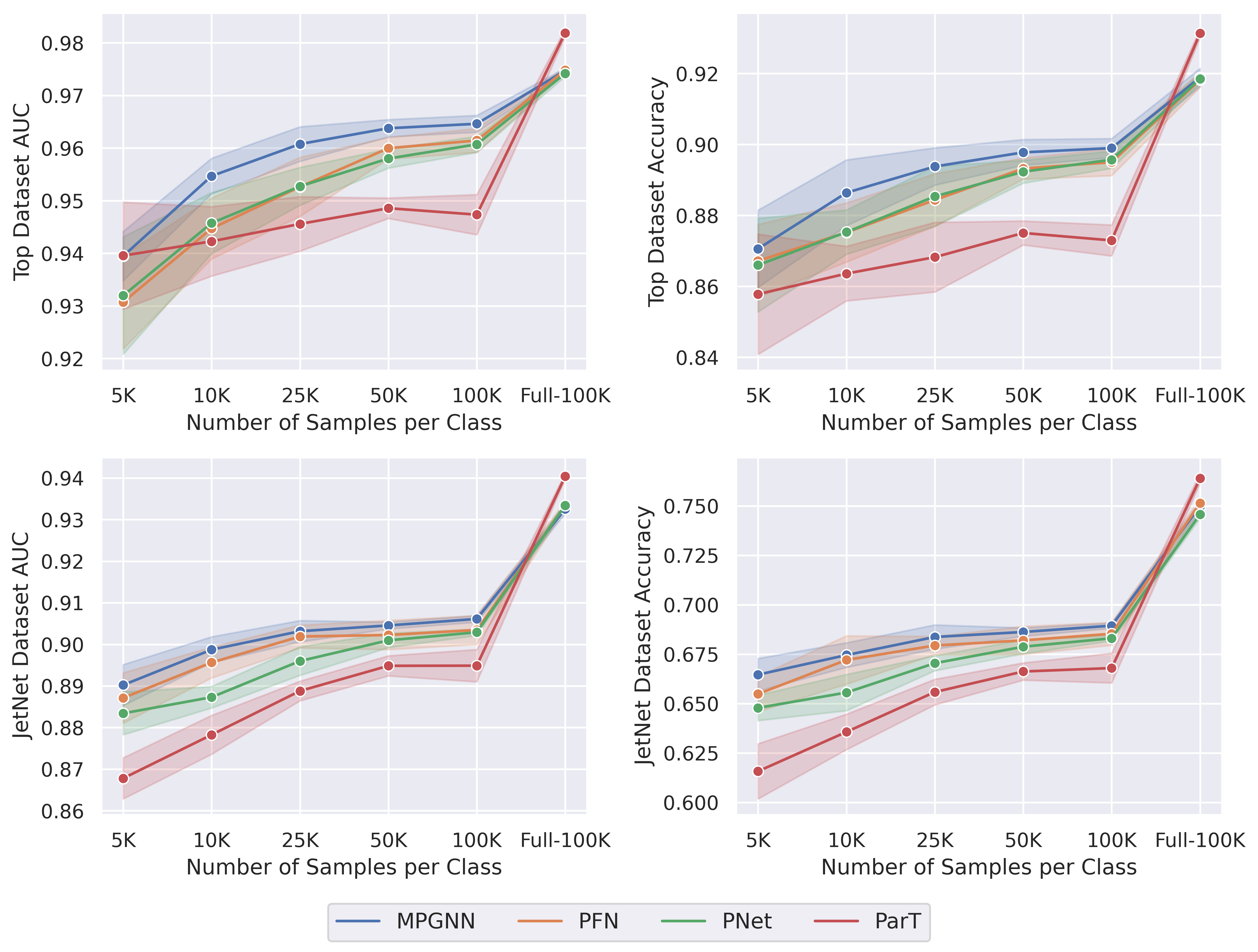}
    \caption{This figure illustrates the performance of state-of-the-art models with varying numbers of training samples. The blue, orange, green, and red lines represent the Particle Flow Network (PFN), Particle Net (PNet), Particle Transformer (ParT), and MPGNN with $n_M=64$, respectively. The upper row displays performance metrics on the \Top dataset, while the lower row shows performance metrics on the \JetNet dataset. The left column presents the area under the curve (AUC), and the right column shows accuracy. The training samples are preprocessed as described in Sec.~\ref{sec_data_setup}, except that 'Full-100K' uses all particle data without applying the transverse momentum cutoff.}
    \label{fig:app_num_train}
\end{figure*}

As described in Sec.~\ref{sec_data_setup}, we selected 25,000 training samples with a maximum of 16 particles per jet for each class. In this appendix, we justify that this setup is sufficient to evaluate the performance of each model. We trained state-of-the-art classical models, including the Particle Flow Network (PFN) \cite{pfn}, Particle Net (PNet) \cite{ptcnet}, Particle Transformer (ParT) \cite{part}, and MPGNN with 64 hidden neurons ($n_M=64$).

The performance of each model on both the \Top and the \JetNet datasets is obtained by training with varying numbers of samples per class, across 5 different random seeds. The results are presented in Fig.~\ref{fig:app_num_train}. The training samples were preprocessed as outlined in Sec.~\ref{sec_data_setup}, using only events with at least 4 and at most 16 particles. The performance of each state-of-the-art model is getting saturated between 25,000 and 50,000 training samples, indicating that the choice of 25,000 samples in Sec.~\ref{sec_data_setup} is almost adequate for demonstrating the model performance. We also conducted experiments with the full-particle jets from the original dataset, without applying the transverse momentum cutoff, using 100,000 samples per class. We found that, when training with a few particles, the simplest MPGNN model performs better than the other models. However, when using the full original dataset, the ParT outperforms the other models.

\section{Number of Parameters in MPGNN and QCGNN} \label{app_num_param}

In this appendix, we compute the number of parameters for the MPGNN and QCGNN models based on the structures outlined in Sec.~\ref{sec_cq_model_setup}. It is important to distinguish these calculations from the total number of parameters reported in Table.~\ref{table:result}, which includes the parameters of the final feed-forward network in both MPGNN and QCGNN.

For MPGNN with $n_M$ hidden neurons in both the hidden and output layers, and an input dimension of 6 (since features of two particles are concatenated), if there are $L_C$ hidden layers, the total number of parameters is given by
\begin{equation}
    \begin{split}
        \#_C &= 6(n_M+1) + n_M L_C (n_M+1)\\
        &= L_C n^2_M + (6+L_C)n_M + 6,
    \end{split}
\end{equation}
where the $+1$ in each parenthesis accounts for the bias term in the linear layers. 

For QCGNN, suppose there are $n_Q$ qubits in the NR with the strongly entangling layers ansatz as depicted in Fig.~\ref{fig:exvqc}. Each strongly entangling layer consists of $n_Q$ rotation gates, with each gate having 3 parameters. If there are $L_Q$ strongly entangling layers and $L_R$ times of re-uploading, the total number of parameters is
\begin{equation}
    \#_Q = 3 L_R L_Q n_Q.
\end{equation}
To ensure that both models have the same output dimension, we set $n_M = n_Q = D$. Assuming $n_Q$ is a multiple of 3, and setting $L_Q = n_Q / 3 = D / 3$, the number of parameters for MPGNN and QCGNN are
\begin{equation}
    \begin{split}
        \#_C &= L_C D^2 + (6+L_C)D + 6,\\
        \#_Q &= L_R D^2.
    \end{split}
\end{equation}
It is evident that by choosing $L_C = L_R = L$, the leading term for both models scales as $O(LD^2)$. In this study, we set $L = 2$. To approximate the linear term $O(LD)$ in MPGNN, one could set $L_R = (n_Q + 1) / 3$, resulting in $\#_Q = LD^2 + LD$. However, this approach was not considered in this study due to the increased simulation time required for longer circuits.

\bibliography{qcgnn}

\begin{thebibliography}{118}%
\makeatletter
\providecommand \@ifxundefined [1]{%
 \@ifx{#1\undefined}
}%
\providecommand \@ifnum [1]{%
 \ifnum #1\expandafter \@firstoftwo
 \else \expandafter \@secondoftwo
 \fi
}%
\providecommand \@ifx [1]{%
 \ifx #1\expandafter \@firstoftwo
 \else \expandafter \@secondoftwo
 \fi
}%
\providecommand \natexlab [1]{#1}%
\providecommand \enquote  [1]{``#1''}%
\providecommand \bibnamefont  [1]{#1}%
\providecommand \bibfnamefont [1]{#1}%
\providecommand \citenamefont [1]{#1}%
\providecommand \href@noop [0]{\@secondoftwo}%
\providecommand \href [0]{\begingroup \@sanitize@url \@href}%
\providecommand \@href[1]{\@@startlink{#1}\@@href}%
\providecommand \@@href[1]{\endgroup#1\@@endlink}%
\providecommand \@sanitize@url [0]{\catcode `\\12\catcode `\$12\catcode `\&12\catcode `\#12\catcode `\^12\catcode `\_12\catcode `\%12\relax}%
\providecommand \@@startlink[1]{}%
\providecommand \@@endlink[0]{}%
\providecommand \url  [0]{\begingroup\@sanitize@url \@url }%
\providecommand \@url [1]{\endgroup\@href {#1}{\urlprefix }}%
\providecommand \urlprefix  [0]{URL }%
\providecommand \Eprint [0]{\href }%
\providecommand \doibase [0]{https://doi.org/}%
\providecommand \selectlanguage [0]{\@gobble}%
\providecommand \bibinfo  [0]{\@secondoftwo}%
\providecommand \bibfield  [0]{\@secondoftwo}%
\providecommand \translation [1]{[#1]}%
\providecommand \BibitemOpen [0]{}%
\providecommand \bibitemStop [0]{}%
\providecommand \bibitemNoStop [0]{.\EOS\space}%
\providecommand \EOS [0]{\spacefactor3000\relax}%
\providecommand \BibitemShut  [1]{\csname bibitem#1\endcsname}%
\let\auto@bib@innerbib\@empty
\bibitem [{\citenamefont {{HEP ML Community}}()}]{hepmllivingreview}%
  \BibitemOpen
  \bibfield  {author} {\bibinfo {author} {\bibnamefont {{HEP ML Community}}},\ }\href {https://iml-wg.github.io/HEPML-LivingReview/} {\bibinfo {title} {{A Living Review of Machine Learning for Particle Physics}}}\BibitemShut {NoStop}%
\bibitem [{\citenamefont {Feickert}\ and\ \citenamefont {Nachman}(2021)}]{hepml1}%
  \BibitemOpen
  \bibfield  {author} {\bibinfo {author} {\bibfnamefont {M.}~\bibnamefont {Feickert}}\ and\ \bibinfo {author} {\bibfnamefont {B.}~\bibnamefont {Nachman}},\ }\href@noop {} {\bibinfo {title} {A living review of machine learning for particle physics}} (\bibinfo {year} {2021}),\ \Eprint {https://arxiv.org/abs/2102.02770} {arXiv:2102.02770 [hep-ph]} \BibitemShut {NoStop}%
\bibitem [{\citenamefont {Radovic}\ \emph {et~al.}(2018)\citenamefont {Radovic}, \citenamefont {Williams}, \citenamefont {Rousseau}, \citenamefont {Kagan}, \citenamefont {Bonacorsi}, \citenamefont {Himmel}, \citenamefont {Aurisano}, \citenamefont {Terao},\ and\ \citenamefont {Wongjirad}}]{hepml2}%
  \BibitemOpen
  \bibfield  {author} {\bibinfo {author} {\bibfnamefont {A.}~\bibnamefont {Radovic}}, \bibinfo {author} {\bibfnamefont {M.}~\bibnamefont {Williams}}, \bibinfo {author} {\bibfnamefont {D.}~\bibnamefont {Rousseau}}, \bibinfo {author} {\bibfnamefont {M.}~\bibnamefont {Kagan}}, \bibinfo {author} {\bibfnamefont {D.}~\bibnamefont {Bonacorsi}}, \bibinfo {author} {\bibfnamefont {A.}~\bibnamefont {Himmel}}, \bibinfo {author} {\bibfnamefont {A.}~\bibnamefont {Aurisano}}, \bibinfo {author} {\bibfnamefont {K.}~\bibnamefont {Terao}},\ and\ \bibinfo {author} {\bibfnamefont {T.}~\bibnamefont {Wongjirad}},\ }\bibfield  {title} {\bibinfo {title} {Machine learning at the energy and intensity frontiers of particle physics},\ }\href {https://doi.org/10.1038/s41586-018-0361-2} {\bibfield  {journal} {\bibinfo  {journal} {Nature}\ }\textbf {\bibinfo {volume} {560}},\ \bibinfo {pages} {41} (\bibinfo {year} {2018})}\BibitemShut {NoStop}%
\bibitem [{\citenamefont {Chen}\ and\ \citenamefont {Chien}(2020)}]{2pcnn}%
  \BibitemOpen
  \bibfield  {author} {\bibinfo {author} {\bibfnamefont {K.-F.}\ \bibnamefont {Chen}}\ and\ \bibinfo {author} {\bibfnamefont {Y.-T.}\ \bibnamefont {Chien}},\ }\bibfield  {title} {\bibinfo {title} {Deep learning jet substructure from two-particle correlations},\ }\href {https://doi.org/10.1103/PhysRevD.101.114025} {\bibfield  {journal} {\bibinfo  {journal} {Phys. Rev. D}\ }\textbf {\bibinfo {volume} {101}},\ \bibinfo {pages} {114025} (\bibinfo {year} {2020})}\BibitemShut {NoStop}%
\bibitem [{\citenamefont {Kheddar}\ \emph {et~al.}(2024)\citenamefont {Kheddar}, \citenamefont {Himeur}, \citenamefont {Amira},\ and\ \citenamefont {Soualah}}]{jet_image1}%
  \BibitemOpen
  \bibfield  {author} {\bibinfo {author} {\bibfnamefont {H.}~\bibnamefont {Kheddar}}, \bibinfo {author} {\bibfnamefont {Y.}~\bibnamefont {Himeur}}, \bibinfo {author} {\bibfnamefont {A.}~\bibnamefont {Amira}},\ and\ \bibinfo {author} {\bibfnamefont {R.}~\bibnamefont {Soualah}},\ }\href {https://arxiv.org/abs/2403.11934} {\bibinfo {title} {Image classification in high-energy physics: A comprehensive survey of applications to jet analysis}} (\bibinfo {year} {2024}),\ \Eprint {https://arxiv.org/abs/2403.11934} {arXiv:2403.11934 [hep-ph]} \BibitemShut {NoStop}%
\bibitem [{\citenamefont {Lee}\ \emph {et~al.}(2019{\natexlab{a}})\citenamefont {Lee}, \citenamefont {Park}, \citenamefont {Watson},\ and\ \citenamefont {Yang}}]{jet_image2}%
  \BibitemOpen
  \bibfield  {author} {\bibinfo {author} {\bibfnamefont {J.~S.~H.}\ \bibnamefont {Lee}}, \bibinfo {author} {\bibfnamefont {I.}~\bibnamefont {Park}}, \bibinfo {author} {\bibfnamefont {I.~J.}\ \bibnamefont {Watson}},\ and\ \bibinfo {author} {\bibfnamefont {S.}~\bibnamefont {Yang}},\ }\bibfield  {title} {\bibinfo {title} {Quark-gluon jet discrimination using convolutional neural networks},\ }\href {https://doi.org/10.3938/jkps.74.219} {\bibfield  {journal} {\bibinfo  {journal} {Journal of the Korean Physical Society}\ }\textbf {\bibinfo {volume} {74}},\ \bibinfo {pages} {219} (\bibinfo {year} {2019}{\natexlab{a}})}\BibitemShut {NoStop}%
\bibitem [{\citenamefont {Li}\ and\ \citenamefont {Sun}(2020)}]{jet_image3}%
  \BibitemOpen
  \bibfield  {author} {\bibinfo {author} {\bibfnamefont {J.}~\bibnamefont {Li}}\ and\ \bibinfo {author} {\bibfnamefont {H.}~\bibnamefont {Sun}},\ }\href {https://arxiv.org/abs/2009.00170} {\bibinfo {title} {An attention based neural network for jet tagging}} (\bibinfo {year} {2020}),\ \Eprint {https://arxiv.org/abs/2009.00170} {arXiv:2009.00170 [hep-ph]} \BibitemShut {NoStop}%
\bibitem [{\citenamefont {Choi}\ \emph {et~al.}(2023)\citenamefont {Choi}, \citenamefont {Li}, \citenamefont {Zhang},\ and\ \citenamefont {Zhang}}]{jet_image4}%
  \BibitemOpen
  \bibfield  {author} {\bibinfo {author} {\bibfnamefont {S.~K.}\ \bibnamefont {Choi}}, \bibinfo {author} {\bibfnamefont {J.}~\bibnamefont {Li}}, \bibinfo {author} {\bibfnamefont {C.}~\bibnamefont {Zhang}},\ and\ \bibinfo {author} {\bibfnamefont {R.}~\bibnamefont {Zhang}},\ }\bibfield  {title} {\bibinfo {title} {Automatic detection of boosted higgs boson and top quark jets in an event image},\ }\href {https://doi.org/10.1103/PhysRevD.108.116002} {\bibfield  {journal} {\bibinfo  {journal} {Phys. Rev. D}\ }\textbf {\bibinfo {volume} {108}},\ \bibinfo {pages} {116002} (\bibinfo {year} {2023})}\BibitemShut {NoStop}%
\bibitem [{\citenamefont {Kasieczka}\ \emph {et~al.}(2017)\citenamefont {Kasieczka}, \citenamefont {Plehn}, \citenamefont {Russell},\ and\ \citenamefont {Schell}}]{jet_image5}%
  \BibitemOpen
  \bibfield  {author} {\bibinfo {author} {\bibfnamefont {G.}~\bibnamefont {Kasieczka}}, \bibinfo {author} {\bibfnamefont {T.}~\bibnamefont {Plehn}}, \bibinfo {author} {\bibfnamefont {M.}~\bibnamefont {Russell}},\ and\ \bibinfo {author} {\bibfnamefont {T.}~\bibnamefont {Schell}},\ }\bibfield  {title} {\bibinfo {title} {Deep-learning top taggers or the end of qcd?},\ }\href {https://doi.org/10.1007/JHEP05(2017)006} {\bibfield  {journal} {\bibinfo  {journal} {Journal of High Energy Physics}\ }\textbf {\bibinfo {volume} {2017}},\ \bibinfo {pages} {6} (\bibinfo {year} {2017})}\BibitemShut {NoStop}%
\bibitem [{\citenamefont {Komiske}\ \emph {et~al.}(2017)\citenamefont {Komiske}, \citenamefont {Metodiev},\ and\ \citenamefont {Schwartz}}]{jet_image6}%
  \BibitemOpen
  \bibfield  {author} {\bibinfo {author} {\bibfnamefont {P.~T.}\ \bibnamefont {Komiske}}, \bibinfo {author} {\bibfnamefont {E.~M.}\ \bibnamefont {Metodiev}},\ and\ \bibinfo {author} {\bibfnamefont {M.~D.}\ \bibnamefont {Schwartz}},\ }\bibfield  {title} {\bibinfo {title} {Deep learning in color: towards automated quark/gluon jet discrimination},\ }\href {https://doi.org/10.1007/JHEP01(2017)110} {\bibfield  {journal} {\bibinfo  {journal} {Journal of High Energy Physics}\ }\textbf {\bibinfo {volume} {2017}},\ \bibinfo {pages} {110} (\bibinfo {year} {2017})}\BibitemShut {NoStop}%
\bibitem [{\citenamefont {Baldi}\ \emph {et~al.}(2016)\citenamefont {Baldi}, \citenamefont {Bauer}, \citenamefont {Eng}, \citenamefont {Sadowski},\ and\ \citenamefont {Whiteson}}]{jet_image7}%
  \BibitemOpen
  \bibfield  {author} {\bibinfo {author} {\bibfnamefont {P.}~\bibnamefont {Baldi}}, \bibinfo {author} {\bibfnamefont {K.}~\bibnamefont {Bauer}}, \bibinfo {author} {\bibfnamefont {C.}~\bibnamefont {Eng}}, \bibinfo {author} {\bibfnamefont {P.}~\bibnamefont {Sadowski}},\ and\ \bibinfo {author} {\bibfnamefont {D.}~\bibnamefont {Whiteson}},\ }\bibfield  {title} {\bibinfo {title} {Jet substructure classification in high-energy physics with deep neural networks},\ }\href {https://doi.org/10.1103/PhysRevD.93.094034} {\bibfield  {journal} {\bibinfo  {journal} {Phys. Rev. D}\ }\textbf {\bibinfo {volume} {93}},\ \bibinfo {pages} {094034} (\bibinfo {year} {2016})}\BibitemShut {NoStop}%
\bibitem [{jet(2017)}]{jet_seq1}%
  \BibitemOpen
  \href {https://cds.cern.ch/record/2255226} {\emph {\bibinfo {title} {{Identification of Jets Containing $b$-Hadrons with Recurrent Neural Networks at the ATLAS Experiment}}}},\ \bibinfo {type} {Tech. Rep.}\ (\bibinfo  {institution} {CERN},\ \bibinfo {address} {Geneva},\ \bibinfo {year} {2017})\ \bibinfo {note} {all figures including auxiliary figures are available at \url{https://atlas.web.cern.ch/Atlas/GROUPS/PHYSICS/PUBNOTES/ATL-PHYS-PUB-2017-003}}\BibitemShut {NoStop}%
\bibitem [{\citenamefont {de~Lima}(2021)}]{jet_seq2}%
  \BibitemOpen
  \bibfield  {author} {\bibinfo {author} {\bibfnamefont {R.~T.}\ \bibnamefont {de~Lima}},\ }\href {https://arxiv.org/abs/2102.06128} {\bibinfo {title} {Sequence-based machine learning models in jet physics}} (\bibinfo {year} {2021}),\ \Eprint {https://arxiv.org/abs/2102.06128} {arXiv:2102.06128 [physics.data-an]} \BibitemShut {NoStop}%
\bibitem [{\citenamefont {Bols}\ \emph {et~al.}(2020)\citenamefont {Bols}, \citenamefont {Kieseler}, \citenamefont {Verzetti}, \citenamefont {Stoye},\ and\ \citenamefont {Stakia}}]{jet_seq3}%
  \BibitemOpen
  \bibfield  {author} {\bibinfo {author} {\bibfnamefont {E.}~\bibnamefont {Bols}}, \bibinfo {author} {\bibfnamefont {J.}~\bibnamefont {Kieseler}}, \bibinfo {author} {\bibfnamefont {M.}~\bibnamefont {Verzetti}}, \bibinfo {author} {\bibfnamefont {M.}~\bibnamefont {Stoye}},\ and\ \bibinfo {author} {\bibfnamefont {A.}~\bibnamefont {Stakia}},\ }\bibfield  {title} {\bibinfo {title} {Jet flavour classification using deepjet},\ }\href {https://doi.org/10.1088/1748-0221/15/12/P12012} {\bibfield  {journal} {\bibinfo  {journal} {Journal of Instrumentation}\ }\textbf {\bibinfo {volume} {15}}\bibinfo  {number} { (12)},\ \bibinfo {pages} {P12012}}\BibitemShut {NoStop}%
\bibitem [{\citenamefont {Guest}\ \emph {et~al.}(2016)\citenamefont {Guest}, \citenamefont {Collado}, \citenamefont {Baldi}, \citenamefont {Hsu}, \citenamefont {Urban},\ and\ \citenamefont {Whiteson}}]{jet_seq4}%
  \BibitemOpen
\bibfield  {number} {  }\bibfield  {author} {\bibinfo {author} {\bibfnamefont {D.}~\bibnamefont {Guest}}, \bibinfo {author} {\bibfnamefont {J.}~\bibnamefont {Collado}}, \bibinfo {author} {\bibfnamefont {P.}~\bibnamefont {Baldi}}, \bibinfo {author} {\bibfnamefont {S.-C.}\ \bibnamefont {Hsu}}, \bibinfo {author} {\bibfnamefont {G.}~\bibnamefont {Urban}},\ and\ \bibinfo {author} {\bibfnamefont {D.}~\bibnamefont {Whiteson}},\ }\bibfield  {title} {\bibinfo {title} {Jet flavor classification in high-energy physics with deep neural networks},\ }\href {https://doi.org/10.1103/PhysRevD.94.112002} {\bibfield  {journal} {\bibinfo  {journal} {Phys. Rev. D}\ }\textbf {\bibinfo {volume} {94}},\ \bibinfo {pages} {112002} (\bibinfo {year} {2016})}\BibitemShut {NoStop}%
\bibitem [{\citenamefont {Lee}\ \emph {et~al.}(2019{\natexlab{b}})\citenamefont {Lee}, \citenamefont {Lee}, \citenamefont {Lee}, \citenamefont {Park}, \citenamefont {Watson},\ and\ \citenamefont {Yang}}]{jet_seq5}%
  \BibitemOpen
  \bibfield  {author} {\bibinfo {author} {\bibfnamefont {J.~S.~H.}\ \bibnamefont {Lee}}, \bibinfo {author} {\bibfnamefont {S.~M.}\ \bibnamefont {Lee}}, \bibinfo {author} {\bibfnamefont {Y.}~\bibnamefont {Lee}}, \bibinfo {author} {\bibfnamefont {I.}~\bibnamefont {Park}}, \bibinfo {author} {\bibfnamefont {I.~J.}\ \bibnamefont {Watson}},\ and\ \bibinfo {author} {\bibfnamefont {S.}~\bibnamefont {Yang}},\ }\bibfield  {title} {\bibinfo {title} {Quark gluon jet discrimination with weakly supervised learning},\ }\href {https://doi.org/10.3938/jkps.75.652} {\bibfield  {journal} {\bibinfo  {journal} {Journal of the Korean Physical Society}\ }\textbf {\bibinfo {volume} {75}},\ \bibinfo {pages} {652} (\bibinfo {year} {2019}{\natexlab{b}})}\BibitemShut {NoStop}%
\bibitem [{\citenamefont {Egan}\ \emph {et~al.}(2017)\citenamefont {Egan}, \citenamefont {Fedorko}, \citenamefont {Lister}, \citenamefont {Pearkes},\ and\ \citenamefont {Gay}}]{jet_seq6}%
  \BibitemOpen
  \bibfield  {author} {\bibinfo {author} {\bibfnamefont {S.}~\bibnamefont {Egan}}, \bibinfo {author} {\bibfnamefont {W.}~\bibnamefont {Fedorko}}, \bibinfo {author} {\bibfnamefont {A.}~\bibnamefont {Lister}}, \bibinfo {author} {\bibfnamefont {J.}~\bibnamefont {Pearkes}},\ and\ \bibinfo {author} {\bibfnamefont {C.}~\bibnamefont {Gay}},\ }\href@noop {} {\bibinfo {title} {Long short-term memory (lstm) networks with jet constituents for boosted top tagging at the lhc}} (\bibinfo {year} {2017}),\ \Eprint {https://arxiv.org/abs/1711.09059} {arXiv:1711.09059 [hep-ex]} \BibitemShut {NoStop}%
\bibitem [{\citenamefont {Pearkes}\ \emph {et~al.}(2017)\citenamefont {Pearkes}, \citenamefont {Fedorko}, \citenamefont {Lister},\ and\ \citenamefont {Gay}}]{jet_seq7}%
  \BibitemOpen
  \bibfield  {author} {\bibinfo {author} {\bibfnamefont {J.}~\bibnamefont {Pearkes}}, \bibinfo {author} {\bibfnamefont {W.}~\bibnamefont {Fedorko}}, \bibinfo {author} {\bibfnamefont {A.}~\bibnamefont {Lister}},\ and\ \bibinfo {author} {\bibfnamefont {C.}~\bibnamefont {Gay}},\ }\href@noop {} {\bibinfo {title} {Jet constituents for deep neural network based top quark tagging}} (\bibinfo {year} {2017}),\ \Eprint {https://arxiv.org/abs/1704.02124} {arXiv:1704.02124 [hep-ex]} \BibitemShut {NoStop}%
\bibitem [{\citenamefont {Cheng}(2018)}]{jet_tree1}%
  \BibitemOpen
  \bibfield  {author} {\bibinfo {author} {\bibfnamefont {T.}~\bibnamefont {Cheng}},\ }\bibfield  {title} {\bibinfo {title} {Recursive neural networks in quark/gluon tagging},\ }\href {https://doi.org/10.1007/s41781-018-0007-y} {\bibfield  {journal} {\bibinfo  {journal} {Computing and Software for Big Science}\ }\textbf {\bibinfo {volume} {2}},\ \bibinfo {pages} {3} (\bibinfo {year} {2018})}\BibitemShut {NoStop}%
\bibitem [{\citenamefont {Louppe}\ \emph {et~al.}(2019)\citenamefont {Louppe}, \citenamefont {Cho}, \citenamefont {Becot},\ and\ \citenamefont {Cranmer}}]{jet_tree2}%
  \BibitemOpen
  \bibfield  {author} {\bibinfo {author} {\bibfnamefont {G.}~\bibnamefont {Louppe}}, \bibinfo {author} {\bibfnamefont {K.}~\bibnamefont {Cho}}, \bibinfo {author} {\bibfnamefont {C.}~\bibnamefont {Becot}},\ and\ \bibinfo {author} {\bibfnamefont {K.}~\bibnamefont {Cranmer}},\ }\bibfield  {title} {\bibinfo {title} {Qcd-aware recursive neural networks for jet physics},\ }\href {https://doi.org/10.1007/JHEP01(2019)057} {\bibfield  {journal} {\bibinfo  {journal} {Journal of High Energy Physics}\ }\textbf {\bibinfo {volume} {2019}},\ \bibinfo {pages} {57} (\bibinfo {year} {2019})}\BibitemShut {NoStop}%
\bibitem [{\citenamefont {Henrion}\ \emph {et~al.}(2017)\citenamefont {Henrion}, \citenamefont {Brehmer}, \citenamefont {Bruna}, \citenamefont {Cho}, \citenamefont {Cranmer}, \citenamefont {Louppe},\ and\ \citenamefont {Rochette}}]{jet_graph1}%
  \BibitemOpen
  \bibfield  {author} {\bibinfo {author} {\bibfnamefont {I.}~\bibnamefont {Henrion}}, \bibinfo {author} {\bibfnamefont {J.}~\bibnamefont {Brehmer}}, \bibinfo {author} {\bibfnamefont {J.}~\bibnamefont {Bruna}}, \bibinfo {author} {\bibfnamefont {K.}~\bibnamefont {Cho}}, \bibinfo {author} {\bibfnamefont {K.}~\bibnamefont {Cranmer}}, \bibinfo {author} {\bibfnamefont {G.}~\bibnamefont {Louppe}},\ and\ \bibinfo {author} {\bibfnamefont {G.}~\bibnamefont {Rochette}},\ }\bibfield  {title} {\bibinfo {title} {Neural message passing for jet physics}\ }(\bibinfo {year} {2017})\BibitemShut {NoStop}%
\bibitem [{\citenamefont {Moreno}\ \emph {et~al.}(2020)\citenamefont {Moreno}, \citenamefont {Cerri}, \citenamefont {Duarte}, \citenamefont {Newman}, \citenamefont {Nguyen}, \citenamefont {Periwal}, \citenamefont {Pierini}, \citenamefont {Serikova}, \citenamefont {Spiropulu},\ and\ \citenamefont {Vlimant}}]{jet_graph2}%
  \BibitemOpen
  \bibfield  {author} {\bibinfo {author} {\bibfnamefont {E.~A.}\ \bibnamefont {Moreno}}, \bibinfo {author} {\bibfnamefont {O.}~\bibnamefont {Cerri}}, \bibinfo {author} {\bibfnamefont {J.~M.}\ \bibnamefont {Duarte}}, \bibinfo {author} {\bibfnamefont {H.~B.}\ \bibnamefont {Newman}}, \bibinfo {author} {\bibfnamefont {T.~Q.}\ \bibnamefont {Nguyen}}, \bibinfo {author} {\bibfnamefont {A.}~\bibnamefont {Periwal}}, \bibinfo {author} {\bibfnamefont {M.}~\bibnamefont {Pierini}}, \bibinfo {author} {\bibfnamefont {A.}~\bibnamefont {Serikova}}, \bibinfo {author} {\bibfnamefont {M.}~\bibnamefont {Spiropulu}},\ and\ \bibinfo {author} {\bibfnamefont {J.-R.}\ \bibnamefont {Vlimant}},\ }\bibfield  {title} {\bibinfo {title} {Jedi-net: a jet identification algorithm based on interaction networks},\ }\href {https://doi.org/10.1140/epjc/s10052-020-7608-4} {\bibfield  {journal} {\bibinfo  {journal} {The European Physical Journal C}\ }\textbf {\bibinfo {volume} {80}},\ \bibinfo {pages} {58} (\bibinfo {year} {2020})}\BibitemShut {NoStop}%
\bibitem [{\citenamefont {Chakraborty}\ \emph {et~al.}(2019)\citenamefont {Chakraborty}, \citenamefont {Lim},\ and\ \citenamefont {Nojiri}}]{jet_graph3}%
  \BibitemOpen
  \bibfield  {author} {\bibinfo {author} {\bibfnamefont {A.}~\bibnamefont {Chakraborty}}, \bibinfo {author} {\bibfnamefont {S.~H.}\ \bibnamefont {Lim}},\ and\ \bibinfo {author} {\bibfnamefont {M.~M.}\ \bibnamefont {Nojiri}},\ }\bibfield  {title} {\bibinfo {title} {Interpretable deep learning for two-prong jet classification with jet spectra},\ }\href {https://doi.org/10.1007/JHEP07(2019)135} {\bibfield  {journal} {\bibinfo  {journal} {Journal of High Energy Physics}\ }\textbf {\bibinfo {volume} {2019}},\ \bibinfo {pages} {135} (\bibinfo {year} {2019})}\BibitemShut {NoStop}%
\bibitem [{\citenamefont {Chakraborty}\ \emph {et~al.}(2020)\citenamefont {Chakraborty}, \citenamefont {Lim}, \citenamefont {Nojiri},\ and\ \citenamefont {Takeuchi}}]{jet_graph4}%
  \BibitemOpen
  \bibfield  {author} {\bibinfo {author} {\bibfnamefont {A.}~\bibnamefont {Chakraborty}}, \bibinfo {author} {\bibfnamefont {S.~H.}\ \bibnamefont {Lim}}, \bibinfo {author} {\bibfnamefont {M.~M.}\ \bibnamefont {Nojiri}},\ and\ \bibinfo {author} {\bibfnamefont {M.}~\bibnamefont {Takeuchi}},\ }\bibfield  {title} {\bibinfo {title} {Neural network-based top tagger with two-point energy correlations and geometry of soft emissions},\ }\href {https://doi.org/10.1007/JHEP07(2020)111} {\bibfield  {journal} {\bibinfo  {journal} {Journal of High Energy Physics}\ }\textbf {\bibinfo {volume} {2020}},\ \bibinfo {pages} {111} (\bibinfo {year} {2020})}\BibitemShut {NoStop}%
\bibitem [{\citenamefont {Shlomi}\ \emph {et~al.}(2020)\citenamefont {Shlomi}, \citenamefont {Battaglia},\ and\ \citenamefont {Vlimant}}]{jet_graph5}%
  \BibitemOpen
  \bibfield  {author} {\bibinfo {author} {\bibfnamefont {J.}~\bibnamefont {Shlomi}}, \bibinfo {author} {\bibfnamefont {P.}~\bibnamefont {Battaglia}},\ and\ \bibinfo {author} {\bibfnamefont {J.-R.}\ \bibnamefont {Vlimant}},\ }\bibfield  {title} {\bibinfo {title} {Graph neural networks in particle physics},\ }\href {https://doi.org/10.1088/2632-2153/abbf9a} {\bibfield  {journal} {\bibinfo  {journal} {Machine Learning: Science and Technology}\ }\textbf {\bibinfo {volume} {2}},\ \bibinfo {pages} {021001} (\bibinfo {year} {2020})}\BibitemShut {NoStop}%
\bibitem [{\citenamefont {Ju}\ and\ \citenamefont {Nachman}(2020)}]{jet_graph6}%
  \BibitemOpen
  \bibfield  {author} {\bibinfo {author} {\bibfnamefont {X.}~\bibnamefont {Ju}}\ and\ \bibinfo {author} {\bibfnamefont {B.}~\bibnamefont {Nachman}},\ }\bibfield  {title} {\bibinfo {title} {Supervised jet clustering with graph neural networks for lorentz boosted bosons},\ }\href {https://doi.org/10.1103/PhysRevD.102.075014} {\bibfield  {journal} {\bibinfo  {journal} {Phys. Rev. D}\ }\textbf {\bibinfo {volume} {102}},\ \bibinfo {pages} {075014} (\bibinfo {year} {2020})}\BibitemShut {NoStop}%
\bibitem [{\citenamefont {Dreyer}\ and\ \citenamefont {Qu}(2021)}]{jet_graph7}%
  \BibitemOpen
  \bibfield  {author} {\bibinfo {author} {\bibfnamefont {F.~A.}\ \bibnamefont {Dreyer}}\ and\ \bibinfo {author} {\bibfnamefont {H.}~\bibnamefont {Qu}},\ }\href {https://arxiv.org/abs/2012.08526} {\bibinfo {title} {Jet tagging in the lund plane with graph networks}} (\bibinfo {year} {2021}),\ \Eprint {https://arxiv.org/abs/2012.08526} {arXiv:2012.08526 [hep-ph]} \BibitemShut {NoStop}%
\bibitem [{\citenamefont {Gong}\ \emph {et~al.}(2022)\citenamefont {Gong}, \citenamefont {Meng}, \citenamefont {Zhang}, \citenamefont {Qu}, \citenamefont {Li}, \citenamefont {Qian}, \citenamefont {Du}, \citenamefont {Ma},\ and\ \citenamefont {Liu}}]{jet_graph8}%
  \BibitemOpen
  \bibfield  {author} {\bibinfo {author} {\bibfnamefont {S.}~\bibnamefont {Gong}}, \bibinfo {author} {\bibfnamefont {Q.}~\bibnamefont {Meng}}, \bibinfo {author} {\bibfnamefont {J.}~\bibnamefont {Zhang}}, \bibinfo {author} {\bibfnamefont {H.}~\bibnamefont {Qu}}, \bibinfo {author} {\bibfnamefont {C.}~\bibnamefont {Li}}, \bibinfo {author} {\bibfnamefont {S.}~\bibnamefont {Qian}}, \bibinfo {author} {\bibfnamefont {W.}~\bibnamefont {Du}}, \bibinfo {author} {\bibfnamefont {Z.-M.}\ \bibnamefont {Ma}},\ and\ \bibinfo {author} {\bibfnamefont {T.-Y.}\ \bibnamefont {Liu}},\ }\bibfield  {title} {\bibinfo {title} {An efficient lorentz equivariant graph neural network for jet tagging},\ }\href {https://doi.org/10.1007/JHEP07(2022)030} {\bibfield  {journal} {\bibinfo  {journal} {Journal of High Energy Physics}\ }\textbf {\bibinfo {volume} {2022}},\ \bibinfo {pages} {30} (\bibinfo {year} {2022})}\BibitemShut {NoStop}%
\bibitem [{\citenamefont {Ma}\ \emph {et~al.}(2023)\citenamefont {Ma}, \citenamefont {Liu},\ and\ \citenamefont {Li}}]{jet_graph9}%
  \BibitemOpen
  \bibfield  {author} {\bibinfo {author} {\bibfnamefont {F.}~\bibnamefont {Ma}}, \bibinfo {author} {\bibfnamefont {F.}~\bibnamefont {Liu}},\ and\ \bibinfo {author} {\bibfnamefont {W.}~\bibnamefont {Li}},\ }\bibfield  {title} {\bibinfo {title} {Jet tagging algorithm of graph network with haar pooling message passing},\ }\href {https://doi.org/10.1103/PhysRevD.108.072007} {\bibfield  {journal} {\bibinfo  {journal} {Phys. Rev. D}\ }\textbf {\bibinfo {volume} {108}},\ \bibinfo {pages} {072007} (\bibinfo {year} {2023})}\BibitemShut {NoStop}%
\bibitem [{\citenamefont {Mokhtar}\ \emph {et~al.}(2022)\citenamefont {Mokhtar}, \citenamefont {Kansal},\ and\ \citenamefont {Duarte}}]{jet_graph10}%
  \BibitemOpen
  \bibfield  {author} {\bibinfo {author} {\bibfnamefont {F.}~\bibnamefont {Mokhtar}}, \bibinfo {author} {\bibfnamefont {R.}~\bibnamefont {Kansal}},\ and\ \bibinfo {author} {\bibfnamefont {J.}~\bibnamefont {Duarte}},\ }\href {https://arxiv.org/abs/2211.09912} {\bibinfo {title} {Do graph neural networks learn traditional jet substructure?}} (\bibinfo {year} {2022}),\ \Eprint {https://arxiv.org/abs/2211.09912} {arXiv:2211.09912 [hep-ex]} \BibitemShut {NoStop}%
\bibitem [{\citenamefont {Murnane}(2023)}]{jet_graph11}%
  \BibitemOpen
  \bibfield  {author} {\bibinfo {author} {\bibfnamefont {D.}~\bibnamefont {Murnane}},\ }\href {https://arxiv.org/abs/2307.16662} {\bibinfo {title} {Graph structure from point clouds: Geometric attention is all you need}} (\bibinfo {year} {2023}),\ \Eprint {https://arxiv.org/abs/2307.16662} {arXiv:2307.16662 [cs.LG]} \BibitemShut {NoStop}%
\bibitem [{\citenamefont {Thais}\ \emph {et~al.}(2022)\citenamefont {Thais}, \citenamefont {Calafiura}, \citenamefont {Chachamis}, \citenamefont {DeZoort}, \citenamefont {Duarte}, \citenamefont {Ganguly}, \citenamefont {Kagan}, \citenamefont {Murnane}, \citenamefont {Neubauer},\ and\ \citenamefont {Terao}}]{jet_graph12}%
  \BibitemOpen
  \bibfield  {author} {\bibinfo {author} {\bibfnamefont {S.}~\bibnamefont {Thais}}, \bibinfo {author} {\bibfnamefont {P.}~\bibnamefont {Calafiura}}, \bibinfo {author} {\bibfnamefont {G.}~\bibnamefont {Chachamis}}, \bibinfo {author} {\bibfnamefont {G.}~\bibnamefont {DeZoort}}, \bibinfo {author} {\bibfnamefont {J.}~\bibnamefont {Duarte}}, \bibinfo {author} {\bibfnamefont {S.}~\bibnamefont {Ganguly}}, \bibinfo {author} {\bibfnamefont {M.}~\bibnamefont {Kagan}}, \bibinfo {author} {\bibfnamefont {D.}~\bibnamefont {Murnane}}, \bibinfo {author} {\bibfnamefont {M.~S.}\ \bibnamefont {Neubauer}},\ and\ \bibinfo {author} {\bibfnamefont {K.}~\bibnamefont {Terao}},\ }\href@noop {} {\bibinfo {title} {Graph neural networks in particle physics: Implementations, innovations, and challenges}} (\bibinfo {year} {2022}),\ \Eprint {https://arxiv.org/abs/2203.12852} {arXiv:2203.12852 [hep-ex]} \BibitemShut {NoStop}%
\bibitem [{\citenamefont {Guo}\ \emph {et~al.}(2021)\citenamefont {Guo}, \citenamefont {Li}, \citenamefont {Li},\ and\ \citenamefont {Zhang}}]{jet_graph13}%
  \BibitemOpen
  \bibfield  {author} {\bibinfo {author} {\bibfnamefont {J.}~\bibnamefont {Guo}}, \bibinfo {author} {\bibfnamefont {J.}~\bibnamefont {Li}}, \bibinfo {author} {\bibfnamefont {T.}~\bibnamefont {Li}},\ and\ \bibinfo {author} {\bibfnamefont {R.}~\bibnamefont {Zhang}},\ }\bibfield  {title} {\bibinfo {title} {Boosted higgs boson jet reconstruction via a graph neural network},\ }\href {https://doi.org/10.1103/PhysRevD.103.116025} {\bibfield  {journal} {\bibinfo  {journal} {Phys. Rev. D}\ }\textbf {\bibinfo {volume} {103}},\ \bibinfo {pages} {116025} (\bibinfo {year} {2021})}\BibitemShut {NoStop}%
\bibitem [{\citenamefont {Komiske}\ \emph {et~al.}(2019)\citenamefont {Komiske}, \citenamefont {Metodiev},\ and\ \citenamefont {Thaler}}]{pfn}%
  \BibitemOpen
  \bibfield  {author} {\bibinfo {author} {\bibfnamefont {P.~T.}\ \bibnamefont {Komiske}}, \bibinfo {author} {\bibfnamefont {E.~M.}\ \bibnamefont {Metodiev}},\ and\ \bibinfo {author} {\bibfnamefont {J.}~\bibnamefont {Thaler}},\ }\bibfield  {title} {\bibinfo {title} {Energy flow networks: deep sets for particle jets},\ }\href {https://doi.org/10.1007/JHEP01(2019)121} {\bibfield  {journal} {\bibinfo  {journal} {Journal of High Energy Physics}\ }\textbf {\bibinfo {volume} {2019}},\ \bibinfo {pages} {121} (\bibinfo {year} {2019})}\BibitemShut {NoStop}%
\bibitem [{\citenamefont {Qu}\ and\ \citenamefont {Gouskos}(2020)}]{ptcnet}%
  \BibitemOpen
  \bibfield  {author} {\bibinfo {author} {\bibfnamefont {H.}~\bibnamefont {Qu}}\ and\ \bibinfo {author} {\bibfnamefont {L.}~\bibnamefont {Gouskos}},\ }\bibfield  {title} {\bibinfo {title} {Jet tagging via particle clouds},\ }\href {https://doi.org/10.1103/PhysRevD.101.056019} {\bibfield  {journal} {\bibinfo  {journal} {Phys. Rev. D}\ }\textbf {\bibinfo {volume} {101}},\ \bibinfo {pages} {056019} (\bibinfo {year} {2020})}\BibitemShut {NoStop}%
\bibitem [{\citenamefont {Qu}\ \emph {et~al.}(2024)\citenamefont {Qu}, \citenamefont {Li},\ and\ \citenamefont {Qian}}]{part}%
  \BibitemOpen
  \bibfield  {author} {\bibinfo {author} {\bibfnamefont {H.}~\bibnamefont {Qu}}, \bibinfo {author} {\bibfnamefont {C.}~\bibnamefont {Li}},\ and\ \bibinfo {author} {\bibfnamefont {S.}~\bibnamefont {Qian}},\ }\href {https://arxiv.org/abs/2202.03772} {\bibinfo {title} {Particle transformer for jet tagging}} (\bibinfo {year} {2024}),\ \Eprint {https://arxiv.org/abs/2202.03772} {arXiv:2202.03772 [hep-ph]} \BibitemShut {NoStop}%
\bibitem [{\citenamefont {Dolan}\ and\ \citenamefont {Ore}(2021)}]{jet_set1}%
  \BibitemOpen
  \bibfield  {author} {\bibinfo {author} {\bibfnamefont {M.~J.}\ \bibnamefont {Dolan}}\ and\ \bibinfo {author} {\bibfnamefont {A.}~\bibnamefont {Ore}},\ }\bibfield  {title} {\bibinfo {title} {Equivariant energy flow networks for jet tagging},\ }\href {https://doi.org/10.1103/PhysRevD.103.074022} {\bibfield  {journal} {\bibinfo  {journal} {Phys. Rev. D}\ }\textbf {\bibinfo {volume} {103}},\ \bibinfo {pages} {074022} (\bibinfo {year} {2021})}\BibitemShut {NoStop}%
\bibitem [{jet(2020)}]{jet_set2}%
  \BibitemOpen
  \href {https://cds.cern.ch/record/2718948} {\emph {\bibinfo {title} {{Deep Sets based Neural Networks for Impact Parameter Flavour Tagging in ATLAS}}}},\ \bibinfo {type} {Tech. Rep.}\ (\bibinfo  {institution} {CERN},\ \bibinfo {address} {Geneva},\ \bibinfo {year} {2020})\ \bibinfo {note} {all figures including auxiliary figures are available at \url{https://atlas.web.cern.ch/Atlas/GROUPS/PHYSICS/PUBNOTES/ATL-PHYS-PUB-2020-014}}\BibitemShut {NoStop}%
\bibitem [{\citenamefont {Käch}\ \emph {et~al.}(2022)\citenamefont {Käch}, \citenamefont {Krücker},\ and\ \citenamefont {Melzer-Pellmann}}]{jet_set3}%
  \BibitemOpen
  \bibfield  {author} {\bibinfo {author} {\bibfnamefont {B.}~\bibnamefont {Käch}}, \bibinfo {author} {\bibfnamefont {D.}~\bibnamefont {Krücker}},\ and\ \bibinfo {author} {\bibfnamefont {I.}~\bibnamefont {Melzer-Pellmann}},\ }\href {https://arxiv.org/abs/2211.13623} {\bibinfo {title} {Point cloud generation using transformer encoders and normalising flows}} (\bibinfo {year} {2022}),\ \Eprint {https://arxiv.org/abs/2211.13623} {arXiv:2211.13623 [hep-ex]} \BibitemShut {NoStop}%
\bibitem [{\citenamefont {Athanasakos}\ \emph {et~al.}(2023)\citenamefont {Athanasakos}, \citenamefont {Larkoski}, \citenamefont {Mulligan}, \citenamefont {Ploskon},\ and\ \citenamefont {Ringer}}]{jet_set4}%
  \BibitemOpen
  \bibfield  {author} {\bibinfo {author} {\bibfnamefont {D.}~\bibnamefont {Athanasakos}}, \bibinfo {author} {\bibfnamefont {A.~J.}\ \bibnamefont {Larkoski}}, \bibinfo {author} {\bibfnamefont {J.}~\bibnamefont {Mulligan}}, \bibinfo {author} {\bibfnamefont {M.}~\bibnamefont {Ploskon}},\ and\ \bibinfo {author} {\bibfnamefont {F.}~\bibnamefont {Ringer}},\ }\href {https://arxiv.org/abs/2305.08979} {\bibinfo {title} {Is infrared-collinear safe information all you need for jet classification?}} (\bibinfo {year} {2023}),\ \Eprint {https://arxiv.org/abs/2305.08979} {arXiv:2305.08979 [hep-ph]} \BibitemShut {NoStop}%
\bibitem [{\citenamefont {Käch}\ and\ \citenamefont {Melzer-Pellmann}(2023)}]{jet_set5}%
  \BibitemOpen
  \bibfield  {author} {\bibinfo {author} {\bibfnamefont {B.}~\bibnamefont {Käch}}\ and\ \bibinfo {author} {\bibfnamefont {I.}~\bibnamefont {Melzer-Pellmann}},\ }\href {https://arxiv.org/abs/2305.15254} {\bibinfo {title} {Attention to mean-fields for particle cloud generation}} (\bibinfo {year} {2023}),\ \Eprint {https://arxiv.org/abs/2305.15254} {arXiv:2305.15254 [hep-ex]} \BibitemShut {NoStop}%
\bibitem [{\citenamefont {Mondal}\ \emph {et~al.}(2023)\citenamefont {Mondal}, \citenamefont {Barone},\ and\ \citenamefont {Schmidt}}]{jet_set6}%
  \BibitemOpen
  \bibfield  {author} {\bibinfo {author} {\bibfnamefont {S.}~\bibnamefont {Mondal}}, \bibinfo {author} {\bibfnamefont {G.}~\bibnamefont {Barone}},\ and\ \bibinfo {author} {\bibfnamefont {A.}~\bibnamefont {Schmidt}},\ }\href {https://arxiv.org/abs/2311.11011} {\bibinfo {title} {Paired jet: A multi-pronged resonance tagging strategy across all lorentz boosts}} (\bibinfo {year} {2023}),\ \Eprint {https://arxiv.org/abs/2311.11011} {arXiv:2311.11011 [hep-ex]} \BibitemShut {NoStop}%
\bibitem [{\citenamefont {Odagiu}\ \emph {et~al.}(2024)\citenamefont {Odagiu}, \citenamefont {Que}, \citenamefont {Duarte}, \citenamefont {Haller}, \citenamefont {Kasieczka}, \citenamefont {Lobanov}, \citenamefont {Loncar}, \citenamefont {Luk}, \citenamefont {Ngadiuba}, \citenamefont {Pierini}, \citenamefont {Rincke}, \citenamefont {Seksaria}, \citenamefont {Summers}, \citenamefont {Sznajder}, \citenamefont {Tapper},\ and\ \citenamefont {Aarrestad}}]{jet_set7}%
  \BibitemOpen
  \bibfield  {author} {\bibinfo {author} {\bibfnamefont {P.}~\bibnamefont {Odagiu}}, \bibinfo {author} {\bibfnamefont {Z.}~\bibnamefont {Que}}, \bibinfo {author} {\bibfnamefont {J.}~\bibnamefont {Duarte}}, \bibinfo {author} {\bibfnamefont {J.}~\bibnamefont {Haller}}, \bibinfo {author} {\bibfnamefont {G.}~\bibnamefont {Kasieczka}}, \bibinfo {author} {\bibfnamefont {A.}~\bibnamefont {Lobanov}}, \bibinfo {author} {\bibfnamefont {V.}~\bibnamefont {Loncar}}, \bibinfo {author} {\bibfnamefont {W.}~\bibnamefont {Luk}}, \bibinfo {author} {\bibfnamefont {J.}~\bibnamefont {Ngadiuba}}, \bibinfo {author} {\bibfnamefont {M.}~\bibnamefont {Pierini}}, \bibinfo {author} {\bibfnamefont {P.}~\bibnamefont {Rincke}}, \bibinfo {author} {\bibfnamefont {A.}~\bibnamefont {Seksaria}}, \bibinfo {author} {\bibfnamefont {S.}~\bibnamefont {Summers}}, \bibinfo {author} {\bibfnamefont {A.}~\bibnamefont {Sznajder}}, \bibinfo {author} {\bibfnamefont {A.}~\bibnamefont {Tapper}},\ and\ \bibinfo {author} {\bibfnamefont {T.~K.}\ \bibnamefont {Aarrestad}},\ }\href {https://arxiv.org/abs/2402.01876} {\bibinfo {title} {Sets are all you need: Ultrafast jet classification on fpgas for hl-lhc}} (\bibinfo {year} {2024}),\ \Eprint {https://arxiv.org/abs/2402.01876} {arXiv:2402.01876 [hep-ex]} \BibitemShut {NoStop}%
\bibitem [{\citenamefont {Gambhir}\ \emph {et~al.}(2024)\citenamefont {Gambhir}, \citenamefont {Osathapan},\ and\ \citenamefont {Thaler}}]{jet_set8}%
  \BibitemOpen
  \bibfield  {author} {\bibinfo {author} {\bibfnamefont {R.}~\bibnamefont {Gambhir}}, \bibinfo {author} {\bibfnamefont {A.}~\bibnamefont {Osathapan}},\ and\ \bibinfo {author} {\bibfnamefont {J.}~\bibnamefont {Thaler}},\ }\href {https://arxiv.org/abs/2403.08854} {\bibinfo {title} {Moments of clarity: Streamlining latent spaces in machine learning using moment pooling}} (\bibinfo {year} {2024}),\ \Eprint {https://arxiv.org/abs/2403.08854} {arXiv:2403.08854 [hep-ph]} \BibitemShut {NoStop}%
\bibitem [{\citenamefont {Biamonte}\ \emph {et~al.}(2017)\citenamefont {Biamonte}, \citenamefont {Wittek}, \citenamefont {Pancotti}, \citenamefont {Rebentrost}, \citenamefont {Wiebe},\ and\ \citenamefont {Lloyd}}]{qml1}%
  \BibitemOpen
  \bibfield  {author} {\bibinfo {author} {\bibfnamefont {J.}~\bibnamefont {Biamonte}}, \bibinfo {author} {\bibfnamefont {P.}~\bibnamefont {Wittek}}, \bibinfo {author} {\bibfnamefont {N.}~\bibnamefont {Pancotti}}, \bibinfo {author} {\bibfnamefont {P.}~\bibnamefont {Rebentrost}}, \bibinfo {author} {\bibfnamefont {N.}~\bibnamefont {Wiebe}},\ and\ \bibinfo {author} {\bibfnamefont {S.}~\bibnamefont {Lloyd}},\ }\bibfield  {title} {\bibinfo {title} {Quantum machine learning},\ }\href {https://doi.org/10.1038/nature23474} {\bibfield  {journal} {\bibinfo  {journal} {Nature}\ }\textbf {\bibinfo {volume} {549}},\ \bibinfo {pages} {195} (\bibinfo {year} {2017})}\BibitemShut {NoStop}%
\bibitem [{\citenamefont {Zeguendry}\ \emph {et~al.}(2023)\citenamefont {Zeguendry}, \citenamefont {Jarir},\ and\ \citenamefont {Quafafou}}]{qml2}%
  \BibitemOpen
  \bibfield  {author} {\bibinfo {author} {\bibfnamefont {A.}~\bibnamefont {Zeguendry}}, \bibinfo {author} {\bibfnamefont {Z.}~\bibnamefont {Jarir}},\ and\ \bibinfo {author} {\bibfnamefont {M.}~\bibnamefont {Quafafou}},\ }\bibfield  {title} {\bibinfo {title} {Quantum machine learning: A review and case studies},\ }\bibfield  {journal} {\bibinfo  {journal} {Entropy}\ }\textbf {\bibinfo {volume} {25}},\ \href {https://doi.org/10.3390/e25020287} {10.3390/e25020287} (\bibinfo {year} {2023})\BibitemShut {NoStop}%
\bibitem [{\citenamefont {García}\ \emph {et~al.}(2022)\citenamefont {García}, \citenamefont {Cruz-Benito},\ and\ \citenamefont {García-Peñalvo}}]{qml3}%
  \BibitemOpen
  \bibfield  {author} {\bibinfo {author} {\bibfnamefont {D.~P.}\ \bibnamefont {García}}, \bibinfo {author} {\bibfnamefont {J.}~\bibnamefont {Cruz-Benito}},\ and\ \bibinfo {author} {\bibfnamefont {F.~J.}\ \bibnamefont {García-Peñalvo}},\ }\href@noop {} {\bibinfo {title} {Systematic literature review: Quantum machine learning and its applications}} (\bibinfo {year} {2022}),\ \Eprint {https://arxiv.org/abs/2201.04093} {arXiv:2201.04093 [quant-ph]} \BibitemShut {NoStop}%
\bibitem [{\citenamefont {Tychola}\ \emph {et~al.}(2023)\citenamefont {Tychola}, \citenamefont {Kalampokas},\ and\ \citenamefont {Papakostas}}]{qml4}%
  \BibitemOpen
  \bibfield  {author} {\bibinfo {author} {\bibfnamefont {K.~A.}\ \bibnamefont {Tychola}}, \bibinfo {author} {\bibfnamefont {T.}~\bibnamefont {Kalampokas}},\ and\ \bibinfo {author} {\bibfnamefont {G.~A.}\ \bibnamefont {Papakostas}},\ }\bibfield  {title} {\bibinfo {title} {Quantum machine learning mdash;an overview},\ }\bibfield  {journal} {\bibinfo  {journal} {Electronics}\ }\textbf {\bibinfo {volume} {12}},\ \href {https://doi.org/10.3390/electronics12112379} {10.3390/electronics12112379} (\bibinfo {year} {2023})\BibitemShut {NoStop}%
\bibitem [{\citenamefont {Schuld}\ and\ \citenamefont {Petruccione}(2021)}]{qml5}%
  \BibitemOpen
  \bibfield  {author} {\bibinfo {author} {\bibfnamefont {M.}~\bibnamefont {Schuld}}\ and\ \bibinfo {author} {\bibfnamefont {F.}~\bibnamefont {Petruccione}},\ }\href {https://doi.org/10.1007/978-3-030-83098-4} {\emph {\bibinfo {title} {Machine Learning with Quantum Computers}}}\ (\bibinfo {year} {2021})\BibitemShut {NoStop}%
\bibitem [{\citenamefont {Guan}\ \emph {et~al.}(2021)\citenamefont {Guan}, \citenamefont {Perdue}, \citenamefont {Pesah}, \citenamefont {Schuld}, \citenamefont {Terashi}, \citenamefont {Vallecorsa},\ and\ \citenamefont {Vlimant}}]{qml_hep1}%
  \BibitemOpen
  \bibfield  {author} {\bibinfo {author} {\bibfnamefont {W.}~\bibnamefont {Guan}}, \bibinfo {author} {\bibfnamefont {G.}~\bibnamefont {Perdue}}, \bibinfo {author} {\bibfnamefont {A.}~\bibnamefont {Pesah}}, \bibinfo {author} {\bibfnamefont {M.}~\bibnamefont {Schuld}}, \bibinfo {author} {\bibfnamefont {K.}~\bibnamefont {Terashi}}, \bibinfo {author} {\bibfnamefont {S.}~\bibnamefont {Vallecorsa}},\ and\ \bibinfo {author} {\bibfnamefont {J.-R.}\ \bibnamefont {Vlimant}},\ }\bibfield  {title} {\bibinfo {title} {Quantum machine learning in high energy physics},\ }\href {https://doi.org/10.1088/2632-2153/abc17d} {\bibfield  {journal} {\bibinfo  {journal} {Machine Learning: Science and Technology}\ }\textbf {\bibinfo {volume} {2}},\ \bibinfo {pages} {011003} (\bibinfo {year} {2021})}\BibitemShut {NoStop}%
\bibitem [{\citenamefont {Araz}\ and\ \citenamefont {Spannowsky}(2021)}]{qml_reco1}%
  \BibitemOpen
  \bibfield  {author} {\bibinfo {author} {\bibfnamefont {J.~Y.}\ \bibnamefont {Araz}}\ and\ \bibinfo {author} {\bibfnamefont {M.}~\bibnamefont {Spannowsky}},\ }\bibfield  {title} {\bibinfo {title} {Quantum-inspired event reconstruction with tensor networks: Matrix product states},\ }\href {https://doi.org/10.1007/JHEP08(2021)112} {\bibfield  {journal} {\bibinfo  {journal} {Journal of High Energy Physics}\ }\textbf {\bibinfo {volume} {2021}},\ \bibinfo {pages} {112} (\bibinfo {year} {2021})}\BibitemShut {NoStop}%
\bibitem [{\citenamefont {Duckett}\ \emph {et~al.}(2024)\citenamefont {Duckett}, \citenamefont {Facini}, \citenamefont {Jastrzebski}, \citenamefont {Malik}, \citenamefont {Scanlon},\ and\ \citenamefont {Rettie}}]{qml_reco2}%
  \BibitemOpen
  \bibfield  {author} {\bibinfo {author} {\bibfnamefont {P.}~\bibnamefont {Duckett}}, \bibinfo {author} {\bibfnamefont {G.}~\bibnamefont {Facini}}, \bibinfo {author} {\bibfnamefont {M.}~\bibnamefont {Jastrzebski}}, \bibinfo {author} {\bibfnamefont {S.}~\bibnamefont {Malik}}, \bibinfo {author} {\bibfnamefont {T.}~\bibnamefont {Scanlon}},\ and\ \bibinfo {author} {\bibfnamefont {S.}~\bibnamefont {Rettie}},\ }\bibfield  {title} {\bibinfo {title} {Reconstructing charged particle track segments with a quantum-enhanced support vector machine},\ }\href {https://doi.org/10.1103/PhysRevD.109.052002} {\bibfield  {journal} {\bibinfo  {journal} {Phys. Rev. D}\ }\textbf {\bibinfo {volume} {109}},\ \bibinfo {pages} {052002} (\bibinfo {year} {2024})}\BibitemShut {NoStop}%
\bibitem [{\citenamefont {{T\"uys\"uz, Cenk}}\ \emph {et~al.}(2020)\citenamefont {{T\"uys\"uz, Cenk}}, \citenamefont {{Carminati, Federico}}, \citenamefont {{Demirk\"oz, Bilge}}, \citenamefont {{Dobos, Daniel}}, \citenamefont {{Fracas, Fabio}}, \citenamefont {{Novotny, Kristiane}}, \citenamefont {{Potamianos, Karolos}}, \citenamefont {{Vallecorsa, Sofia}},\ and\ \citenamefont {{Vlimant, Jean-Roch}}}]{qml_reco3}%
  \BibitemOpen
  \bibfield  {author} {\bibinfo {author} {\bibnamefont {{T\"uys\"uz, Cenk}}}, \bibinfo {author} {\bibnamefont {{Carminati, Federico}}}, \bibinfo {author} {\bibnamefont {{Demirk\"oz, Bilge}}}, \bibinfo {author} {\bibnamefont {{Dobos, Daniel}}}, \bibinfo {author} {\bibnamefont {{Fracas, Fabio}}}, \bibinfo {author} {\bibnamefont {{Novotny, Kristiane}}}, \bibinfo {author} {\bibnamefont {{Potamianos, Karolos}}}, \bibinfo {author} {\bibnamefont {{Vallecorsa, Sofia}}},\ and\ \bibinfo {author} {\bibnamefont {{Vlimant, Jean-Roch}}},\ }\bibfield  {title} {\bibinfo {title} {Particle track reconstruction with quantum algorithms},\ }\href {https://doi.org/10.1051/epjconf/202024509013} {\bibfield  {journal} {\bibinfo  {journal} {EPJ Web Conf.}\ }\textbf {\bibinfo {volume} {245}},\ \bibinfo {pages} {09013} (\bibinfo {year} {2020})}\BibitemShut {NoStop}%
\bibitem [{\citenamefont {Blance}\ and\ \citenamefont {Spannowsky}(2021{\natexlab{a}})}]{qml_cf1}%
  \BibitemOpen
  \bibfield  {author} {\bibinfo {author} {\bibfnamefont {A.}~\bibnamefont {Blance}}\ and\ \bibinfo {author} {\bibfnamefont {M.}~\bibnamefont {Spannowsky}},\ }\bibfield  {title} {\bibinfo {title} {Quantum machine learning for particle physics using a variational quantum classifier},\ }\href {https://doi.org/10.1007/JHEP02(2021)212} {\bibfield  {journal} {\bibinfo  {journal} {Journal of High Energy Physics}\ }\textbf {\bibinfo {volume} {2021}},\ \bibinfo {pages} {212} (\bibinfo {year} {2021}{\natexlab{a}})}\BibitemShut {NoStop}%
\bibitem [{\citenamefont {Terashi}\ \emph {et~al.}(2021)\citenamefont {Terashi}, \citenamefont {Kaneda}, \citenamefont {Kishimoto}, \citenamefont {Saito}, \citenamefont {Sawada},\ and\ \citenamefont {Tanaka}}]{qml_cf2}%
  \BibitemOpen
  \bibfield  {author} {\bibinfo {author} {\bibfnamefont {K.}~\bibnamefont {Terashi}}, \bibinfo {author} {\bibfnamefont {M.}~\bibnamefont {Kaneda}}, \bibinfo {author} {\bibfnamefont {T.}~\bibnamefont {Kishimoto}}, \bibinfo {author} {\bibfnamefont {M.}~\bibnamefont {Saito}}, \bibinfo {author} {\bibfnamefont {R.}~\bibnamefont {Sawada}},\ and\ \bibinfo {author} {\bibfnamefont {J.}~\bibnamefont {Tanaka}},\ }\bibfield  {title} {\bibinfo {title} {Event classification with quantum machine learning in high-energy physics},\ }\href {https://doi.org/10.1007/s41781-020-00047-7} {\bibfield  {journal} {\bibinfo  {journal} {Computing and Software for Big Science}\ }\textbf {\bibinfo {volume} {5}},\ \bibinfo {pages} {2} (\bibinfo {year} {2021})}\BibitemShut {NoStop}%
\bibitem [{\citenamefont {Chen}\ \emph {et~al.}(2020)\citenamefont {Chen}, \citenamefont {Wei}, \citenamefont {Zhang}, \citenamefont {Yu},\ and\ \citenamefont {Yoo}}]{qml_cf3}%
  \BibitemOpen
  \bibfield  {author} {\bibinfo {author} {\bibfnamefont {S.~Y.-C.}\ \bibnamefont {Chen}}, \bibinfo {author} {\bibfnamefont {T.-C.}\ \bibnamefont {Wei}}, \bibinfo {author} {\bibfnamefont {C.}~\bibnamefont {Zhang}}, \bibinfo {author} {\bibfnamefont {H.}~\bibnamefont {Yu}},\ and\ \bibinfo {author} {\bibfnamefont {S.}~\bibnamefont {Yoo}},\ }\href {https://arxiv.org/abs/2012.12177} {\bibinfo {title} {Quantum convolutional neural networks for high energy physics data analysis}} (\bibinfo {year} {2020}),\ \Eprint {https://arxiv.org/abs/2012.12177} {arXiv:2012.12177 [cs.LG]} \BibitemShut {NoStop}%
\bibitem [{\citenamefont {Wu}\ \emph {et~al.}(2021{\natexlab{a}})\citenamefont {Wu}, \citenamefont {Chan}, \citenamefont {Guan}, \citenamefont {Sun}, \citenamefont {Wang}, \citenamefont {Zhou}, \citenamefont {Livny}, \citenamefont {Carminati}, \citenamefont {Meglio}, \citenamefont {Li}, \citenamefont {Lykken}, \citenamefont {Spentzouris}, \citenamefont {Chen}, \citenamefont {Yoo},\ and\ \citenamefont {Wei}}]{qml_cf4}%
  \BibitemOpen
  \bibfield  {author} {\bibinfo {author} {\bibfnamefont {S.~L.}\ \bibnamefont {Wu}}, \bibinfo {author} {\bibfnamefont {J.}~\bibnamefont {Chan}}, \bibinfo {author} {\bibfnamefont {W.}~\bibnamefont {Guan}}, \bibinfo {author} {\bibfnamefont {S.}~\bibnamefont {Sun}}, \bibinfo {author} {\bibfnamefont {A.}~\bibnamefont {Wang}}, \bibinfo {author} {\bibfnamefont {C.}~\bibnamefont {Zhou}}, \bibinfo {author} {\bibfnamefont {M.}~\bibnamefont {Livny}}, \bibinfo {author} {\bibfnamefont {F.}~\bibnamefont {Carminati}}, \bibinfo {author} {\bibfnamefont {A.~D.}\ \bibnamefont {Meglio}}, \bibinfo {author} {\bibfnamefont {A.~C.~Y.}\ \bibnamefont {Li}}, \bibinfo {author} {\bibfnamefont {J.}~\bibnamefont {Lykken}}, \bibinfo {author} {\bibfnamefont {P.}~\bibnamefont {Spentzouris}}, \bibinfo {author} {\bibfnamefont {S.~Y.-C.}\ \bibnamefont {Chen}}, \bibinfo {author} {\bibfnamefont {S.}~\bibnamefont {Yoo}},\ and\ \bibinfo {author} {\bibfnamefont {T.-C.}\ \bibnamefont {Wei}},\ }\bibfield  {title} {\bibinfo {title} {Application of quantum machine learning using the quantum variational classifier method to high energy physics analysis at the lhc on ibm quantum computer simulator and hardware with 10 qubits},\ }\href {https://doi.org/10.1088/1361-6471/ac1391} {\bibfield  {journal} {\bibinfo  {journal} {Journal of Physics G: Nuclear and Particle Physics}\ }\textbf {\bibinfo {volume} {48}},\ \bibinfo {pages} {125003} (\bibinfo {year} {2021}{\natexlab{a}})}\BibitemShut {NoStop}%
\bibitem [{\citenamefont {Chen}\ \emph {et~al.}(2021)\citenamefont {Chen}, \citenamefont {Wei}, \citenamefont {Zhang}, \citenamefont {Yu},\ and\ \citenamefont {Yoo}}]{qml_cf5}%
  \BibitemOpen
  \bibfield  {author} {\bibinfo {author} {\bibfnamefont {S.~Y.-C.}\ \bibnamefont {Chen}}, \bibinfo {author} {\bibfnamefont {T.-C.}\ \bibnamefont {Wei}}, \bibinfo {author} {\bibfnamefont {C.}~\bibnamefont {Zhang}}, \bibinfo {author} {\bibfnamefont {H.}~\bibnamefont {Yu}},\ and\ \bibinfo {author} {\bibfnamefont {S.}~\bibnamefont {Yoo}},\ }\href {https://arxiv.org/abs/2101.06189} {\bibinfo {title} {Hybrid quantum-classical graph convolutional network}} (\bibinfo {year} {2021}),\ \Eprint {https://arxiv.org/abs/2101.06189} {arXiv:2101.06189 [cs.LG]} \BibitemShut {NoStop}%
\bibitem [{\citenamefont {Blance}\ and\ \citenamefont {Spannowsky}(2021{\natexlab{b}})}]{qml_cf6}%
  \BibitemOpen
  \bibfield  {author} {\bibinfo {author} {\bibfnamefont {A.}~\bibnamefont {Blance}}\ and\ \bibinfo {author} {\bibfnamefont {M.}~\bibnamefont {Spannowsky}},\ }\bibfield  {title} {\bibinfo {title} {Unsupervised event classification with graphs on classical and photonic quantum computers},\ }\href {https://doi.org/10.1007/JHEP08(2021)170} {\bibfield  {journal} {\bibinfo  {journal} {Journal of High Energy Physics}\ }\textbf {\bibinfo {volume} {2021}},\ \bibinfo {pages} {170} (\bibinfo {year} {2021}{\natexlab{b}})}\BibitemShut {NoStop}%
\bibitem [{\citenamefont {Heredge}\ \emph {et~al.}(2021)\citenamefont {Heredge}, \citenamefont {Hill}, \citenamefont {Hollenberg},\ and\ \citenamefont {Sevior}}]{qml_cf7}%
  \BibitemOpen
  \bibfield  {author} {\bibinfo {author} {\bibfnamefont {J.}~\bibnamefont {Heredge}}, \bibinfo {author} {\bibfnamefont {C.}~\bibnamefont {Hill}}, \bibinfo {author} {\bibfnamefont {L.}~\bibnamefont {Hollenberg}},\ and\ \bibinfo {author} {\bibfnamefont {M.}~\bibnamefont {Sevior}},\ }\href {https://arxiv.org/abs/2103.12257} {\bibinfo {title} {Quantum support vector machines for continuum suppression in b meson decays}} (\bibinfo {year} {2021}),\ \Eprint {https://arxiv.org/abs/2103.12257} {arXiv:2103.12257 [quant-ph]} \BibitemShut {NoStop}%
\bibitem [{\citenamefont {Wu}\ \emph {et~al.}(2021{\natexlab{b}})\citenamefont {Wu}, \citenamefont {Sun}, \citenamefont {Guan}, \citenamefont {Zhou}, \citenamefont {Chan}, \citenamefont {Cheng}, \citenamefont {Pham}, \citenamefont {Qian}, \citenamefont {Wang}, \citenamefont {Zhang}, \citenamefont {Livny}, \citenamefont {Glick}, \citenamefont {Barkoutsos}, \citenamefont {Woerner}, \citenamefont {Tavernelli}, \citenamefont {Carminati}, \citenamefont {Di~Meglio}, \citenamefont {Li}, \citenamefont {Lykken}, \citenamefont {Spentzouris}, \citenamefont {Chen}, \citenamefont {Yoo},\ and\ \citenamefont {Wei}}]{qml_cf8}%
  \BibitemOpen
  \bibfield  {author} {\bibinfo {author} {\bibfnamefont {S.~L.}\ \bibnamefont {Wu}}, \bibinfo {author} {\bibfnamefont {S.}~\bibnamefont {Sun}}, \bibinfo {author} {\bibfnamefont {W.}~\bibnamefont {Guan}}, \bibinfo {author} {\bibfnamefont {C.}~\bibnamefont {Zhou}}, \bibinfo {author} {\bibfnamefont {J.}~\bibnamefont {Chan}}, \bibinfo {author} {\bibfnamefont {C.~L.}\ \bibnamefont {Cheng}}, \bibinfo {author} {\bibfnamefont {T.}~\bibnamefont {Pham}}, \bibinfo {author} {\bibfnamefont {Y.}~\bibnamefont {Qian}}, \bibinfo {author} {\bibfnamefont {A.~Z.}\ \bibnamefont {Wang}}, \bibinfo {author} {\bibfnamefont {R.}~\bibnamefont {Zhang}}, \bibinfo {author} {\bibfnamefont {M.}~\bibnamefont {Livny}}, \bibinfo {author} {\bibfnamefont {J.}~\bibnamefont {Glick}}, \bibinfo {author} {\bibfnamefont {P.~K.}\ \bibnamefont {Barkoutsos}}, \bibinfo {author} {\bibfnamefont {S.}~\bibnamefont {Woerner}}, \bibinfo {author} {\bibfnamefont {I.}~\bibnamefont {Tavernelli}}, \bibinfo {author} {\bibfnamefont {F.}~\bibnamefont {Carminati}}, \bibinfo {author} {\bibfnamefont {A.}~\bibnamefont {Di~Meglio}}, \bibinfo {author} {\bibfnamefont {A.~C.~Y.}\ \bibnamefont {Li}}, \bibinfo {author} {\bibfnamefont {J.}~\bibnamefont {Lykken}}, \bibinfo {author} {\bibfnamefont {P.}~\bibnamefont {Spentzouris}}, \bibinfo {author} {\bibfnamefont {S.~Y.-C.}\ \bibnamefont {Chen}}, \bibinfo {author} {\bibfnamefont {S.}~\bibnamefont {Yoo}},\ and\ \bibinfo {author} {\bibfnamefont {T.-C.}\ \bibnamefont {Wei}},\ }\bibfield  {title} {\bibinfo {title} {Application of quantum machine learning using the quantum kernel algorithm on high energy physics analysis at the lhc},\ }\href {https://doi.org/10.1103/PhysRevResearch.3.033221} {\bibfield  {journal} {\bibinfo  {journal} {Phys. Rev. Res.}\ }\textbf {\bibinfo {volume} {3}},\ \bibinfo {pages} {033221} (\bibinfo {year} {2021}{\natexlab{b}})}\BibitemShut {NoStop}%
\bibitem [{\citenamefont {Belis}\ \emph {et~al.}(2021)\citenamefont {Belis}, \citenamefont {González-Castillo}, \citenamefont {Reissel}, \citenamefont {Vallecorsa}, \citenamefont {Combarro}, \citenamefont {Dissertori},\ and\ \citenamefont {Reiter}}]{qml_cf9}%
  \BibitemOpen
  \bibfield  {author} {\bibinfo {author} {\bibfnamefont {V.}~\bibnamefont {Belis}}, \bibinfo {author} {\bibfnamefont {S.}~\bibnamefont {González-Castillo}}, \bibinfo {author} {\bibfnamefont {C.}~\bibnamefont {Reissel}}, \bibinfo {author} {\bibfnamefont {S.}~\bibnamefont {Vallecorsa}}, \bibinfo {author} {\bibfnamefont {E.~F.}\ \bibnamefont {Combarro}}, \bibinfo {author} {\bibfnamefont {G.}~\bibnamefont {Dissertori}},\ and\ \bibinfo {author} {\bibfnamefont {F.}~\bibnamefont {Reiter}},\ }\bibfield  {title} {\bibinfo {title} {Higgs analysis with quantum classifiers},\ }\href {https://doi.org/10.1051/epjconf/202125103070} {\bibfield  {journal} {\bibinfo  {journal} {EPJ Web of Conferences}\ }\textbf {\bibinfo {volume} {251}},\ \bibinfo {pages} {03070} (\bibinfo {year} {2021})}\BibitemShut {NoStop}%
\bibitem [{\citenamefont {Gianelle}\ \emph {et~al.}(2022)\citenamefont {Gianelle}, \citenamefont {Koppenburg}, \citenamefont {Lucchesi}, \citenamefont {Nicotra}, \citenamefont {Rodrigues}, \citenamefont {Sestini}, \citenamefont {de~Vries},\ and\ \citenamefont {Zuliani}}]{qml_cf10}%
  \BibitemOpen
  \bibfield  {author} {\bibinfo {author} {\bibfnamefont {A.}~\bibnamefont {Gianelle}}, \bibinfo {author} {\bibfnamefont {P.}~\bibnamefont {Koppenburg}}, \bibinfo {author} {\bibfnamefont {D.}~\bibnamefont {Lucchesi}}, \bibinfo {author} {\bibfnamefont {D.}~\bibnamefont {Nicotra}}, \bibinfo {author} {\bibfnamefont {E.}~\bibnamefont {Rodrigues}}, \bibinfo {author} {\bibfnamefont {L.}~\bibnamefont {Sestini}}, \bibinfo {author} {\bibfnamefont {J.}~\bibnamefont {de~Vries}},\ and\ \bibinfo {author} {\bibfnamefont {D.}~\bibnamefont {Zuliani}},\ }\bibfield  {title} {\bibinfo {title} {Quantum machine learning for b-jet charge identification},\ }\href {https://doi.org/10.1007/JHEP08(2022)014} {\bibfield  {journal} {\bibinfo  {journal} {Journal of High Energy Physics}\ }\textbf {\bibinfo {volume} {2022}},\ \bibinfo {pages} {14} (\bibinfo {year} {2022})}\BibitemShut {NoStop}%
\bibitem [{\citenamefont {Abel}\ \emph {et~al.}(2022)\citenamefont {Abel}, \citenamefont {Criado},\ and\ \citenamefont {Spannowsky}}]{qml_cf11}%
  \BibitemOpen
  \bibfield  {author} {\bibinfo {author} {\bibfnamefont {S.}~\bibnamefont {Abel}}, \bibinfo {author} {\bibfnamefont {J.~C.}\ \bibnamefont {Criado}},\ and\ \bibinfo {author} {\bibfnamefont {M.}~\bibnamefont {Spannowsky}},\ }\bibfield  {title} {\bibinfo {title} {Completely quantum neural networks},\ }\href {https://doi.org/10.1103/PhysRevA.106.022601} {\bibfield  {journal} {\bibinfo  {journal} {Phys. Rev. A}\ }\textbf {\bibinfo {volume} {106}},\ \bibinfo {pages} {022601} (\bibinfo {year} {2022})}\BibitemShut {NoStop}%
\bibitem [{\citenamefont {Araz}\ and\ \citenamefont {Spannowsky}(2022)}]{qml_cf12}%
  \BibitemOpen
  \bibfield  {author} {\bibinfo {author} {\bibfnamefont {J.~Y.}\ \bibnamefont {Araz}}\ and\ \bibinfo {author} {\bibfnamefont {M.}~\bibnamefont {Spannowsky}},\ }\bibfield  {title} {\bibinfo {title} {Classical versus quantum: Comparing tensor-network-based quantum circuits on large hadron collider data},\ }\bibfield  {journal} {\bibinfo  {journal} {Physical Review A}\ }\textbf {\bibinfo {volume} {106}},\ \href {https://doi.org/10.1103/physreva.106.062423} {10.1103/physreva.106.062423} (\bibinfo {year} {2022})\BibitemShut {NoStop}%
\bibitem [{\citenamefont {Peixoto}\ \emph {et~al.}(2023)\citenamefont {Peixoto}, \citenamefont {Castro}, \citenamefont {Crispim~Romão}, \citenamefont {Oliveira},\ and\ \citenamefont {Ochoa}}]{qml_cf13}%
  \BibitemOpen
  \bibfield  {author} {\bibinfo {author} {\bibfnamefont {M.~C.}\ \bibnamefont {Peixoto}}, \bibinfo {author} {\bibfnamefont {N.~F.}\ \bibnamefont {Castro}}, \bibinfo {author} {\bibfnamefont {M.}~\bibnamefont {Crispim~Romão}}, \bibinfo {author} {\bibfnamefont {M.~G.~J.}\ \bibnamefont {Oliveira}},\ and\ \bibinfo {author} {\bibfnamefont {I.}~\bibnamefont {Ochoa}},\ }\bibfield  {title} {\bibinfo {title} {Fitting a collider in a quantum computer: tackling the challenges of quantum machine learning for big datasets},\ }\bibfield  {journal} {\bibinfo  {journal} {Frontiers in Artificial Intelligence}\ }\textbf {\bibinfo {volume} {6}},\ \href {https://doi.org/10.3389/frai.2023.1268852} {10.3389/frai.2023.1268852} (\bibinfo {year} {2023})\BibitemShut {NoStop}%
\bibitem [{\citenamefont {Hammad}\ \emph {et~al.}(2023)\citenamefont {Hammad}, \citenamefont {Kong}, \citenamefont {Park},\ and\ \citenamefont {Shim}}]{qml_cf14}%
  \BibitemOpen
  \bibfield  {author} {\bibinfo {author} {\bibfnamefont {A.}~\bibnamefont {Hammad}}, \bibinfo {author} {\bibfnamefont {K.}~\bibnamefont {Kong}}, \bibinfo {author} {\bibfnamefont {M.}~\bibnamefont {Park}},\ and\ \bibinfo {author} {\bibfnamefont {S.}~\bibnamefont {Shim}},\ }\href {https://arxiv.org/abs/2311.16866} {\bibinfo {title} {Quantum metric learning for new physics searches at the lhc}} (\bibinfo {year} {2023}),\ \Eprint {https://arxiv.org/abs/2311.16866} {arXiv:2311.16866 [hep-ph]} \BibitemShut {NoStop}%
\bibitem [{\citenamefont {Ngairangbam}\ \emph {et~al.}(2022)\citenamefont {Ngairangbam}, \citenamefont {Spannowsky},\ and\ \citenamefont {Takeuchi}}]{qml_ad1}%
  \BibitemOpen
  \bibfield  {author} {\bibinfo {author} {\bibfnamefont {V.~S.}\ \bibnamefont {Ngairangbam}}, \bibinfo {author} {\bibfnamefont {M.}~\bibnamefont {Spannowsky}},\ and\ \bibinfo {author} {\bibfnamefont {M.}~\bibnamefont {Takeuchi}},\ }\bibfield  {title} {\bibinfo {title} {Anomaly detection in high-energy physics using a quantum autoencoder},\ }\href {https://doi.org/10.1103/PhysRevD.105.095004} {\bibfield  {journal} {\bibinfo  {journal} {Phys. Rev. D}\ }\textbf {\bibinfo {volume} {105}},\ \bibinfo {pages} {095004} (\bibinfo {year} {2022})}\BibitemShut {NoStop}%
\bibitem [{\citenamefont {Alvi}\ \emph {et~al.}(2023)\citenamefont {Alvi}, \citenamefont {Bauer},\ and\ \citenamefont {Nachman}}]{qml_ad2}%
  \BibitemOpen
  \bibfield  {author} {\bibinfo {author} {\bibfnamefont {S.}~\bibnamefont {Alvi}}, \bibinfo {author} {\bibfnamefont {C.~W.}\ \bibnamefont {Bauer}},\ and\ \bibinfo {author} {\bibfnamefont {B.}~\bibnamefont {Nachman}},\ }\bibfield  {title} {\bibinfo {title} {Quantum anomaly detection for collider physics},\ }\href {https://doi.org/10.1007/JHEP02(2023)220} {\bibfield  {journal} {\bibinfo  {journal} {Journal of High Energy Physics}\ }\textbf {\bibinfo {volume} {2023}},\ \bibinfo {pages} {220} (\bibinfo {year} {2023})}\BibitemShut {NoStop}%
\bibitem [{\citenamefont {Araz}\ and\ \citenamefont {Spannowsky}(2023)}]{qml_ad3}%
  \BibitemOpen
  \bibfield  {author} {\bibinfo {author} {\bibfnamefont {J.~Y.}\ \bibnamefont {Araz}}\ and\ \bibinfo {author} {\bibfnamefont {M.}~\bibnamefont {Spannowsky}},\ }\bibfield  {title} {\bibinfo {title} {Quantum-probabilistic hamiltonian learning for generative modeling and anomaly detection},\ }\href {https://doi.org/10.1103/PhysRevA.108.062422} {\bibfield  {journal} {\bibinfo  {journal} {Phys. Rev. A}\ }\textbf {\bibinfo {volume} {108}},\ \bibinfo {pages} {062422} (\bibinfo {year} {2023})}\BibitemShut {NoStop}%
\bibitem [{\citenamefont {Woźniak}\ \emph {et~al.}(2023)\citenamefont {Woźniak}, \citenamefont {Belis}, \citenamefont {Puljak}, \citenamefont {Barkoutsos}, \citenamefont {Dissertori}, \citenamefont {Grossi}, \citenamefont {Pierini}, \citenamefont {Reiter}, \citenamefont {Tavernelli},\ and\ \citenamefont {Vallecorsa}}]{qml_ad4}%
  \BibitemOpen
  \bibfield  {author} {\bibinfo {author} {\bibfnamefont {K.~A.}\ \bibnamefont {Woźniak}}, \bibinfo {author} {\bibfnamefont {V.}~\bibnamefont {Belis}}, \bibinfo {author} {\bibfnamefont {E.}~\bibnamefont {Puljak}}, \bibinfo {author} {\bibfnamefont {P.}~\bibnamefont {Barkoutsos}}, \bibinfo {author} {\bibfnamefont {G.}~\bibnamefont {Dissertori}}, \bibinfo {author} {\bibfnamefont {M.}~\bibnamefont {Grossi}}, \bibinfo {author} {\bibfnamefont {M.}~\bibnamefont {Pierini}}, \bibinfo {author} {\bibfnamefont {F.}~\bibnamefont {Reiter}}, \bibinfo {author} {\bibfnamefont {I.}~\bibnamefont {Tavernelli}},\ and\ \bibinfo {author} {\bibfnamefont {S.}~\bibnamefont {Vallecorsa}},\ }\href {https://arxiv.org/abs/2301.10780} {\bibinfo {title} {Quantum anomaly detection in the latent space of proton collision events at the lhc}} (\bibinfo {year} {2023}),\ \Eprint {https://arxiv.org/abs/2301.10780} {arXiv:2301.10780 [quant-ph]} \BibitemShut {NoStop}%
\bibitem [{\citenamefont {Schuhmacher}\ \emph {et~al.}(2023)\citenamefont {Schuhmacher}, \citenamefont {Boggia}, \citenamefont {Belis}, \citenamefont {Puljak}, \citenamefont {Grossi}, \citenamefont {Pierini}, \citenamefont {Vallecorsa}, \citenamefont {Tacchino}, \citenamefont {Barkoutsos},\ and\ \citenamefont {Tavernelli}}]{qml_ad5}%
  \BibitemOpen
  \bibfield  {author} {\bibinfo {author} {\bibfnamefont {J.}~\bibnamefont {Schuhmacher}}, \bibinfo {author} {\bibfnamefont {L.}~\bibnamefont {Boggia}}, \bibinfo {author} {\bibfnamefont {V.}~\bibnamefont {Belis}}, \bibinfo {author} {\bibfnamefont {E.}~\bibnamefont {Puljak}}, \bibinfo {author} {\bibfnamefont {M.}~\bibnamefont {Grossi}}, \bibinfo {author} {\bibfnamefont {M.}~\bibnamefont {Pierini}}, \bibinfo {author} {\bibfnamefont {S.}~\bibnamefont {Vallecorsa}}, \bibinfo {author} {\bibfnamefont {F.}~\bibnamefont {Tacchino}}, \bibinfo {author} {\bibfnamefont {P.}~\bibnamefont {Barkoutsos}},\ and\ \bibinfo {author} {\bibfnamefont {I.}~\bibnamefont {Tavernelli}},\ }\bibfield  {title} {\bibinfo {title} {Unravelling physics beyond the standard model with classical and quantum anomaly detection},\ }\href {https://doi.org/10.1088/2632-2153/ad07f7} {\bibfield  {journal} {\bibinfo  {journal} {Machine Learning: Science and Technology}\ }\textbf {\bibinfo {volume} {4}},\ \bibinfo {pages} {045031} (\bibinfo {year} {2023})}\BibitemShut {NoStop}%
\bibitem [{\citenamefont {Bravo-Prieto}\ \emph {et~al.}(2022)\citenamefont {Bravo-Prieto}, \citenamefont {Baglio}, \citenamefont {C{\`{e}}}, \citenamefont {Francis}, \citenamefont {Grabowska},\ and\ \citenamefont {Carrazza}}]{qml_gen1}%
  \BibitemOpen
  \bibfield  {author} {\bibinfo {author} {\bibfnamefont {C.}~\bibnamefont {Bravo-Prieto}}, \bibinfo {author} {\bibfnamefont {J.}~\bibnamefont {Baglio}}, \bibinfo {author} {\bibfnamefont {M.}~\bibnamefont {C{\`{e}}}}, \bibinfo {author} {\bibfnamefont {A.}~\bibnamefont {Francis}}, \bibinfo {author} {\bibfnamefont {D.~M.}\ \bibnamefont {Grabowska}},\ and\ \bibinfo {author} {\bibfnamefont {S.}~\bibnamefont {Carrazza}},\ }\bibfield  {title} {\bibinfo {title} {Style-based quantum generative adversarial networks for {M}onte {C}arlo events},\ }\href {https://doi.org/10.22331/q-2022-08-17-777} {\bibfield  {journal} {\bibinfo  {journal} {{Quantum}}\ }\textbf {\bibinfo {volume} {6}},\ \bibinfo {pages} {777} (\bibinfo {year} {2022})}\BibitemShut {NoStop}%
\bibitem [{\citenamefont {Delgado}\ and\ \citenamefont {Hamilton}(2022)}]{qml_gen2}%
  \BibitemOpen
  \bibfield  {author} {\bibinfo {author} {\bibfnamefont {A.}~\bibnamefont {Delgado}}\ and\ \bibinfo {author} {\bibfnamefont {K.~E.}\ \bibnamefont {Hamilton}},\ }\bibfield  {title} {\bibinfo {title} {Unsupervised quantum circuit learning in high energy physics},\ }\bibfield  {journal} {\bibinfo  {journal} {Physical Review D}\ }\textbf {\bibinfo {volume} {106}},\ \href {https://doi.org/10.1103/physrevd.106.096006} {10.1103/physrevd.106.096006} (\bibinfo {year} {2022})\BibitemShut {NoStop}%
\bibitem [{\citenamefont {Rousselot}\ and\ \citenamefont {Spannowsky}(2024)}]{qml_gen3}%
  \BibitemOpen
  \bibfield  {author} {\bibinfo {author} {\bibfnamefont {A.}~\bibnamefont {Rousselot}}\ and\ \bibinfo {author} {\bibfnamefont {M.}~\bibnamefont {Spannowsky}},\ }\bibfield  {title} {\bibinfo {title} {{Generative invertible quantum neural networks}},\ }\href {https://doi.org/10.21468/SciPostPhys.16.6.146} {\bibfield  {journal} {\bibinfo  {journal} {SciPost Phys.}\ }\textbf {\bibinfo {volume} {16}},\ \bibinfo {pages} {146} (\bibinfo {year} {2024})}\BibitemShut {NoStop}%
\bibitem [{\citenamefont {Rehm}\ \emph {et~al.}(2023)\citenamefont {Rehm}, \citenamefont {Vallecorsa}, \citenamefont {Borras}, \citenamefont {Krücker}, \citenamefont {Grossi},\ and\ \citenamefont {Varo}}]{qml_gen4}%
  \BibitemOpen
  \bibfield  {author} {\bibinfo {author} {\bibfnamefont {F.}~\bibnamefont {Rehm}}, \bibinfo {author} {\bibfnamefont {S.}~\bibnamefont {Vallecorsa}}, \bibinfo {author} {\bibfnamefont {K.}~\bibnamefont {Borras}}, \bibinfo {author} {\bibfnamefont {D.}~\bibnamefont {Krücker}}, \bibinfo {author} {\bibfnamefont {M.}~\bibnamefont {Grossi}},\ and\ \bibinfo {author} {\bibfnamefont {V.}~\bibnamefont {Varo}},\ }\bibfield  {title} {\bibinfo {title} {Precise image generation on current noisy quantum computing devices},\ }\href {https://doi.org/10.1088/2058-9565/ad0389} {\bibfield  {journal} {\bibinfo  {journal} {Quantum Science and Technology}\ }\textbf {\bibinfo {volume} {9}},\ \bibinfo {pages} {015009} (\bibinfo {year} {2023})}\BibitemShut {NoStop}%
\bibitem [{\citenamefont {Hoque}\ \emph {et~al.}(2024)\citenamefont {Hoque}, \citenamefont {Jia}, \citenamefont {Abhishek}, \citenamefont {Fadaie}, \citenamefont {Toledo-Marín}, \citenamefont {Vale}, \citenamefont {Melko}, \citenamefont {Swiatlowski},\ and\ \citenamefont {Fedorko}}]{qml_gen5}%
  \BibitemOpen
  \bibfield  {author} {\bibinfo {author} {\bibfnamefont {S.}~\bibnamefont {Hoque}}, \bibinfo {author} {\bibfnamefont {H.}~\bibnamefont {Jia}}, \bibinfo {author} {\bibfnamefont {A.}~\bibnamefont {Abhishek}}, \bibinfo {author} {\bibfnamefont {M.}~\bibnamefont {Fadaie}}, \bibinfo {author} {\bibfnamefont {J.~Q.}\ \bibnamefont {Toledo-Marín}}, \bibinfo {author} {\bibfnamefont {T.}~\bibnamefont {Vale}}, \bibinfo {author} {\bibfnamefont {R.~G.}\ \bibnamefont {Melko}}, \bibinfo {author} {\bibfnamefont {M.}~\bibnamefont {Swiatlowski}},\ and\ \bibinfo {author} {\bibfnamefont {W.~T.}\ \bibnamefont {Fedorko}},\ }\href {https://arxiv.org/abs/2312.03179} {\bibinfo {title} {Caloqvae : Simulating high-energy particle-calorimeter interactions using hybrid quantum-classical generative models}} (\bibinfo {year} {2024}),\ \Eprint {https://arxiv.org/abs/2312.03179} {arXiv:2312.03179 [hep-ex]} \BibitemShut {NoStop}%
\bibitem [{\citenamefont {Cerezo}\ \emph {et~al.}(2021)\citenamefont {Cerezo}, \citenamefont {Arrasmith}, \citenamefont {Babbush}, \citenamefont {Benjamin}, \citenamefont {Endo}, \citenamefont {Fujii}, \citenamefont {McClean}, \citenamefont {Mitarai}, \citenamefont {Yuan}, \citenamefont {Cincio},\ and\ \citenamefont {Coles}}]{vqc1}%
  \BibitemOpen
  \bibfield  {author} {\bibinfo {author} {\bibfnamefont {M.}~\bibnamefont {Cerezo}}, \bibinfo {author} {\bibfnamefont {A.}~\bibnamefont {Arrasmith}}, \bibinfo {author} {\bibfnamefont {R.}~\bibnamefont {Babbush}}, \bibinfo {author} {\bibfnamefont {S.~C.}\ \bibnamefont {Benjamin}}, \bibinfo {author} {\bibfnamefont {S.}~\bibnamefont {Endo}}, \bibinfo {author} {\bibfnamefont {K.}~\bibnamefont {Fujii}}, \bibinfo {author} {\bibfnamefont {J.~R.}\ \bibnamefont {McClean}}, \bibinfo {author} {\bibfnamefont {K.}~\bibnamefont {Mitarai}}, \bibinfo {author} {\bibfnamefont {X.}~\bibnamefont {Yuan}}, \bibinfo {author} {\bibfnamefont {L.}~\bibnamefont {Cincio}},\ and\ \bibinfo {author} {\bibfnamefont {P.~J.}\ \bibnamefont {Coles}},\ }\bibfield  {title} {\bibinfo {title} {Variational quantum algorithms},\ }\href {https://doi.org/10.1038/s42254-021-00348-9} {\bibfield  {journal} {\bibinfo  {journal} {Nature Reviews Physics}\ }\textbf {\bibinfo {volume} {3}},\ \bibinfo {pages} {625} (\bibinfo {year} {2021})}\BibitemShut {NoStop}%
\bibitem [{\citenamefont {Peruzzo}\ \emph {et~al.}(2014)\citenamefont {Peruzzo}, \citenamefont {McClean}, \citenamefont {Shadbolt}, \citenamefont {Yung}, \citenamefont {Zhou}, \citenamefont {Love}, \citenamefont {Aspuru-Guzik},\ and\ \citenamefont {O'Brien}}]{vqc2}%
  \BibitemOpen
  \bibfield  {author} {\bibinfo {author} {\bibfnamefont {A.}~\bibnamefont {Peruzzo}}, \bibinfo {author} {\bibfnamefont {J.}~\bibnamefont {McClean}}, \bibinfo {author} {\bibfnamefont {P.}~\bibnamefont {Shadbolt}}, \bibinfo {author} {\bibfnamefont {M.-H.}\ \bibnamefont {Yung}}, \bibinfo {author} {\bibfnamefont {X.-Q.}\ \bibnamefont {Zhou}}, \bibinfo {author} {\bibfnamefont {P.~J.}\ \bibnamefont {Love}}, \bibinfo {author} {\bibfnamefont {A.}~\bibnamefont {Aspuru-Guzik}},\ and\ \bibinfo {author} {\bibfnamefont {J.~L.}\ \bibnamefont {O'Brien}},\ }\bibfield  {title} {\bibinfo {title} {A variational eigenvalue solver on a photonic quantum processor},\ }\href {https://doi.org/10.1038/ncomms5213} {\bibfield  {journal} {\bibinfo  {journal} {Nature Communications}\ }\textbf {\bibinfo {volume} {5}},\ \bibinfo {pages} {4213} (\bibinfo {year} {2014})}\BibitemShut {NoStop}%
\bibitem [{\citenamefont {McClean}\ \emph {et~al.}(2016)\citenamefont {McClean}, \citenamefont {Romero}, \citenamefont {Babbush},\ and\ \citenamefont {Aspuru-Guzik}}]{vqc3}%
  \BibitemOpen
  \bibfield  {author} {\bibinfo {author} {\bibfnamefont {J.~R.}\ \bibnamefont {McClean}}, \bibinfo {author} {\bibfnamefont {J.}~\bibnamefont {Romero}}, \bibinfo {author} {\bibfnamefont {R.}~\bibnamefont {Babbush}},\ and\ \bibinfo {author} {\bibfnamefont {A.}~\bibnamefont {Aspuru-Guzik}},\ }\bibfield  {title} {\bibinfo {title} {The theory of variational hybrid quantum-classical algorithms},\ }\href {https://doi.org/10.1088/1367-2630/18/2/023023} {\bibfield  {journal} {\bibinfo  {journal} {New Journal of Physics}\ }\textbf {\bibinfo {volume} {18}},\ \bibinfo {pages} {023023} (\bibinfo {year} {2016})}\BibitemShut {NoStop}%
\bibitem [{\citenamefont {Weisstein}()}]{completegraph}%
  \BibitemOpen
  \bibfield  {author} {\bibinfo {author} {\bibfnamefont {E.~W.}\ \bibnamefont {Weisstein}},\ }\href {https://mathworld.wolfram.com/CompleteGraph.html} {\bibinfo {title} {"complete graph." from mathworld--a wolfram web resource}},\ \bibinfo {note} {last visited on 2/11/2023}\BibitemShut {NoStop}%
\bibitem [{\citenamefont {Kasieczka}\ \emph {et~al.}(2019{\natexlab{a}})\citenamefont {Kasieczka}, \citenamefont {Plehn}, \citenamefont {Thompson},\ and\ \citenamefont {Russel}}]{zenodo_top}%
  \BibitemOpen
  \bibfield  {author} {\bibinfo {author} {\bibfnamefont {G.}~\bibnamefont {Kasieczka}}, \bibinfo {author} {\bibfnamefont {T.}~\bibnamefont {Plehn}}, \bibinfo {author} {\bibfnamefont {J.}~\bibnamefont {Thompson}},\ and\ \bibinfo {author} {\bibfnamefont {M.}~\bibnamefont {Russel}},\ }\bibfield  {title} {\bibinfo {title} {Top quark tagging reference dataset},\ }\href {https://doi.org/10.5281/zenodo.2603256} {10.5281/zenodo.2603256} (\bibinfo {year} {2019}{\natexlab{a}})\BibitemShut {NoStop}%
\bibitem [{\citenamefont {Kasieczka}\ \emph {et~al.}(2019{\natexlab{b}})\citenamefont {Kasieczka}, \citenamefont {Plehn}, \citenamefont {Butter}, \citenamefont {Cranmer}, \citenamefont {Debnath}, \citenamefont {Dillon}, \citenamefont {Fairbairn}, \citenamefont {Faroughy}, \citenamefont {Fedorko}, \citenamefont {Gay}, \citenamefont {Gouskos}, \citenamefont {Kamenik}, \citenamefont {Komiske}, \citenamefont {Leiss}, \citenamefont {Lister}, \citenamefont {Macaluso}, \citenamefont {Metodiev}, \citenamefont {Moore}, \citenamefont {Nachman}, \citenamefont {Nordström}, \citenamefont {Pearkes}, \citenamefont {Qu}, \citenamefont {Rath}, \citenamefont {Rieger}, \citenamefont {Shih}, \citenamefont {Thompson},\ and\ \citenamefont {Varma}}]{dataset_top}%
  \BibitemOpen
  \bibfield  {author} {\bibinfo {author} {\bibfnamefont {G.}~\bibnamefont {Kasieczka}}, \bibinfo {author} {\bibfnamefont {T.}~\bibnamefont {Plehn}}, \bibinfo {author} {\bibfnamefont {A.}~\bibnamefont {Butter}}, \bibinfo {author} {\bibfnamefont {K.}~\bibnamefont {Cranmer}}, \bibinfo {author} {\bibfnamefont {D.}~\bibnamefont {Debnath}}, \bibinfo {author} {\bibfnamefont {B.~M.}\ \bibnamefont {Dillon}}, \bibinfo {author} {\bibfnamefont {M.}~\bibnamefont {Fairbairn}}, \bibinfo {author} {\bibfnamefont {D.~A.}\ \bibnamefont {Faroughy}}, \bibinfo {author} {\bibfnamefont {W.}~\bibnamefont {Fedorko}}, \bibinfo {author} {\bibfnamefont {C.}~\bibnamefont {Gay}}, \bibinfo {author} {\bibfnamefont {L.}~\bibnamefont {Gouskos}}, \bibinfo {author} {\bibfnamefont {J.~F.}\ \bibnamefont {Kamenik}}, \bibinfo {author} {\bibfnamefont {P.}~\bibnamefont {Komiske}}, \bibinfo {author} {\bibfnamefont {S.}~\bibnamefont {Leiss}}, \bibinfo {author} {\bibfnamefont {A.}~\bibnamefont {Lister}}, \bibinfo {author} {\bibfnamefont {S.}~\bibnamefont {Macaluso}}, \bibinfo {author} {\bibfnamefont {E.}~\bibnamefont {Metodiev}}, \bibinfo {author} {\bibfnamefont {L.}~\bibnamefont {Moore}}, \bibinfo {author} {\bibfnamefont {B.}~\bibnamefont {Nachman}}, \bibinfo {author} {\bibfnamefont {K.}~\bibnamefont {Nordström}}, \bibinfo {author} {\bibfnamefont {J.}~\bibnamefont {Pearkes}}, \bibinfo {author} {\bibfnamefont {H.}~\bibnamefont {Qu}}, \bibinfo {author} {\bibfnamefont {Y.}~\bibnamefont {Rath}}, \bibinfo {author} {\bibfnamefont {M.}~\bibnamefont {Rieger}}, \bibinfo {author} {\bibfnamefont {D.}~\bibnamefont {Shih}}, \bibinfo {author} {\bibfnamefont {J.}~\bibnamefont {Thompson}},\ and\ \bibinfo {author} {\bibfnamefont {S.}~\bibnamefont {Varma}},\ }\bibfield  {title} {\bibinfo {title} {The machine learning landscape of top taggers},\ }\bibfield  {journal} {\bibinfo  {journal} {SciPost Physics}\ }\textbf {\bibinfo {volume} {7}},\ \href {https://doi.org/10.21468/scipostphys.7.1.014} {10.21468/scipostphys.7.1.014} (\bibinfo {year} {2019}{\natexlab{b}})\BibitemShut {NoStop}%
\bibitem [{\citenamefont {Kansal}\ \emph {et~al.}(2022{\natexlab{a}})\citenamefont {Kansal}, \citenamefont {Duarte}, \citenamefont {Su}, \citenamefont {Orzari}, \citenamefont {Tomei}, \citenamefont {Pierini}, \citenamefont {Touranakou}, \citenamefont {Vlimant},\ and\ \citenamefont {Gunopulos}}]{zenodo_jetnet}%
  \BibitemOpen
  \bibfield  {author} {\bibinfo {author} {\bibfnamefont {R.}~\bibnamefont {Kansal}}, \bibinfo {author} {\bibfnamefont {J.}~\bibnamefont {Duarte}}, \bibinfo {author} {\bibfnamefont {H.}~\bibnamefont {Su}}, \bibinfo {author} {\bibfnamefont {B.}~\bibnamefont {Orzari}}, \bibinfo {author} {\bibfnamefont {T.}~\bibnamefont {Tomei}}, \bibinfo {author} {\bibfnamefont {M.}~\bibnamefont {Pierini}}, \bibinfo {author} {\bibfnamefont {M.}~\bibnamefont {Touranakou}}, \bibinfo {author} {\bibfnamefont {J.-R.}\ \bibnamefont {Vlimant}},\ and\ \bibinfo {author} {\bibfnamefont {D.}~\bibnamefont {Gunopulos}},\ }\bibfield  {title} {\bibinfo {title} {Jetnet},\ }\href {https://doi.org/10.5281/zenodo.6975118} {10.5281/zenodo.6975118} (\bibinfo {year} {2022}{\natexlab{a}})\BibitemShut {NoStop}%
\bibitem [{\citenamefont {Kansal}\ \emph {et~al.}(2022{\natexlab{b}})\citenamefont {Kansal}, \citenamefont {Duarte}, \citenamefont {Su}, \citenamefont {Orzari}, \citenamefont {Tomei}, \citenamefont {Pierini}, \citenamefont {Touranakou}, \citenamefont {Vlimant},\ and\ \citenamefont {Gunopulos}}]{dataset_jetnet}%
  \BibitemOpen
  \bibfield  {author} {\bibinfo {author} {\bibfnamefont {R.}~\bibnamefont {Kansal}}, \bibinfo {author} {\bibfnamefont {J.}~\bibnamefont {Duarte}}, \bibinfo {author} {\bibfnamefont {H.}~\bibnamefont {Su}}, \bibinfo {author} {\bibfnamefont {B.}~\bibnamefont {Orzari}}, \bibinfo {author} {\bibfnamefont {T.}~\bibnamefont {Tomei}}, \bibinfo {author} {\bibfnamefont {M.}~\bibnamefont {Pierini}}, \bibinfo {author} {\bibfnamefont {M.}~\bibnamefont {Touranakou}}, \bibinfo {author} {\bibfnamefont {J.-R.}\ \bibnamefont {Vlimant}},\ and\ \bibinfo {author} {\bibfnamefont {D.}~\bibnamefont {Gunopulos}},\ }\href {https://arxiv.org/abs/2106.11535} {\bibinfo {title} {Particle cloud generation with message passing generative adversarial networks}} (\bibinfo {year} {2022}{\natexlab{b}}),\ \Eprint {https://arxiv.org/abs/2106.11535} {arXiv:2106.11535 [cs.LG]} \BibitemShut {NoStop}%
\bibitem [{\citenamefont {Fan}\ \emph {et~al.}(2019)\citenamefont {Fan}, \citenamefont {Ma}, \citenamefont {Li}, \citenamefont {He}, \citenamefont {Zhao}, \citenamefont {Tang},\ and\ \citenamefont {Yin}}]{gnnex1}%
  \BibitemOpen
  \bibfield  {author} {\bibinfo {author} {\bibfnamefont {W.}~\bibnamefont {Fan}}, \bibinfo {author} {\bibfnamefont {Y.}~\bibnamefont {Ma}}, \bibinfo {author} {\bibfnamefont {Q.}~\bibnamefont {Li}}, \bibinfo {author} {\bibfnamefont {Y.}~\bibnamefont {He}}, \bibinfo {author} {\bibfnamefont {E.}~\bibnamefont {Zhao}}, \bibinfo {author} {\bibfnamefont {J.}~\bibnamefont {Tang}},\ and\ \bibinfo {author} {\bibfnamefont {D.}~\bibnamefont {Yin}},\ }\href@noop {} {\bibinfo {title} {Graph neural networks for social recommendation}} (\bibinfo {year} {2019}),\ \Eprint {https://arxiv.org/abs/1902.07243} {arXiv:1902.07243 [cs.IR]} \BibitemShut {NoStop}%
\bibitem [{\citenamefont {Zhang}\ \emph {et~al.}(2021)\citenamefont {Zhang}, \citenamefont {Liang}, \citenamefont {Liu},\ and\ \citenamefont {Tang}}]{gnnex2}%
  \BibitemOpen
  \bibfield  {author} {\bibinfo {author} {\bibfnamefont {X.-M.}\ \bibnamefont {Zhang}}, \bibinfo {author} {\bibfnamefont {L.}~\bibnamefont {Liang}}, \bibinfo {author} {\bibfnamefont {L.}~\bibnamefont {Liu}},\ and\ \bibinfo {author} {\bibfnamefont {M.-J.}\ \bibnamefont {Tang}},\ }\bibfield  {title} {\bibinfo {title} {Graph neural networks and their current applications in bioinformatics},\ }\bibfield  {journal} {\bibinfo  {journal} {Frontiers in Genetics}\ }\textbf {\bibinfo {volume} {12}},\ \href {https://doi.org/10.3389/fgene.2021.690049} {10.3389/fgene.2021.690049} (\bibinfo {year} {2021})\BibitemShut {NoStop}%
\bibitem [{\citenamefont {Reiser}\ \emph {et~al.}(2022)\citenamefont {Reiser}, \citenamefont {Neubert}, \citenamefont {Eberhard}, \citenamefont {Torresi}, \citenamefont {Zhou}, \citenamefont {Shao}, \citenamefont {Metni}, \citenamefont {van Hoesel}, \citenamefont {Schopmans}, \citenamefont {Sommer},\ and\ \citenamefont {Friederich}}]{gnnex3}%
  \BibitemOpen
  \bibfield  {author} {\bibinfo {author} {\bibfnamefont {P.}~\bibnamefont {Reiser}}, \bibinfo {author} {\bibfnamefont {M.}~\bibnamefont {Neubert}}, \bibinfo {author} {\bibfnamefont {A.}~\bibnamefont {Eberhard}}, \bibinfo {author} {\bibfnamefont {L.}~\bibnamefont {Torresi}}, \bibinfo {author} {\bibfnamefont {C.}~\bibnamefont {Zhou}}, \bibinfo {author} {\bibfnamefont {C.}~\bibnamefont {Shao}}, \bibinfo {author} {\bibfnamefont {H.}~\bibnamefont {Metni}}, \bibinfo {author} {\bibfnamefont {C.}~\bibnamefont {van Hoesel}}, \bibinfo {author} {\bibfnamefont {H.}~\bibnamefont {Schopmans}}, \bibinfo {author} {\bibfnamefont {T.}~\bibnamefont {Sommer}},\ and\ \bibinfo {author} {\bibfnamefont {P.}~\bibnamefont {Friederich}},\ }\bibfield  {title} {\bibinfo {title} {Graph neural networks for materials science and chemistry},\ }\href {https://doi.org/10.1038/s43246-022-00315-6} {\bibfield  {journal} {\bibinfo  {journal} {Communications Materials}\ }\textbf {\bibinfo {volume} {3}},\ \bibinfo {pages} {93} (\bibinfo {year} {2022})}\BibitemShut {NoStop}%
\bibitem [{\citenamefont {Zaheer}\ \emph {et~al.}(2018)\citenamefont {Zaheer}, \citenamefont {Kottur}, \citenamefont {Ravanbakhsh}, \citenamefont {Poczos}, \citenamefont {Salakhutdinov},\ and\ \citenamefont {Smola}}]{deepset}%
  \BibitemOpen
  \bibfield  {author} {\bibinfo {author} {\bibfnamefont {M.}~\bibnamefont {Zaheer}}, \bibinfo {author} {\bibfnamefont {S.}~\bibnamefont {Kottur}}, \bibinfo {author} {\bibfnamefont {S.}~\bibnamefont {Ravanbakhsh}}, \bibinfo {author} {\bibfnamefont {B.}~\bibnamefont {Poczos}}, \bibinfo {author} {\bibfnamefont {R.}~\bibnamefont {Salakhutdinov}},\ and\ \bibinfo {author} {\bibfnamefont {A.}~\bibnamefont {Smola}},\ }\href {https://arxiv.org/abs/1703.06114} {\bibinfo {title} {Deep sets}} (\bibinfo {year} {2018}),\ \Eprint {https://arxiv.org/abs/1703.06114} {arXiv:1703.06114 [cs.LG]} \BibitemShut {NoStop}%
\bibitem [{\citenamefont {Gilmer}\ \emph {et~al.}(2017)\citenamefont {Gilmer}, \citenamefont {Schoenholz}, \citenamefont {Riley}, \citenamefont {Vinyals},\ and\ \citenamefont {Dahl}}]{mpn1}%
  \BibitemOpen
  \bibfield  {author} {\bibinfo {author} {\bibfnamefont {J.}~\bibnamefont {Gilmer}}, \bibinfo {author} {\bibfnamefont {S.~S.}\ \bibnamefont {Schoenholz}}, \bibinfo {author} {\bibfnamefont {P.~F.}\ \bibnamefont {Riley}}, \bibinfo {author} {\bibfnamefont {O.}~\bibnamefont {Vinyals}},\ and\ \bibinfo {author} {\bibfnamefont {G.~E.}\ \bibnamefont {Dahl}},\ }\href@noop {} {\bibinfo {title} {Neural message passing for quantum chemistry}} (\bibinfo {year} {2017}),\ \Eprint {https://arxiv.org/abs/1704.01212} {arXiv:1704.01212 [cs.LG]} \BibitemShut {NoStop}%
\bibitem [{\citenamefont {P{\'{e}}rez-Salinas}\ \emph {et~al.}(2020)\citenamefont {P{\'{e}}rez-Salinas}, \citenamefont {Cervera-Lierta}, \citenamefont {Gil-Fuster},\ and\ \citenamefont {Latorre}}]{reupload}%
  \BibitemOpen
  \bibfield  {author} {\bibinfo {author} {\bibfnamefont {A.}~\bibnamefont {P{\'{e}}rez-Salinas}}, \bibinfo {author} {\bibfnamefont {A.}~\bibnamefont {Cervera-Lierta}}, \bibinfo {author} {\bibfnamefont {E.}~\bibnamefont {Gil-Fuster}},\ and\ \bibinfo {author} {\bibfnamefont {J.~I.}\ \bibnamefont {Latorre}},\ }\bibfield  {title} {\bibinfo {title} {Data re-uploading for a universal quantum classifier},\ }\href {https://doi.org/10.22331/q-2020-02-06-226} {\bibfield  {journal} {\bibinfo  {journal} {{Quantum}}\ }\textbf {\bibinfo {volume} {4}},\ \bibinfo {pages} {226} (\bibinfo {year} {2020})}\BibitemShut {NoStop}%
\bibitem [{\citenamefont {Preskill}(2018)}]{nisq}%
  \BibitemOpen
  \bibfield  {author} {\bibinfo {author} {\bibfnamefont {J.}~\bibnamefont {Preskill}},\ }\bibfield  {title} {\bibinfo {title} {Quantum {C}omputing in the {NISQ} era and beyond},\ }\href {https://doi.org/10.22331/q-2018-08-06-79} {\bibfield  {journal} {\bibinfo  {journal} {{Quantum}}\ }\textbf {\bibinfo {volume} {2}},\ \bibinfo {pages} {79} (\bibinfo {year} {2018})}\BibitemShut {NoStop}%
\bibitem [{\citenamefont {Mitarai}\ \emph {et~al.}(2018)\citenamefont {Mitarai}, \citenamefont {Negoro}, \citenamefont {Kitagawa},\ and\ \citenamefont {Fujii}}]{psr1}%
  \BibitemOpen
  \bibfield  {author} {\bibinfo {author} {\bibfnamefont {K.}~\bibnamefont {Mitarai}}, \bibinfo {author} {\bibfnamefont {M.}~\bibnamefont {Negoro}}, \bibinfo {author} {\bibfnamefont {M.}~\bibnamefont {Kitagawa}},\ and\ \bibinfo {author} {\bibfnamefont {K.}~\bibnamefont {Fujii}},\ }\bibfield  {title} {\bibinfo {title} {Quantum circuit learning},\ }\href {https://doi.org/10.1103/PhysRevA.98.032309} {\bibfield  {journal} {\bibinfo  {journal} {Phys. Rev. A}\ }\textbf {\bibinfo {volume} {98}},\ \bibinfo {pages} {032309} (\bibinfo {year} {2018})}\BibitemShut {NoStop}%
\bibitem [{\citenamefont {Schuld}\ \emph {et~al.}(2019)\citenamefont {Schuld}, \citenamefont {Bergholm}, \citenamefont {Gogolin}, \citenamefont {Izaac},\ and\ \citenamefont {Killoran}}]{psr2}%
  \BibitemOpen
  \bibfield  {author} {\bibinfo {author} {\bibfnamefont {M.}~\bibnamefont {Schuld}}, \bibinfo {author} {\bibfnamefont {V.}~\bibnamefont {Bergholm}}, \bibinfo {author} {\bibfnamefont {C.}~\bibnamefont {Gogolin}}, \bibinfo {author} {\bibfnamefont {J.}~\bibnamefont {Izaac}},\ and\ \bibinfo {author} {\bibfnamefont {N.}~\bibnamefont {Killoran}},\ }\bibfield  {title} {\bibinfo {title} {Evaluating analytic gradients on quantum hardware},\ }\href {https://doi.org/10.1103/PhysRevA.99.032331} {\bibfield  {journal} {\bibinfo  {journal} {Phys. Rev. A}\ }\textbf {\bibinfo {volume} {99}},\ \bibinfo {pages} {032331} (\bibinfo {year} {2019})}\BibitemShut {NoStop}%
\bibitem [{\citenamefont {Wierichs}\ \emph {et~al.}(2022)\citenamefont {Wierichs}, \citenamefont {Izaac}, \citenamefont {Wang},\ and\ \citenamefont {Lin}}]{psr3}%
  \BibitemOpen
  \bibfield  {author} {\bibinfo {author} {\bibfnamefont {D.}~\bibnamefont {Wierichs}}, \bibinfo {author} {\bibfnamefont {J.}~\bibnamefont {Izaac}}, \bibinfo {author} {\bibfnamefont {C.}~\bibnamefont {Wang}},\ and\ \bibinfo {author} {\bibfnamefont {C.~Y.-Y.}\ \bibnamefont {Lin}},\ }\bibfield  {title} {\bibinfo {title} {General parameter-shift rules for quantum gradients},\ }\href {https://doi.org/10.22331/q-2022-03-30-677} {\bibfield  {journal} {\bibinfo  {journal} {{Quantum}}\ }\textbf {\bibinfo {volume} {6}},\ \bibinfo {pages} {677} (\bibinfo {year} {2022})}\BibitemShut {NoStop}%
\bibitem [{\citenamefont {Crooks}(2019)}]{psr4}%
  \BibitemOpen
  \bibfield  {author} {\bibinfo {author} {\bibfnamefont {G.~E.}\ \bibnamefont {Crooks}},\ }\href@noop {} {\bibinfo {title} {Gradients of parameterized quantum gates using the parameter-shift rule and gate decomposition}} (\bibinfo {year} {2019}),\ \Eprint {https://arxiv.org/abs/1905.13311} {arXiv:1905.13311 [quant-ph]} \BibitemShut {NoStop}%
\bibitem [{\citenamefont {Shukla}\ and\ \citenamefont {Vedula}(2024)}]{uso1}%
  \BibitemOpen
  \bibfield  {author} {\bibinfo {author} {\bibfnamefont {A.}~\bibnamefont {Shukla}}\ and\ \bibinfo {author} {\bibfnamefont {P.}~\bibnamefont {Vedula}},\ }\bibfield  {title} {\bibinfo {title} {An efficient quantum algorithm for preparation of uniform quantum superposition states},\ }\bibfield  {journal} {\bibinfo  {journal} {Quantum Information Processing}\ }\textbf {\bibinfo {volume} {23}},\ \href {https://doi.org/10.1007/s11128-024-04258-4} {10.1007/s11128-024-04258-4} (\bibinfo {year} {2024})\BibitemShut {NoStop}%
\bibitem [{\citenamefont {Babbush}\ \emph {et~al.}(2018)\citenamefont {Babbush}, \citenamefont {Gidney}, \citenamefont {Berry}, \citenamefont {Wiebe}, \citenamefont {McClean}, \citenamefont {Paler}, \citenamefont {Fowler},\ and\ \citenamefont {Neven}}]{uso2}%
  \BibitemOpen
  \bibfield  {author} {\bibinfo {author} {\bibfnamefont {R.}~\bibnamefont {Babbush}}, \bibinfo {author} {\bibfnamefont {C.}~\bibnamefont {Gidney}}, \bibinfo {author} {\bibfnamefont {D.~W.}\ \bibnamefont {Berry}}, \bibinfo {author} {\bibfnamefont {N.}~\bibnamefont {Wiebe}}, \bibinfo {author} {\bibfnamefont {J.}~\bibnamefont {McClean}}, \bibinfo {author} {\bibfnamefont {A.}~\bibnamefont {Paler}}, \bibinfo {author} {\bibfnamefont {A.}~\bibnamefont {Fowler}},\ and\ \bibinfo {author} {\bibfnamefont {H.}~\bibnamefont {Neven}},\ }\bibfield  {title} {\bibinfo {title} {Encoding electronic spectra in quantum circuits with linear t complexity},\ }\bibfield  {journal} {\bibinfo  {journal} {Physical Review X}\ }\textbf {\bibinfo {volume} {8}},\ \href {https://doi.org/10.1103/physrevx.8.041015} {10.1103/physrevx.8.041015} (\bibinfo {year} {2018})\BibitemShut {NoStop}%
\bibitem [{\citenamefont {Vale}\ \emph {et~al.}(2023)\citenamefont {Vale}, \citenamefont {Azevedo}, \citenamefont {Araújo}, \citenamefont {Araujo},\ and\ \citenamefont {da~Silva}}]{q_decomp1}%
  \BibitemOpen
  \bibfield  {author} {\bibinfo {author} {\bibfnamefont {R.}~\bibnamefont {Vale}}, \bibinfo {author} {\bibfnamefont {T.~M.~D.}\ \bibnamefont {Azevedo}}, \bibinfo {author} {\bibfnamefont {I.~C.~S.}\ \bibnamefont {Araújo}}, \bibinfo {author} {\bibfnamefont {I.~F.}\ \bibnamefont {Araujo}},\ and\ \bibinfo {author} {\bibfnamefont {A.~J.}\ \bibnamefont {da~Silva}},\ }\href {https://arxiv.org/abs/2302.06377} {\bibinfo {title} {Decomposition of multi-controlled special unitary single-qubit gates}} (\bibinfo {year} {2023}),\ \Eprint {https://arxiv.org/abs/2302.06377} {arXiv:2302.06377 [quant-ph]} \BibitemShut {NoStop}%
\bibitem [{\citenamefont {Saeedi}\ and\ \citenamefont {Pedram}(2013)}]{q_decomp2}%
  \BibitemOpen
  \bibfield  {author} {\bibinfo {author} {\bibfnamefont {M.}~\bibnamefont {Saeedi}}\ and\ \bibinfo {author} {\bibfnamefont {M.}~\bibnamefont {Pedram}},\ }\bibfield  {title} {\bibinfo {title} {Linear-depth quantum circuits for $n$-qubit toffoli gates with no ancilla},\ }\href {https://doi.org/10.1103/PhysRevA.87.062318} {\bibfield  {journal} {\bibinfo  {journal} {Phys. Rev. A}\ }\textbf {\bibinfo {volume} {87}},\ \bibinfo {pages} {062318} (\bibinfo {year} {2013})}\BibitemShut {NoStop}%
\bibitem [{\citenamefont {da~Silva}\ and\ \citenamefont {Park}(2022)}]{q_decomp3}%
  \BibitemOpen
  \bibfield  {author} {\bibinfo {author} {\bibfnamefont {A.~J.}\ \bibnamefont {da~Silva}}\ and\ \bibinfo {author} {\bibfnamefont {D.~K.}\ \bibnamefont {Park}},\ }\bibfield  {title} {\bibinfo {title} {Linear-depth quantum circuits for multiqubit controlled gates},\ }\bibfield  {journal} {\bibinfo  {journal} {Physical Review A}\ }\textbf {\bibinfo {volume} {106}},\ \href {https://doi.org/10.1103/physreva.106.042602} {10.1103/physreva.106.042602} (\bibinfo {year} {2022})\BibitemShut {NoStop}%
\bibitem [{\citenamefont {Nielsen}\ and\ \citenamefont {Chuang}(2007)}]{NC}%
  \BibitemOpen
  \bibfield  {author} {\bibinfo {author} {\bibfnamefont {M.~A.}\ \bibnamefont {Nielsen}}\ and\ \bibinfo {author} {\bibfnamefont {I.~L.}\ \bibnamefont {Chuang}},\ }\bibinfo {title} {Controlled operations},\ in\ \href@noop {} {\emph {\bibinfo {booktitle} {Quantum Computation and Quantum Information}}}\ (\bibinfo  {publisher} {Cambridge University Press},\ \bibinfo {year} {2007})\ Chap.\ \bibinfo {chapter} {4.3 Controlled operations - Figure 4.10}\BibitemShut {NoStop}%
\bibitem [{\citenamefont {Havl{\'\i}{\v c}ek}\ \emph {et~al.}(2019)\citenamefont {Havl{\'\i}{\v c}ek}, \citenamefont {C{\'o}rcoles}, \citenamefont {Temme}, \citenamefont {Harrow}, \citenamefont {Kandala}, \citenamefont {Chow},\ and\ \citenamefont {Gambetta}}]{q_kernel1}%
  \BibitemOpen
  \bibfield  {author} {\bibinfo {author} {\bibfnamefont {V.}~\bibnamefont {Havl{\'\i}{\v c}ek}}, \bibinfo {author} {\bibfnamefont {A.~D.}\ \bibnamefont {C{\'o}rcoles}}, \bibinfo {author} {\bibfnamefont {K.}~\bibnamefont {Temme}}, \bibinfo {author} {\bibfnamefont {A.~W.}\ \bibnamefont {Harrow}}, \bibinfo {author} {\bibfnamefont {A.}~\bibnamefont {Kandala}}, \bibinfo {author} {\bibfnamefont {J.~M.}\ \bibnamefont {Chow}},\ and\ \bibinfo {author} {\bibfnamefont {J.~M.}\ \bibnamefont {Gambetta}},\ }\bibfield  {title} {\bibinfo {title} {Supervised learning with quantum-enhanced feature spaces},\ }\href {https://doi.org/10.1038/s41586-019-0980-2} {\bibfield  {journal} {\bibinfo  {journal} {Nature}\ }\textbf {\bibinfo {volume} {567}},\ \bibinfo {pages} {209} (\bibinfo {year} {2019})}\BibitemShut {NoStop}%
\bibitem [{\citenamefont {Schuld}(2021)}]{q_kernel2}%
  \BibitemOpen
  \bibfield  {author} {\bibinfo {author} {\bibfnamefont {M.}~\bibnamefont {Schuld}},\ }\href {https://arxiv.org/abs/2101.11020} {\bibinfo {title} {Supervised quantum machine learning models are kernel methods}} (\bibinfo {year} {2021}),\ \Eprint {https://arxiv.org/abs/2101.11020} {arXiv:2101.11020 [quant-ph]} \BibitemShut {NoStop}%
\bibitem [{\citenamefont {Alwall}\ \emph {et~al.}(2014)\citenamefont {Alwall}, \citenamefont {Frederix}, \citenamefont {Frixione}, \citenamefont {Hirschi}, \citenamefont {Maltoni}, \citenamefont {Mattelaer}, \citenamefont {Shao}, \citenamefont {Stelzer}, \citenamefont {Torrielli},\ and\ \citenamefont {Zaro}}]{mg5}%
  \BibitemOpen
  \bibfield  {author} {\bibinfo {author} {\bibfnamefont {J.}~\bibnamefont {Alwall}}, \bibinfo {author} {\bibfnamefont {R.}~\bibnamefont {Frederix}}, \bibinfo {author} {\bibfnamefont {S.}~\bibnamefont {Frixione}}, \bibinfo {author} {\bibfnamefont {V.}~\bibnamefont {Hirschi}}, \bibinfo {author} {\bibfnamefont {F.}~\bibnamefont {Maltoni}}, \bibinfo {author} {\bibfnamefont {O.}~\bibnamefont {Mattelaer}}, \bibinfo {author} {\bibfnamefont {H.~S.}\ \bibnamefont {Shao}}, \bibinfo {author} {\bibfnamefont {T.}~\bibnamefont {Stelzer}}, \bibinfo {author} {\bibfnamefont {P.}~\bibnamefont {Torrielli}},\ and\ \bibinfo {author} {\bibfnamefont {M.}~\bibnamefont {Zaro}},\ }\bibfield  {title} {\bibinfo {title} {{The automated computation of tree-level and next-to-leading order differential cross sections, and their matching to parton shower simulations}},\ }\href {https://doi.org/10.1007/JHEP07(2014)079} {\bibfield  {journal} {\bibinfo  {journal} {JHEP}\ }\textbf {\bibinfo {volume} {07}},\ \bibinfo {pages} {079}},\ \Eprint {https://arxiv.org/abs/1405.0301} {arXiv:1405.0301 [hep-ph]} \BibitemShut {NoStop}%
\bibitem [{\citenamefont {Ovyn}\ \emph {et~al.}(2010)\citenamefont {Ovyn}, \citenamefont {Rouby},\ and\ \citenamefont {Lemaitre}}]{delphes1}%
  \BibitemOpen
  \bibfield  {author} {\bibinfo {author} {\bibfnamefont {S.}~\bibnamefont {Ovyn}}, \bibinfo {author} {\bibfnamefont {X.}~\bibnamefont {Rouby}},\ and\ \bibinfo {author} {\bibfnamefont {V.}~\bibnamefont {Lemaitre}},\ }\href@noop {} {\bibinfo {title} {Delphes, a framework for fast simulation of a generic collider experiment}} (\bibinfo {year} {2010}),\ \Eprint {https://arxiv.org/abs/0903.2225} {arXiv:0903.2225 [hep-ph]} \BibitemShut {NoStop}%
\bibitem [{\citenamefont {de~Favereau}\ \emph {et~al.}(2014)\citenamefont {de~Favereau}, \citenamefont {Delaere}, \citenamefont {Demin}, \citenamefont {Giammanco}, \citenamefont {Lema\^\i{}tre}, \citenamefont {Mertens},\ and\ \citenamefont {Selvaggi}}]{delphes2}%
  \BibitemOpen
  \bibfield  {author} {\bibinfo {author} {\bibfnamefont {J.}~\bibnamefont {de~Favereau}}, \bibinfo {author} {\bibfnamefont {C.}~\bibnamefont {Delaere}}, \bibinfo {author} {\bibfnamefont {P.}~\bibnamefont {Demin}}, \bibinfo {author} {\bibfnamefont {A.}~\bibnamefont {Giammanco}}, \bibinfo {author} {\bibfnamefont {V.}~\bibnamefont {Lema\^\i{}tre}}, \bibinfo {author} {\bibfnamefont {A.}~\bibnamefont {Mertens}},\ and\ \bibinfo {author} {\bibfnamefont {M.}~\bibnamefont {Selvaggi}} (\bibinfo {collaboration} {DELPHES 3}),\ }\bibfield  {title} {\bibinfo {title} {{DELPHES 3, A modular framework for fast simulation of a generic collider experiment}},\ }\href {https://doi.org/10.1007/JHEP02(2014)057} {\bibfield  {journal} {\bibinfo  {journal} {JHEP}\ }\textbf {\bibinfo {volume} {02}},\ \bibinfo {pages} {057}},\ \Eprint {https://arxiv.org/abs/1307.6346} {arXiv:1307.6346 [hep-ex]} \BibitemShut {NoStop}%
\bibitem [{\citenamefont {Bierlich}\ \emph {et~al.}(2022)\citenamefont {Bierlich}, \citenamefont {Chakraborty}, \citenamefont {Desai}, \citenamefont {Gellersen}, \citenamefont {Helenius}, \citenamefont {Ilten}, \citenamefont {Lönnblad}, \citenamefont {Mrenna}, \citenamefont {Prestel}, \citenamefont {Preuss}, \citenamefont {Sjöstrand}, \citenamefont {Skands}, \citenamefont {Utheim},\ and\ \citenamefont {Verheyen}}]{pythia1}%
  \BibitemOpen
  \bibfield  {author} {\bibinfo {author} {\bibfnamefont {C.}~\bibnamefont {Bierlich}}, \bibinfo {author} {\bibfnamefont {S.}~\bibnamefont {Chakraborty}}, \bibinfo {author} {\bibfnamefont {N.}~\bibnamefont {Desai}}, \bibinfo {author} {\bibfnamefont {L.}~\bibnamefont {Gellersen}}, \bibinfo {author} {\bibfnamefont {I.}~\bibnamefont {Helenius}}, \bibinfo {author} {\bibfnamefont {P.}~\bibnamefont {Ilten}}, \bibinfo {author} {\bibfnamefont {L.}~\bibnamefont {Lönnblad}}, \bibinfo {author} {\bibfnamefont {S.}~\bibnamefont {Mrenna}}, \bibinfo {author} {\bibfnamefont {S.}~\bibnamefont {Prestel}}, \bibinfo {author} {\bibfnamefont {C.~T.}\ \bibnamefont {Preuss}}, \bibinfo {author} {\bibfnamefont {T.}~\bibnamefont {Sjöstrand}}, \bibinfo {author} {\bibfnamefont {P.}~\bibnamefont {Skands}}, \bibinfo {author} {\bibfnamefont {M.}~\bibnamefont {Utheim}},\ and\ \bibinfo {author} {\bibfnamefont {R.}~\bibnamefont {Verheyen}},\ }\href@noop {} {\bibinfo {title} {A comprehensive guide to the physics and usage of pythia 8.3}} (\bibinfo {year} {2022}),\ \Eprint {https://arxiv.org/abs/2203.11601} {arXiv:2203.11601 [hep-ph]} \BibitemShut {NoStop}%
\bibitem [{\citenamefont {Pappadopulo}\ \emph {et~al.}(2014)\citenamefont {Pappadopulo}, \citenamefont {Thamm}, \citenamefont {Torre},\ and\ \citenamefont {Wulzer}}]{hvt}%
  \BibitemOpen
  \bibfield  {author} {\bibinfo {author} {\bibfnamefont {D.}~\bibnamefont {Pappadopulo}}, \bibinfo {author} {\bibfnamefont {A.}~\bibnamefont {Thamm}}, \bibinfo {author} {\bibfnamefont {R.}~\bibnamefont {Torre}},\ and\ \bibinfo {author} {\bibfnamefont {A.}~\bibnamefont {Wulzer}},\ }\bibfield  {title} {\bibinfo {title} {{Heavy Vector Triplets: Bridging Theory and Data}},\ }\href {https://doi.org/10.1007/JHEP09(2014)060} {\bibfield  {journal} {\bibinfo  {journal} {JHEP}\ }\textbf {\bibinfo {volume} {09}},\ \bibinfo {pages} {060}},\ \Eprint {https://arxiv.org/abs/1402.4431} {arXiv:1402.4431 [hep-ph]} \BibitemShut {NoStop}%
\bibitem [{\citenamefont {Cacciari}\ \emph {et~al.}(2008)\citenamefont {Cacciari}, \citenamefont {Salam},\ and\ \citenamefont {Soyez}}]{antikt}%
  \BibitemOpen
  \bibfield  {author} {\bibinfo {author} {\bibfnamefont {M.}~\bibnamefont {Cacciari}}, \bibinfo {author} {\bibfnamefont {G.~P.}\ \bibnamefont {Salam}},\ and\ \bibinfo {author} {\bibfnamefont {G.}~\bibnamefont {Soyez}},\ }\bibfield  {title} {\bibinfo {title} {The anti-kt jet clustering algorithm},\ }\href {https://doi.org/10.1088/1126-6708/2008/04/063} {\bibfield  {journal} {\bibinfo  {journal} {Journal of High Energy Physics}\ }\textbf {\bibinfo {volume} {2008}},\ \bibinfo {pages} {063} (\bibinfo {year} {2008})}\BibitemShut {NoStop}%
\bibitem [{\citenamefont {Cacciari}\ \emph {et~al.}(2012)\citenamefont {Cacciari}, \citenamefont {Salam},\ and\ \citenamefont {Soyez}}]{fastjet}%
  \BibitemOpen
  \bibfield  {author} {\bibinfo {author} {\bibfnamefont {M.}~\bibnamefont {Cacciari}}, \bibinfo {author} {\bibfnamefont {G.~P.}\ \bibnamefont {Salam}},\ and\ \bibinfo {author} {\bibfnamefont {G.}~\bibnamefont {Soyez}},\ }\bibfield  {title} {\bibinfo {title} {Fastjet user manual},\ }\href {https://doi.org/10.1140/epjc/s10052-012-1896-2} {\bibfield  {journal} {\bibinfo  {journal} {The European Physical Journal C}\ }\textbf {\bibinfo {volume} {72}},\ \bibinfo {pages} {1896} (\bibinfo {year} {2012})}\BibitemShut {NoStop}%
\bibitem [{\citenamefont {Schuld}\ \emph {et~al.}(2020)\citenamefont {Schuld}, \citenamefont {Bocharov}, \citenamefont {Svore},\ and\ \citenamefont {Wiebe}}]{strong_ent}%
  \BibitemOpen
  \bibfield  {author} {\bibinfo {author} {\bibfnamefont {M.}~\bibnamefont {Schuld}}, \bibinfo {author} {\bibfnamefont {A.}~\bibnamefont {Bocharov}}, \bibinfo {author} {\bibfnamefont {K.~M.}\ \bibnamefont {Svore}},\ and\ \bibinfo {author} {\bibfnamefont {N.}~\bibnamefont {Wiebe}},\ }\bibfield  {title} {\bibinfo {title} {Circuit-centric quantum classifiers},\ }\href {https://doi.org/10.1103/PhysRevA.101.032308} {\bibfield  {journal} {\bibinfo  {journal} {Phys. Rev. A}\ }\textbf {\bibinfo {volume} {101}},\ \bibinfo {pages} {032308} (\bibinfo {year} {2020})}\BibitemShut {NoStop}%
\bibitem [{\citenamefont {Agarap}(2019)}]{relu}%
  \BibitemOpen
  \bibfield  {author} {\bibinfo {author} {\bibfnamefont {A.~F.}\ \bibnamefont {Agarap}},\ }\href {https://arxiv.org/abs/1803.08375} {\bibinfo {title} {Deep learning using rectified linear units (relu)}} (\bibinfo {year} {2019}),\ \Eprint {https://arxiv.org/abs/1803.08375} {arXiv:1803.08375 [cs.NE]} \BibitemShut {NoStop}%
\bibitem [{\citenamefont {Team}()}]{pytorch}%
  \BibitemOpen
  \bibfield  {author} {\bibinfo {author} {\bibfnamefont {P.}~\bibnamefont {Team}},\ }\href {https://pytorch.org} {\bibinfo {title} {Pytorch}}\BibitemShut {NoStop}%
\bibitem [{\citenamefont {Fey}\ and\ \citenamefont {Lenssen}(2019)}]{pygeo}%
  \BibitemOpen
  \bibfield  {author} {\bibinfo {author} {\bibfnamefont {M.}~\bibnamefont {Fey}}\ and\ \bibinfo {author} {\bibfnamefont {J.~E.}\ \bibnamefont {Lenssen}},\ }\bibfield  {title} {\bibinfo {title} {Fast graph representation learning with {PyTorch Geometric}},\ }in\ \href@noop {} {\emph {\bibinfo {booktitle} {ICLR Workshop on Representation Learning on Graphs and Manifolds}}}\ (\bibinfo {year} {2019})\BibitemShut {NoStop}%
\bibitem [{\citenamefont {Bergholm}\ \emph {et~al.}(2022)\citenamefont {Bergholm}, \citenamefont {Izaac}, \citenamefont {Schuld}, \citenamefont {Gogolin}, \citenamefont {Ahmed}, \citenamefont {Ajith}, \citenamefont {Alam}, \citenamefont {Alonso-Linaje}, \citenamefont {AkashNarayanan}, \citenamefont {Asadi}, \citenamefont {Arrazola}, \citenamefont {Azad}, \citenamefont {Banning}, \citenamefont {Blank}, \citenamefont {Bromley}, \citenamefont {Cordier}, \citenamefont {Ceroni}, \citenamefont {Delgado}, \citenamefont {Matteo}, \citenamefont {Dusko}, \citenamefont {Garg}, \citenamefont {Guala}, \citenamefont {Hayes}, \citenamefont {Hill}, \citenamefont {Ijaz}, \citenamefont {Isacsson}, \citenamefont {Ittah}, \citenamefont {Jahangiri}, \citenamefont {Jain}, \citenamefont {Jiang}, \citenamefont {Khandelwal}, \citenamefont {Kottmann}, \citenamefont {Lang}, \citenamefont {Lee}, \citenamefont {Loke}, \citenamefont {Lowe}, \citenamefont {McKiernan}, \citenamefont {Meyer}, \citenamefont {Montañez-Barrera}, \citenamefont {Moyard}, \citenamefont {Niu}, \citenamefont {O'Riordan}, \citenamefont {Oud}, \citenamefont {Panigrahi}, \citenamefont {Park}, \citenamefont {Polatajko}, \citenamefont {Quesada}, \citenamefont {Roberts}, \citenamefont {Sá}, \citenamefont {Schoch}, \citenamefont {Shi}, \citenamefont {Shu}, \citenamefont {Sim}, \citenamefont {Singh}, \citenamefont {Strandberg}, \citenamefont {Soni}, \citenamefont {Száva}, \citenamefont {Thabet}, \citenamefont {Vargas-Hernández}, \citenamefont {Vincent}, \citenamefont {Vitucci}, \citenamefont {Weber}, \citenamefont {Wierichs}, \citenamefont {Wiersema}, \citenamefont {Willmann}, \citenamefont {Wong}, \citenamefont {Zhang},\ and\ \citenamefont {Killoran}}]{pennylane}%
  \BibitemOpen
  \bibfield  {author} {\bibinfo {author} {\bibfnamefont {V.}~\bibnamefont {Bergholm}}, \bibinfo {author} {\bibfnamefont {J.}~\bibnamefont {Izaac}}, \bibinfo {author} {\bibfnamefont {M.}~\bibnamefont {Schuld}}, \bibinfo {author} {\bibfnamefont {C.}~\bibnamefont {Gogolin}}, \bibinfo {author} {\bibfnamefont {S.}~\bibnamefont {Ahmed}}, \bibinfo {author} {\bibfnamefont {V.}~\bibnamefont {Ajith}}, \bibinfo {author} {\bibfnamefont {M.~S.}\ \bibnamefont {Alam}}, \bibinfo {author} {\bibfnamefont {G.}~\bibnamefont {Alonso-Linaje}}, \bibinfo {author} {\bibfnamefont {B.}~\bibnamefont {AkashNarayanan}}, \bibinfo {author} {\bibfnamefont {A.}~\bibnamefont {Asadi}}, \bibinfo {author} {\bibfnamefont {J.~M.}\ \bibnamefont {Arrazola}}, \bibinfo {author} {\bibfnamefont {U.}~\bibnamefont {Azad}}, \bibinfo {author} {\bibfnamefont {S.}~\bibnamefont {Banning}}, \bibinfo {author} {\bibfnamefont {C.}~\bibnamefont {Blank}}, \bibinfo {author} {\bibfnamefont {T.~R.}\ \bibnamefont {Bromley}}, \bibinfo {author} {\bibfnamefont {B.~A.}\ \bibnamefont {Cordier}}, \bibinfo {author} {\bibfnamefont {J.}~\bibnamefont {Ceroni}}, \bibinfo {author} {\bibfnamefont {A.}~\bibnamefont {Delgado}}, \bibinfo {author} {\bibfnamefont {O.~D.}\ \bibnamefont {Matteo}}, \bibinfo {author} {\bibfnamefont {A.}~\bibnamefont {Dusko}}, \bibinfo {author} {\bibfnamefont {T.}~\bibnamefont {Garg}}, \bibinfo {author} {\bibfnamefont {D.}~\bibnamefont {Guala}}, \bibinfo {author} {\bibfnamefont {A.}~\bibnamefont {Hayes}}, \bibinfo {author} {\bibfnamefont {R.}~\bibnamefont {Hill}}, \bibinfo {author} {\bibfnamefont {A.}~\bibnamefont {Ijaz}}, \bibinfo {author} {\bibfnamefont {T.}~\bibnamefont {Isacsson}}, \bibinfo {author} {\bibfnamefont {D.}~\bibnamefont {Ittah}}, \bibinfo {author} {\bibfnamefont {S.}~\bibnamefont {Jahangiri}}, \bibinfo {author} {\bibfnamefont {P.}~\bibnamefont {Jain}}, \bibinfo {author} {\bibfnamefont {E.}~\bibnamefont {Jiang}}, \bibinfo {author} {\bibfnamefont {A.}~\bibnamefont {Khandelwal}}, \bibinfo {author} {\bibfnamefont {K.}~\bibnamefont {Kottmann}}, \bibinfo {author} {\bibfnamefont {R.~A.}\ \bibnamefont {Lang}}, \bibinfo {author} {\bibfnamefont {C.}~\bibnamefont {Lee}}, \bibinfo {author} {\bibfnamefont {T.}~\bibnamefont {Loke}}, \bibinfo {author} {\bibfnamefont {A.}~\bibnamefont {Lowe}}, \bibinfo {author} {\bibfnamefont {K.}~\bibnamefont {McKiernan}}, \bibinfo {author} {\bibfnamefont {J.~J.}\ \bibnamefont {Meyer}}, \bibinfo {author} {\bibfnamefont {J.~A.}\ \bibnamefont {Montañez-Barrera}}, \bibinfo {author} {\bibfnamefont {R.}~\bibnamefont {Moyard}}, \bibinfo {author} {\bibfnamefont {Z.}~\bibnamefont {Niu}}, \bibinfo {author} {\bibfnamefont {L.~J.}\ \bibnamefont {O'Riordan}}, \bibinfo {author} {\bibfnamefont {S.}~\bibnamefont {Oud}}, \bibinfo {author} {\bibfnamefont {A.}~\bibnamefont {Panigrahi}}, \bibinfo {author} {\bibfnamefont {C.-Y.}\ \bibnamefont {Park}}, \bibinfo {author} {\bibfnamefont {D.}~\bibnamefont {Polatajko}}, \bibinfo {author} {\bibfnamefont {N.}~\bibnamefont {Quesada}}, \bibinfo {author} {\bibfnamefont {C.}~\bibnamefont {Roberts}}, \bibinfo {author} {\bibfnamefont {N.}~\bibnamefont {Sá}}, \bibinfo {author} {\bibfnamefont {I.}~\bibnamefont {Schoch}}, \bibinfo {author} {\bibfnamefont {B.}~\bibnamefont {Shi}}, \bibinfo {author} {\bibfnamefont {S.}~\bibnamefont {Shu}}, \bibinfo {author} {\bibfnamefont {S.}~\bibnamefont {Sim}}, \bibinfo {author} {\bibfnamefont {A.}~\bibnamefont {Singh}}, \bibinfo {author} {\bibfnamefont {I.}~\bibnamefont {Strandberg}}, \bibinfo {author} {\bibfnamefont {J.}~\bibnamefont {Soni}}, \bibinfo {author} {\bibfnamefont {A.}~\bibnamefont {Száva}}, \bibinfo {author} {\bibfnamefont {S.}~\bibnamefont {Thabet}}, \bibinfo {author} {\bibfnamefont {R.~A.}\ \bibnamefont {Vargas-Hernández}}, \bibinfo {author} {\bibfnamefont {T.}~\bibnamefont {Vincent}}, \bibinfo {author} {\bibfnamefont {N.}~\bibnamefont {Vitucci}}, \bibinfo {author} {\bibfnamefont {M.}~\bibnamefont {Weber}}, \bibinfo {author} {\bibfnamefont {D.}~\bibnamefont {Wierichs}}, \bibinfo {author} {\bibfnamefont {R.}~\bibnamefont {Wiersema}}, \bibinfo {author} {\bibfnamefont {M.}~\bibnamefont {Willmann}}, \bibinfo {author} {\bibfnamefont {V.}~\bibnamefont {Wong}}, \bibinfo {author} {\bibfnamefont {S.}~\bibnamefont {Zhang}},\ and\ \bibinfo {author} {\bibfnamefont {N.}~\bibnamefont {Killoran}},\ }\href@noop {} {\bibinfo {title} {Pennylane: Automatic differentiation of hybrid quantum-classical computations}} (\bibinfo {year} {2022}),\ \Eprint {https://arxiv.org/abs/1811.04968} {arXiv:1811.04968 [quant-ph]} \BibitemShut {NoStop}%
\bibitem [{\citenamefont {Kingma}\ and\ \citenamefont {Ba}(2017)}]{adam}%
  \BibitemOpen
  \bibfield  {author} {\bibinfo {author} {\bibfnamefont {D.~P.}\ \bibnamefont {Kingma}}\ and\ \bibinfo {author} {\bibfnamefont {J.}~\bibnamefont {Ba}},\ }\href {https://arxiv.org/abs/1412.6980} {\bibinfo {title} {Adam: A method for stochastic optimization}} (\bibinfo {year} {2017}),\ \Eprint {https://arxiv.org/abs/1412.6980} {arXiv:1412.6980 [cs.LG]} \BibitemShut {NoStop}%
\bibitem [{\citenamefont {Javadi-Abhari}\ \emph {et~al.}(2024)\citenamefont {Javadi-Abhari}, \citenamefont {Treinish}, \citenamefont {Krsulich}, \citenamefont {Wood}, \citenamefont {Lishman}, \citenamefont {Gacon}, \citenamefont {Martiel}, \citenamefont {Nation}, \citenamefont {Bishop}, \citenamefont {Cross}, \citenamefont {Johnson},\ and\ \citenamefont {Gambetta}}]{ibmq}%
  \BibitemOpen
  \bibfield  {author} {\bibinfo {author} {\bibfnamefont {A.}~\bibnamefont {Javadi-Abhari}}, \bibinfo {author} {\bibfnamefont {M.}~\bibnamefont {Treinish}}, \bibinfo {author} {\bibfnamefont {K.}~\bibnamefont {Krsulich}}, \bibinfo {author} {\bibfnamefont {C.~J.}\ \bibnamefont {Wood}}, \bibinfo {author} {\bibfnamefont {J.}~\bibnamefont {Lishman}}, \bibinfo {author} {\bibfnamefont {J.}~\bibnamefont {Gacon}}, \bibinfo {author} {\bibfnamefont {S.}~\bibnamefont {Martiel}}, \bibinfo {author} {\bibfnamefont {P.~D.}\ \bibnamefont {Nation}}, \bibinfo {author} {\bibfnamefont {L.~S.}\ \bibnamefont {Bishop}}, \bibinfo {author} {\bibfnamefont {A.~W.}\ \bibnamefont {Cross}}, \bibinfo {author} {\bibfnamefont {B.~R.}\ \bibnamefont {Johnson}},\ and\ \bibinfo {author} {\bibfnamefont {J.~M.}\ \bibnamefont {Gambetta}},\ }\href {https://doi.org/10.48550/arXiv.2405.08810} {\bibinfo {title} {Quantum computing with {Q}iskit}} (\bibinfo {year} {2024}),\ \Eprint {https://arxiv.org/abs/2405.08810} {arXiv:2405.08810 [quant-ph]} \BibitemShut {NoStop}%
\end{thebibliography}%

\end{document}